\documentclass[a4paper]{article}
\usepackage{graphicx}

\begin{document}

\begin{center}
{\Large \bf Improved Systematic of $pp$ Elastic Scattering Data}
\end{center}

\begin{center}
{V. Uzhinsky\footnote{On leave of LIT, JINR, Dubna, Russia} and
 A. Galoyan\footnote{On leave of VBLHEP, JINR, Dubna, Russia}}
\end{center}

\begin{center}
{CERN, Geneva, Switzerland}
\end{center}

\begin{center}
\begin{minipage}{12cm}
Unified systematic of elastic scattering data (USESD) proposed by the authors (arxiv:1111.4984 [hep-ph])
is based on symmetrized 2-dimensional Fermi distribution for $pp$ elastic scattering amplitude in
the impact parameter representation. It allows to describe differential cross sections of the
reactions up to $|t|\sim$ 1.75 (GeV/c)$^2$. To extend it to higher $|t|$ values we consider
a two coherent exponential parametrization of the cross sections and show that it cannot describe
the cross sections at small $|t|$ at $P_{lab}>$ 10 GeV/c. We extract a description of high $|t|$
region from the parameterization and couple it with USESD. As a result, we obtained a good
description of all $pp$ elastic scattering data at $P_{lab}>$ 10 GeV/c. It can be easily used
in Glauber Monte Carlo codes for calculations of nucleus-nucleus interaction properties.
\end{minipage}
\end{center}

\section*{Introduction}
Recently, authors of the paper \cite{Asymptotica} remembered the old parameterization \cite{TwoExp}
of $pp$ elastic scattering amplitude:
\begin{equation}
f(s,t)=i\left[ A_1\ e^{B_1t/2}\ + \ A_2\ e^{i\phi}e^{B_2t/2}\right],
\label{Eq1}
\end{equation}
and applied it for a fitting of the Totem Collaboration data \cite{Totem} on elastic $pp$ scattering at
$\sqrt{s_{pp}}=7$ TeV. The amplitude was proposed in 1973 by R.J.N. Phillips and V.D. Barger.
Nearly at the same time, it was independently proposed and tested in papers
\cite{PbarP1,PbarP2,PbarP3} where it was applied for a fitting  of $\bar pp$ elastic scattering.

The authors of the paper \cite{TwoExp} analyzed
only $pp$ experimental data at $P_{lab}=$12, 14.2, 19.2, 24, 29.7 GeV/c and at $\sqrt{s_{pp}}=53$ GeV.
No $\chi^2/NoF$ and the parameter's errors were given by them. So,
a quality of the parameterization is unknown. It's predictive power is unclear.

The authors of the papers \cite{PbarP1,PbarP2,PbarP3} fitted antiproton-proton scattering
experimental data at $P_{lab}=$1.11 -- 16 \cite{PbarP3} GeV/c (twenty-eight sets of data). A fitting
of eight selected data sets gave smallest errors. Three coherent exponentials was proposed and used in
the paper \cite{PbarP4}. There is not any reclamation to technical details of the fittings.
We repeated the work and published the results in the paper \cite{OurPaperPbarP} where
we proposed energy dependencies of the parameters.

An advantage of Eq.~\ref{Eq1} is that it can be easily applied in Glauber model calculations of hadron-nucleus
and nucleus-nucleus interaction properties at high and super high energies, and it can improve
an exactness of the calculations. Usually, only one exponent is used in the calculations. The aim of our
present paper is a checking of Eq.~\ref{Eq1} in a wide range of energies.

There is another parameterization of high energy $pp$ elastic scattering data proposed by us in
the paper \cite{USESD}:
\begin{equation}
f(s,t)=i\ A\left[R^2\frac{\pi d q}{sinh(\pi d q)}\frac{J_1(R q)}{R q}\ +\
\frac{1}{2q^2}\frac{\pi d q}{sinh(\pi d q)}\left( \frac{\pi d q}{tanh(\pi d q)}-1\right)\ J_0(R q)+...\right],
\ \ \ q=\sqrt{-t}.
\label{Eq2}
\end{equation}
It is a Fourier-Bessel transform of a symmetrized 2-dimensional Fermi-function \cite{SFermi},
\begin{equation}
\gamma(\vec b)=A\left[ \frac{1}{1+e^{(b-R)/d}} + \frac{1}{1+e^{-(b+R)/d}}-1\right],
\label{Eq3}
\end{equation}
where $\gamma(\vec b)$ is the elastic scattering amplitude in the impact parameter representation.

A possibility to describe a high $P_T$ elastic $pp$ scattering was considered in Ref.~\cite{USESD}
following papers \cite{DL,Martynov}, but we were not satisfied by results. Thus, below we undertake
an effort to combine our approach with two coherent exponential one. As a result, we have obtained
an improved parameterization of $pp$ elastic scattering data.

\section{Validation of two coherent exponential expression}
A direct application of Eq.~\ref{Eq1} is complicated by strong parameter correlation. Thus, we
write a differential elastic scattering cross section using Eq.~\ref{Eq1} as:
\begin{equation}
\frac{d\sigma}{dt}=A_1\left( e^{B_1t/2}-A_2\ e^{B_2t/2}\right)^2\ + \ A_3\ e^{B_2t},
\label{Eq4}
\end{equation}
and fit experimental data at $P_{lab}>$ 10 (GeV/c). We took only 45 from 64 sets of experimental data
which gave meaningful results. Fit results of the selected data sets are presented in Fig.~1, 2.
\begin{figure}[cbth]
\includegraphics[width=160mm,height=45mm,clip]{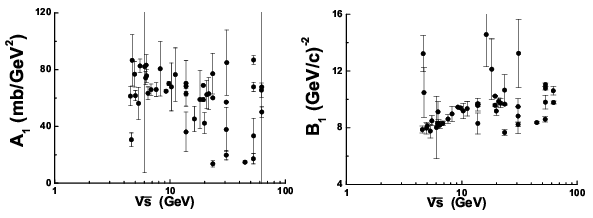}
\caption{Energy dependencies of the parameters $A_1$ and $B_1$.}
\label{Fig1}
\vspace{5mm}
\includegraphics[width=160mm,height=45mm,clip]{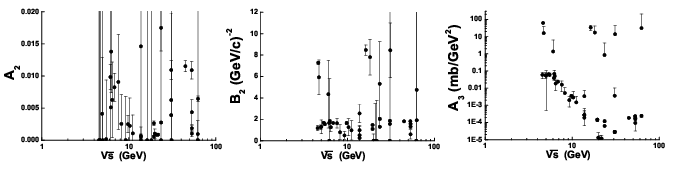}
\caption{Energy dependencies of the parameters $A_2$, $B_2$ and $A_3$.}
\label{Fig2}
\end{figure}

As seen, the results are rather unstable. Though, $\chi^2/NoF =$2572/1803$\simeq$1.43.
Of course, selecting only some experimental data one can obtain stable values of the parameters,
as it was done in Ref. \cite{TwoExp}.

More serious drawback of the parameterization \ref{Eq1} is that it cannot describe small angle
scattering data, especially, total cross sections, $\sigma_{tot}$, and ratio of real and imaginary
parts of elastic scattering amplitude at $t=0$, $\rho(0)=Re f(0,s)/Im f(0,s)$.
\begin{figure}[cbth]
\includegraphics[width=160mm,height=35mm,clip]{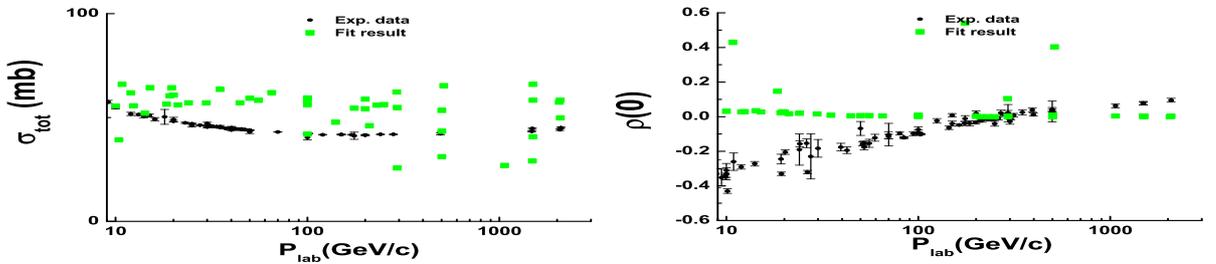}
\caption{$\sigma_{tot}$ and $Re f/Im f$ as functions of energy. Black points are experimental data from
PDG data base \protect\cite{PDG}. Squares are calculation results without error bars.}
\label{Fig3}
\end{figure}

Thus, we cannot recommend to use the parameterization for practical calculations.

\section{Combination of the approaches}
Another situation takes place with the parameterization \ref{Eq2}. It describe small angle scattering
data rather well, but falls down at large $|t|$. At the same time, the parameterization \ref{Eq1}
describes large $|t|$ data quite well. Thus, taking into account that $\phi \sim \pi$  in Eq.~\ref{Eq1}
we can combine the parameterizations as:
\begin{equation}
Im f(s,t)=A_1\left[ R^2\frac{\pi d q}{sinh(\pi d q)}\frac{J_1(R q)}{R q}\ +\
\frac{1}{2q^2}\frac{\pi d q}{sinh(\pi d q)}\left( \frac{\pi d q}{tanh(\pi d q)}-1\right)\ J_0(R q)
\right],
\label{Eq5}
\end{equation}
\begin{equation}
Re f(s,t)=A_1 \cdot \rho \cdot (R^2/2+\pi^2d^2/6)\frac{\pi d q}{sinh(\pi d q)}J_0(R q)\ +\
A_2\ e^{-B_2q^2/2},
\label{Eq6}
\end{equation}
\begin{equation}
\rho = 0.135-3/\sqrt{s}+4/s+80/s^3,
\label{Eq7}
\end{equation}
\begin{equation}
\frac{d\sigma}{dt}=10\cdot 25.68185 \cdot \pi \cdot (Imf^2+ Ref^2) \ \ \ [mb/(GeV/c)^2],
\label{Eq8}
\end{equation}
where $q=\sqrt{-t\ 25.68185}$ $[fm^{-1}]$, $t$ is 4-momentum transfer $[(GeV/c)^2]$,
$R$ and $d$ are in $[fm]$, $A_2$ is in $[fm^2]$. The parameter $A_1$ was introduced in order
to take into account uncertainties in absolute normalization of experimental data. Experimental
data on $\rho$ \cite{PDG} were approximated by Eq.~\ref{Eq7}.

There are a lot of experimental data, but most of them are small angle ones. They do not allow
an unambiguous determination of the parameters. Thus, we select data at $p_{lab}=$14.2, 19.2, 24, 200,
293, 501 GeV/c \cite{Plab}, and $\sqrt{s}=$44.7, 52.9, 62.5, and 7000 GeV \cite{SqrtS,Totem} in which regions
of minimum at $|t|\sim$ 1.5 (GeV/c)$^2$ and high $|t|$ tails are presented.

Results of the fit ($\chi^2/NoF=1.51$) of the parameters $A_2$ and $B_2$ are presented in Fig.~4.
\begin{figure}[cbth]
\includegraphics[width=160mm,height=50mm,clip]{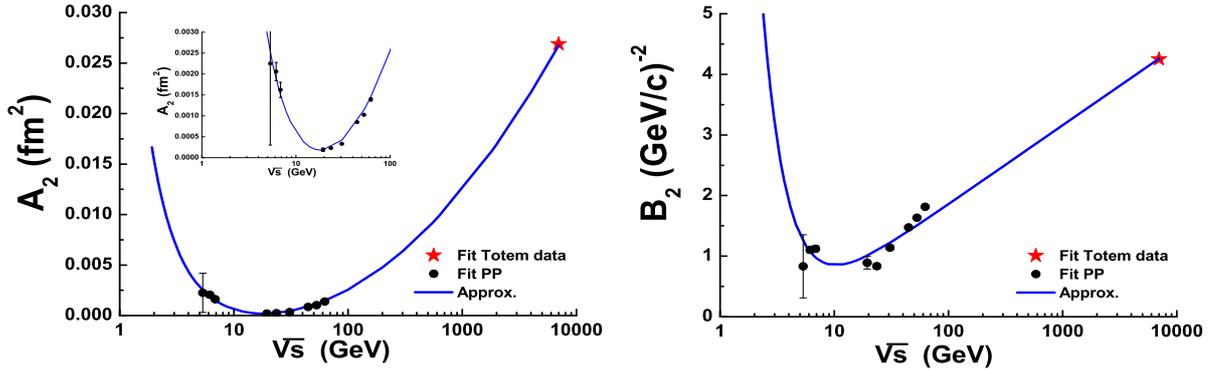}
\caption{Points are fit results for $A_2$ and $B_2$. Lines are approximations of the energy dependencies.}
\label{Fig4}
\end{figure}

The energy dependencies of the parameters can be approximated as:
\begin{equation}
A_2=1.77\ 10^{-4}\ [\log{(s/225)}]^2\ + \ 0.05/s, \ \ \
B_2=0.283*\log{(s)}\ +\ 30/s\ - \ 0.75.
\label{Eq9}
\end{equation}
A change of the sign of $A_2$ in Eq.~\ref{Eq6} from "+" to "-" makes the fit worse.

Having the result for high $|t|$ values we can now determine "soft" parameters ($A_1$, $R$ and $d$)
more exactly than it was done before. In our paper \cite{USESD} we found the parameters at an
artificial restriction on "soft" amplitude application region (Eq.~\ref{Eq5}), $|t| <$ 1.75 (GeV/c)$^2$.
Now we fix $A_2$ and $B_2$ by Eq.~\ref{Eq9}, and fit other parameters using 64 sets of experimental data
(see references in \cite{USESD}). Results of the fitting are presented in Fig.~5 ($\chi^2/NoF=$1.96).

As seen, energy dependencies of $R$ and $d$ are determined rather well. They can be parameterized as:
\begin{equation}
R=0.9/s^{0.25} + 0.053 \log(s), \ \ \ d=0.379 - 0.26/s^{0.25}.
\label{Eq10}
\end{equation}

Fitted values of $A_1$ scatter rather strong especially at low energies. We propose the following
parameterization of its energy dependence:
\begin{equation}
A_1=1.077 - 0.175\ e^{-0.001 s^{0.5}} +0.45/s^{0.25},
\label{Eq11}
\end{equation}

According to Eq.~\ref{Eq2} total cross section is given as:
\begin{equation}
\sigma_{tot}=2\pi\ A_1\ (R^2+\pi^2d^2/3).
\label{Eq12}
\end{equation}
If we use only fitting results for $A_1$, $R$, and $d$, we obtain green points in Fig.~6
shown without error bars. As seen, they are scattered rather strong.

\begin{figure}[cbth]
\includegraphics[width=160mm,height=50mm,clip]{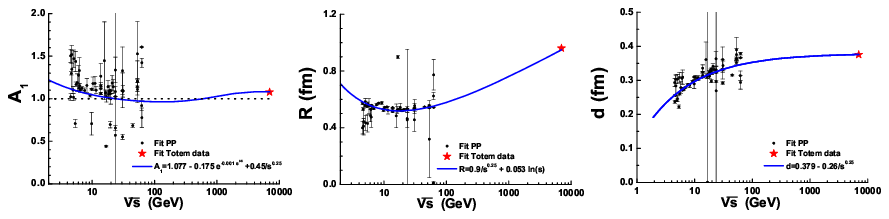}
\caption{Points are fit results. Lines are approximations of the energy dependencies.}
\label{Fig5}
\vspace{5mm}
\includegraphics[width=160mm,height=50mm,clip]{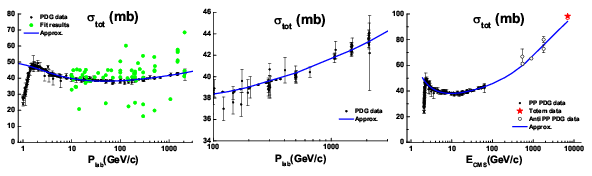}
\caption{Open and closed points are experimental data from PDG \cite{PDG} data base.
         Lines are calculation results using approximations of $A_1$(see text).}
\label{Fig6}
\end{figure}

Using the approximations given by Eqs. \ref{Eq7}, \ref{Eq9}, and \ref{Eq10} ($A_1$ was not fixed)
we have $\chi^2/NoF=8179/2256\simeq 4.79$. A fixing of $A_1$ by Eq.~\ref{Eq11} leads to an increasing  %3.63
of $\chi^2/NoF$ in 2 times. A quality of the experimental data descriptions is presented in Appendix A.

\section{Application of USESD in Glauber Monte Carlo codes}
Glauber Monte Carlo codes \cite{PhobosMC,Polyaki} (see review in \cite{GLmc}) calculate
various properties of inelastic nucleus-nucleus interaction such as: number of participating nucleons,
multiplicity of intra-nuclear collision, impact parameter distributions and so on. Mainly they use
so-called "nucleon inelastic overlap function", $g(b)$, in the simplest form:
\begin{equation}
g(b)=\theta(R_{int}-b), \ \ \ R_{int}=\sqrt{\sigma_{in}/\pi}.
\label{Eq14}
\end{equation}

In the Glauber theory the function is given as:
\begin{equation}
g(b)=\gamma(b)+\gamma^*(b)-|\gamma(b)|^2.
\label{Eq15}
\end{equation}

According to Eqs.~\ref{Eq5}, \ref{Eq6},
\begin{equation}
\gamma(b)=A_1\left\{ \frac{e^{-(b-\tilde{R})/d}}{1+e^{-(b-\tilde{R})/d}} +
                    \frac{1}                   {1+e^{-(b+\tilde{R})/d}}-1\right\}-
\label{Eq16}
\end{equation}
$$
-i\ \rho A_1\frac{\tilde{R}^2/2+\pi^2d^2/6}{\tilde{R}d}
        \left\{ \frac{e^{-(b-\tilde{R})/d}}{\left[1+e^{-(b-\tilde{R})/d}\right]^2} +
                 \frac{e^{-(b+\tilde{R})/d}}{\left[1+e^{-(b+\tilde{R})/d}\right]^2}
    \right\}-
$$
$$
-i\ \frac{A_2}{2\pi B_2\ 25.64} e^{-b^2/(2\ B_2\ 25.64)}
$$
\begin{equation}
\tilde{R}=R\ + \ (0.07+d+0.2d^2)\ e^{-1.2 R/d}.
\label{Eq17}
\end{equation}

A reason of the introduction of $\tilde{R}$ is that Eq.~\ref{Eq2} was obtained from Eq.~\ref{Eq3}
assuming $d/R << 1$. As seen from the fitting results it is not true for $pp$ interactions. Thus,
the correction of Eq.~\ref{Eq17} was found at numerical investigation.

\section*{Conclusion}
USESD is enlarged by the simple description of high $|t|$ elastic scattering. The new parameters
have been determined. Exact formulae are presented.

\noindent{\bf Appendix A: Comparison of experimental data on $pp$-interactions with USESD parameterization}

\begin{figure}[cbth]
%-------------------------------------------------------
\includegraphics[width=75mm,height=70mm,clip]{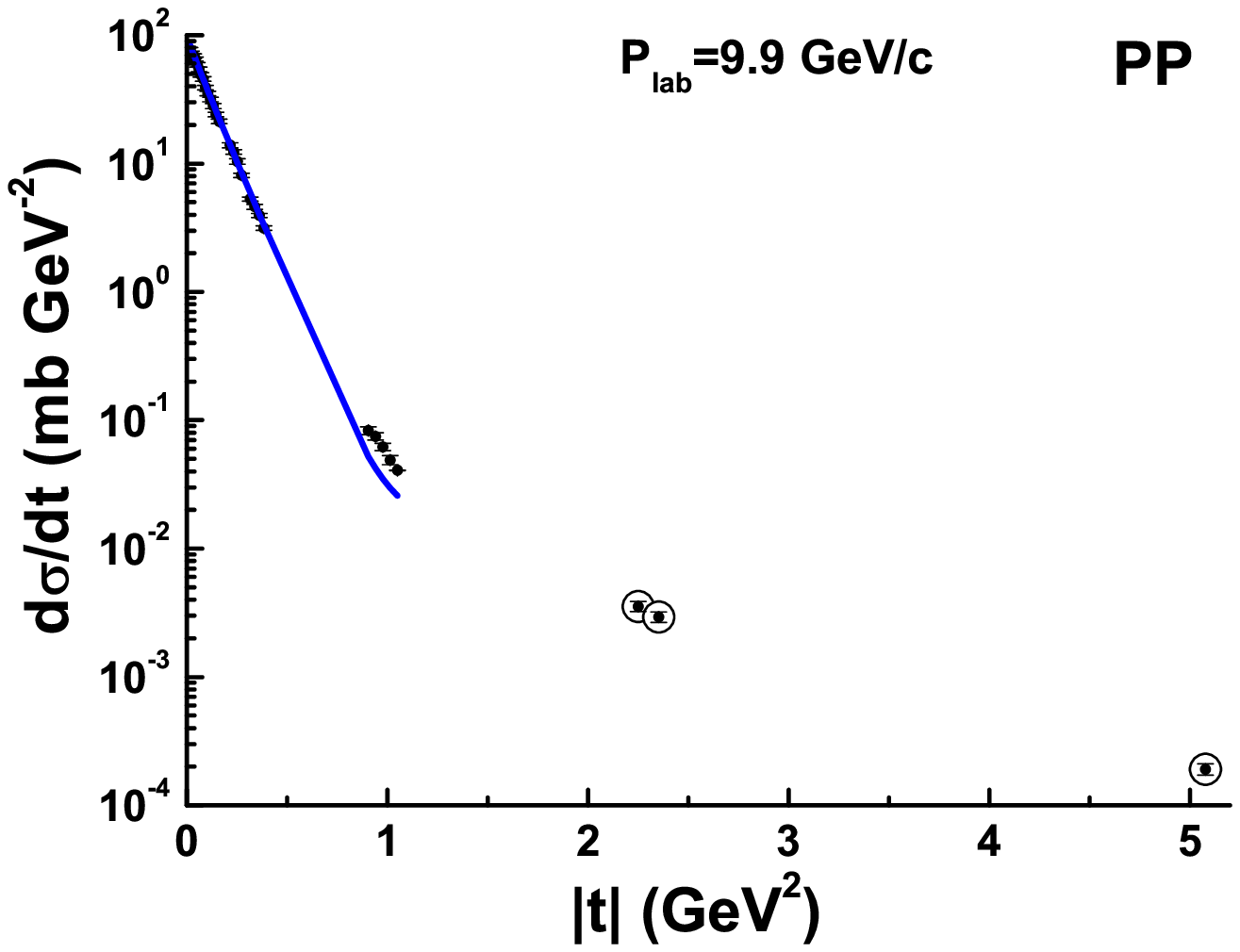}\hspace{5mm}\includegraphics[width=75mm,height=70mm,clip]{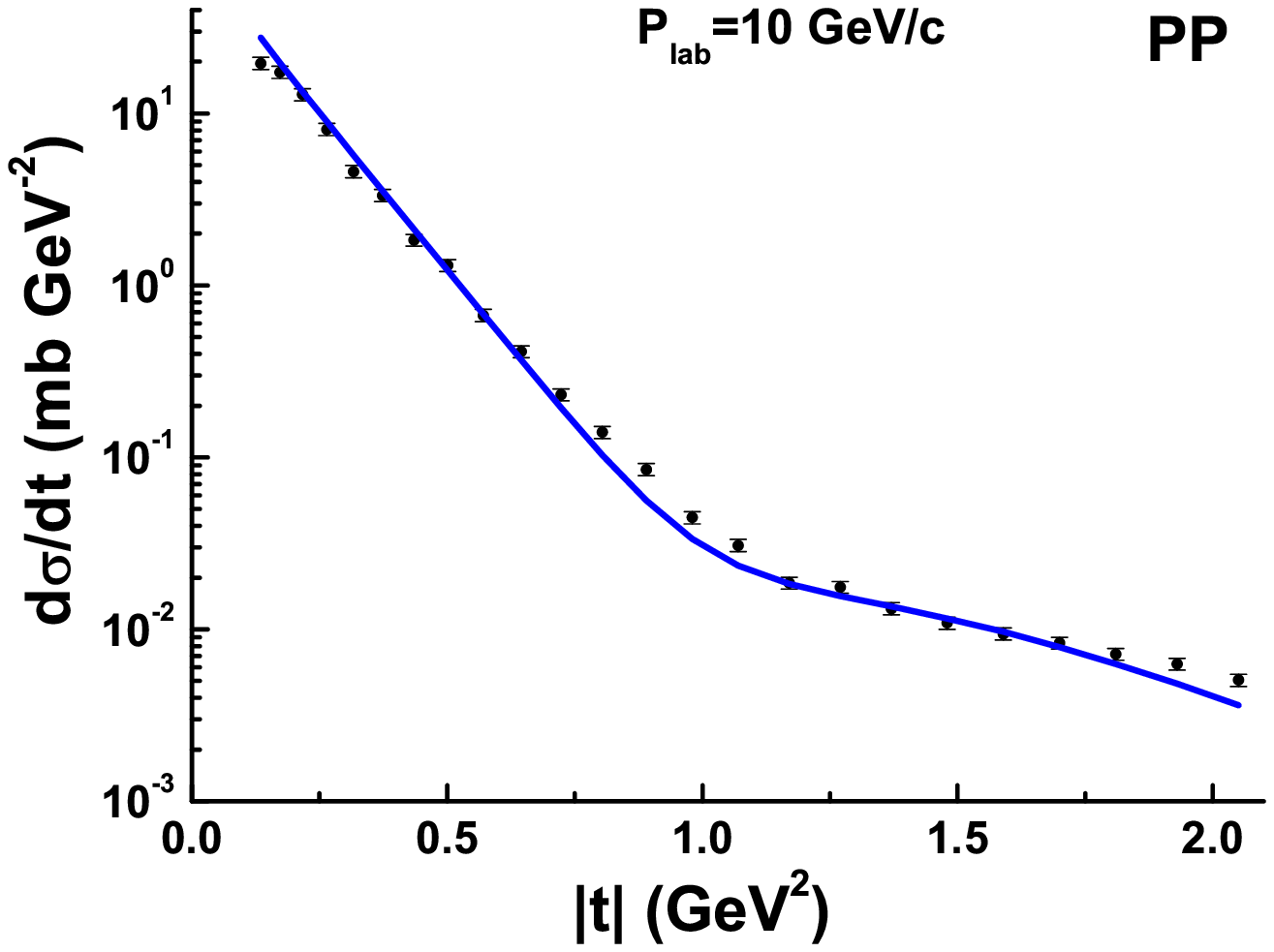}
\begin{minipage}{75mm}
{
\caption{The points are the experimental data by R.M. Edelstein et al., Phys. Rev. {\bf D5} (1972) 1073.}
}
\end{minipage}
\hspace{5mm}
\begin{minipage}{75mm}
{
\caption{The points are the experimental data by J.V. Allaby et al., Nucl. Phys. {\bf B52} (1973) 316.}
}
\end{minipage}
%-------------------------------------------------------
\includegraphics[width=75mm,height=70mm,clip]{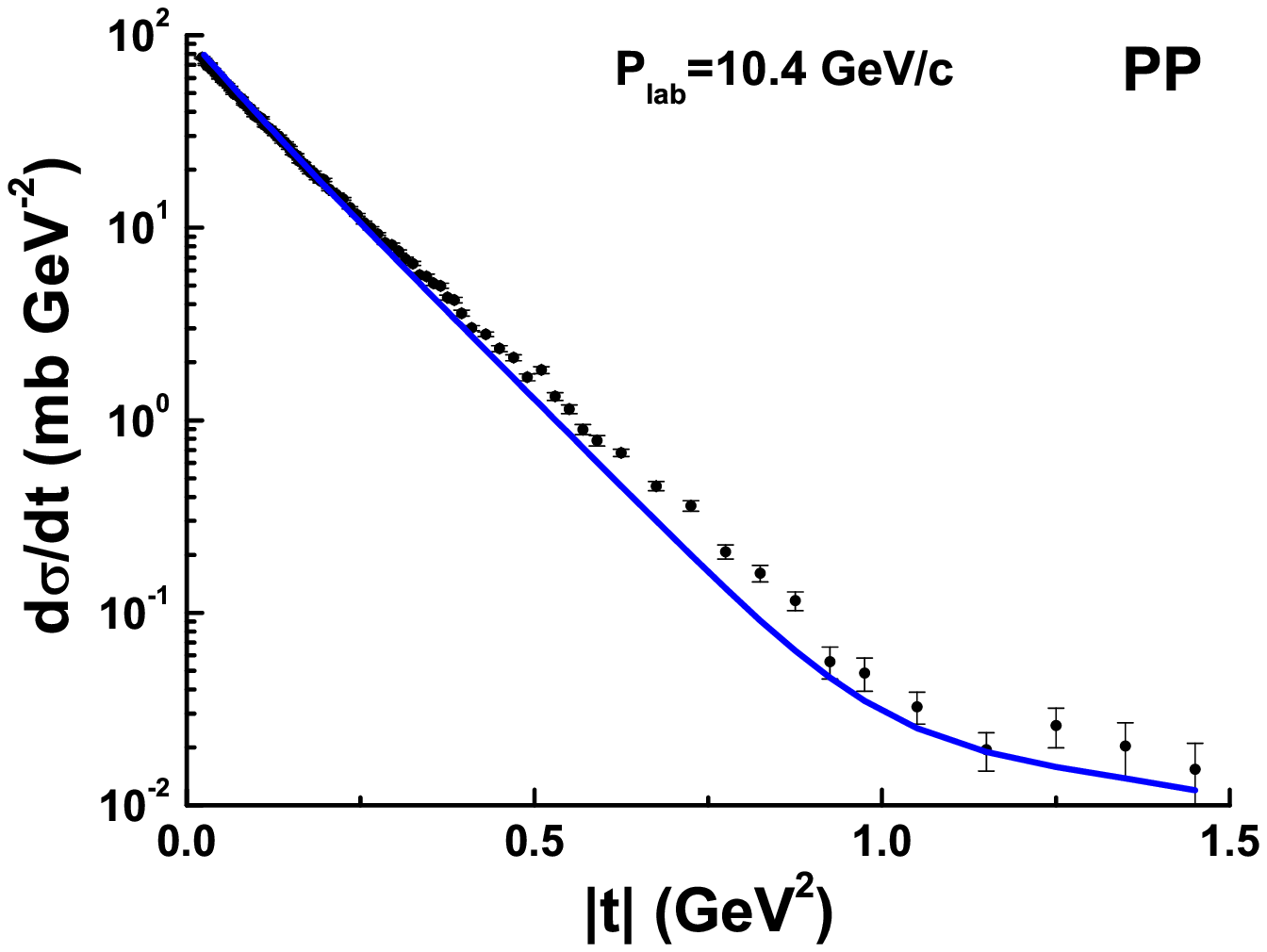}\hspace{5mm}\includegraphics[width=75mm,height=70mm,clip]{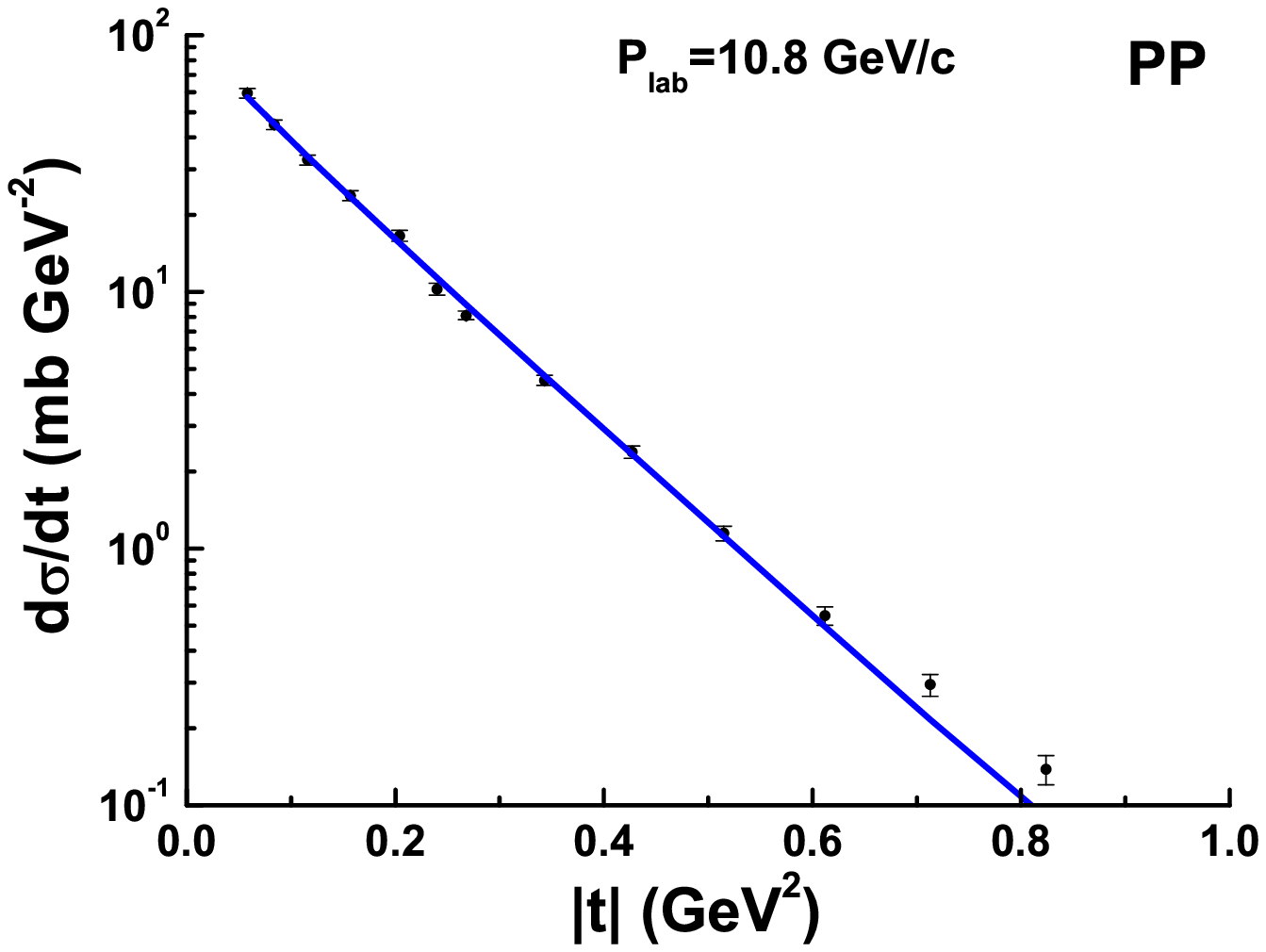}
\begin{minipage}{75mm}
{
\caption{The points are the experimental data by G.W. Brandenburg et al., Phys. Lett. {\bf 58B} (1975) 367.}
}
\end{minipage}
\hspace{5mm}
\begin{minipage}{75mm}
{
\caption{The points are the experimental data by K.J. Foley et al., Phys. Rev. Lett. {\bf 11} (1963) 425.}
}
\end{minipage}

\end{figure}

%%%%%%%%%%%%%%%%%%%%%%%%%%%%%%%%%%%
\begin{figure}[cbth]

%-------------------------------------------------------
\includegraphics[width=75mm,height=66mm,clip]{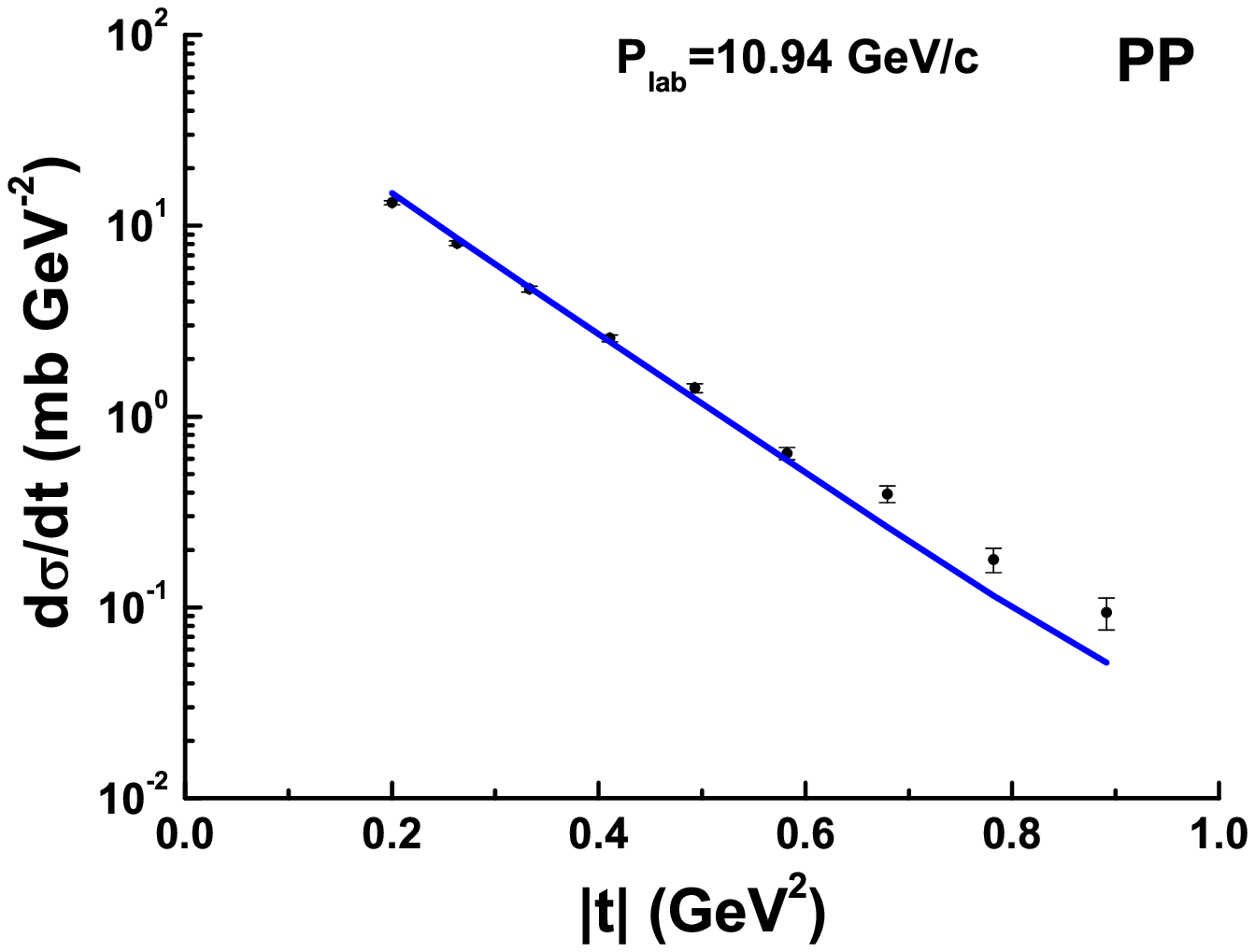}\hspace{5mm}\includegraphics[width=75mm,height=66mm,clip]{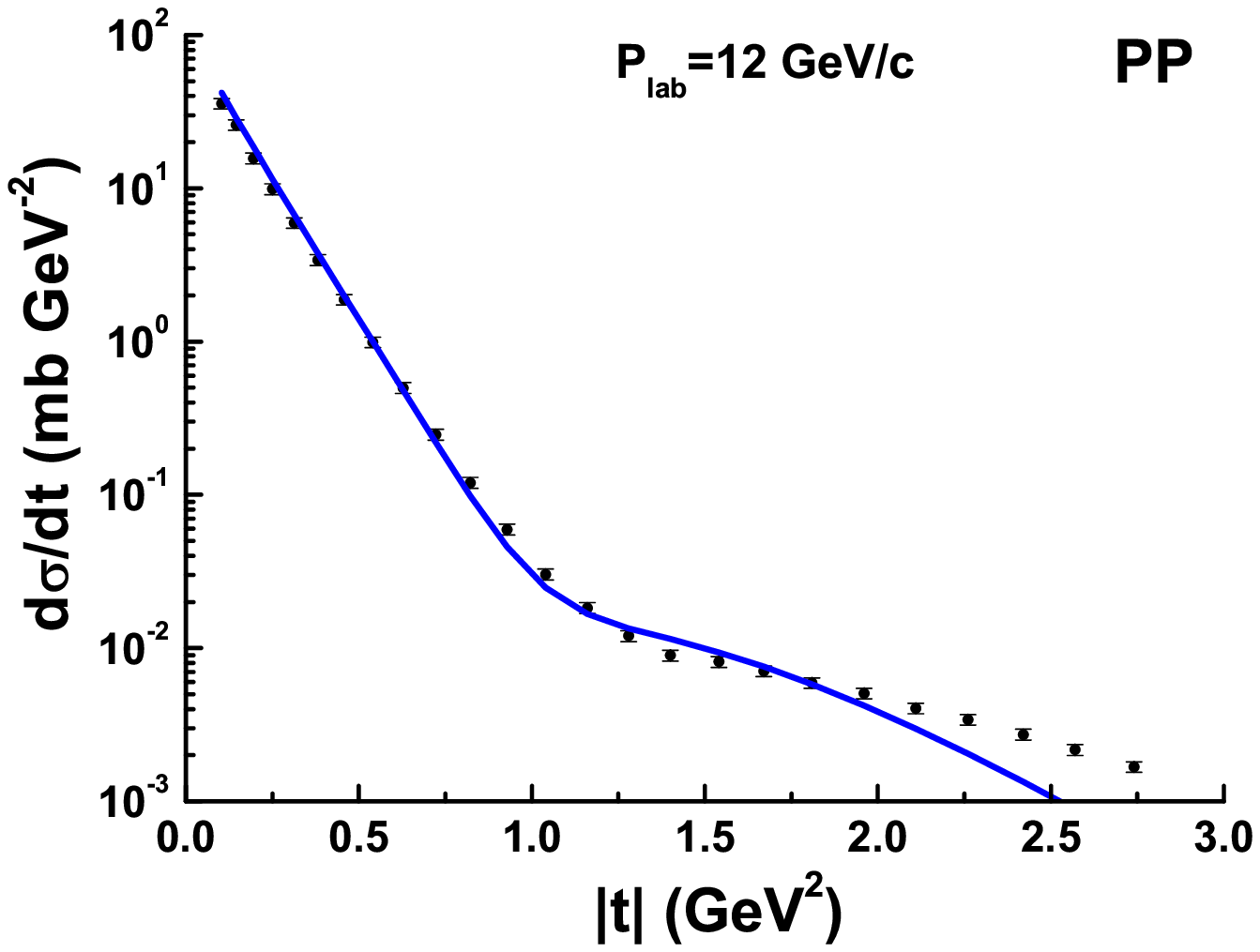}
\begin{minipage}{75mm}
{
\caption{The points are the experimental data by K.J. Foley et al., Phys. Rev. Lett. {\bf 15} (1965) 45.}
}
\end{minipage}
\hspace{5mm}
\begin{minipage}{75mm}
{
\caption{The points are the experimental data by J.V. Allaby et al., Nucl. Phys. {\bf B52} (1973) 316.}
}
\end{minipage}
%-------------------------------------------------------
\includegraphics[width=75mm,height=66mm,clip]{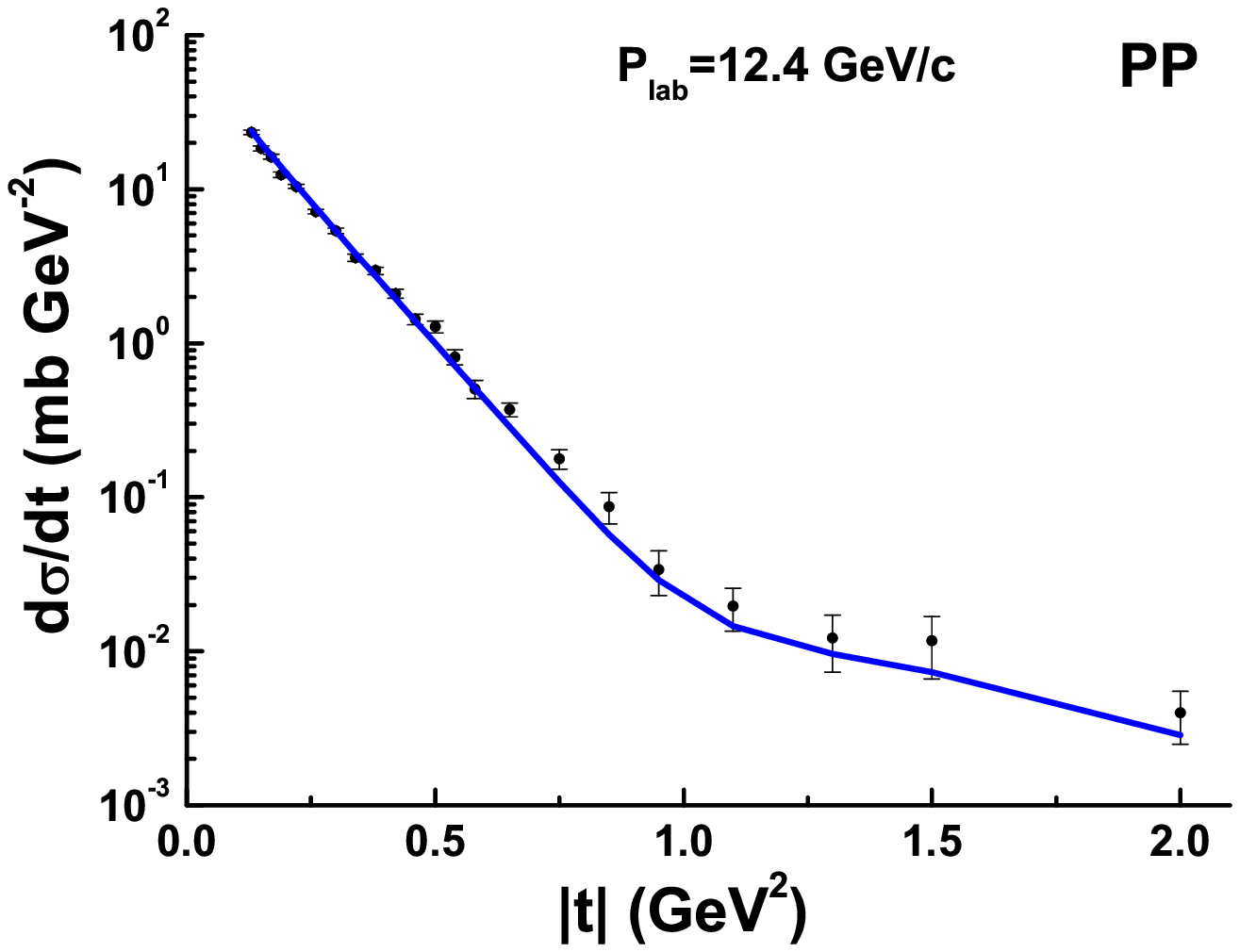}\hspace{5mm}\includegraphics[width=75mm,height=66mm,clip]{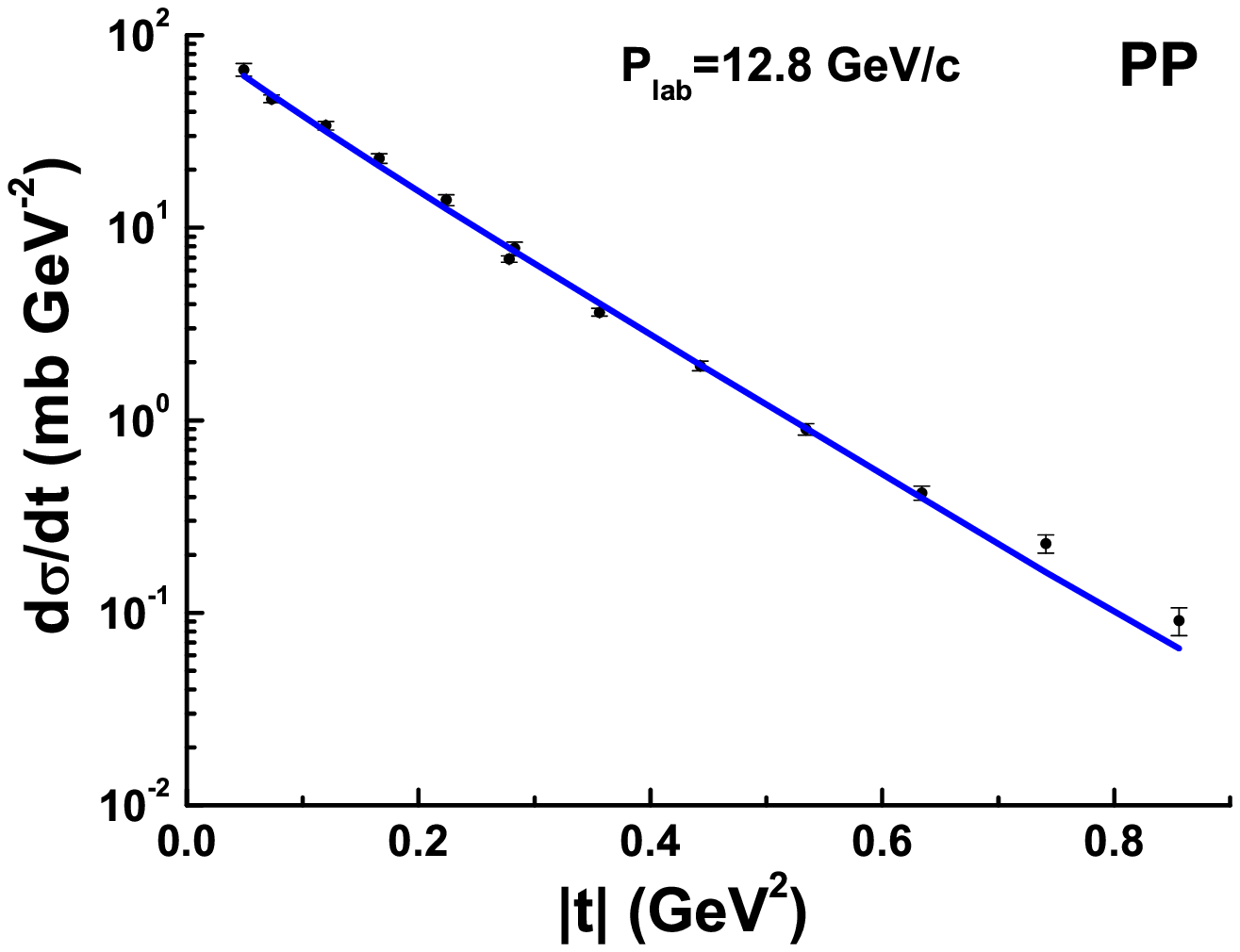}
\begin{minipage}{75mm}
{
\caption{The points are the experimental data by D. Harting, Nuovo Cimento {\bf 38} (1965) 60.}
}
\end{minipage}
\hspace{5mm}
\begin{minipage}{75mm}
{
\caption{The points are the experimental data by K.J. Foley et al., Phys. Rev. Lett. {\bf 11} (1963) 425.}
}
\end{minipage}
%-------------------------------------------------------
\includegraphics[width=75mm,height=66mm,clip]{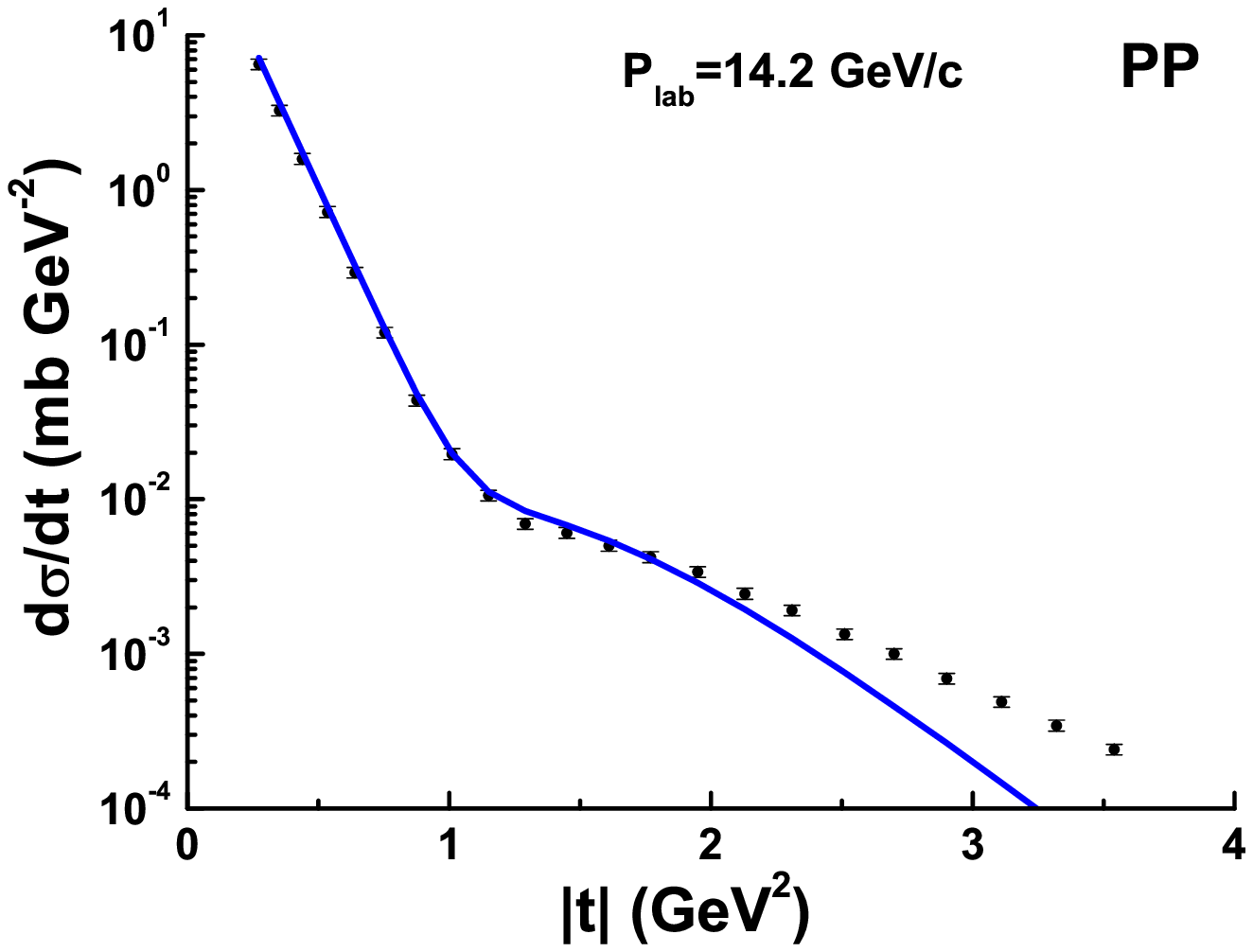}\hspace{5mm}\includegraphics[width=75mm,height=66mm,clip]{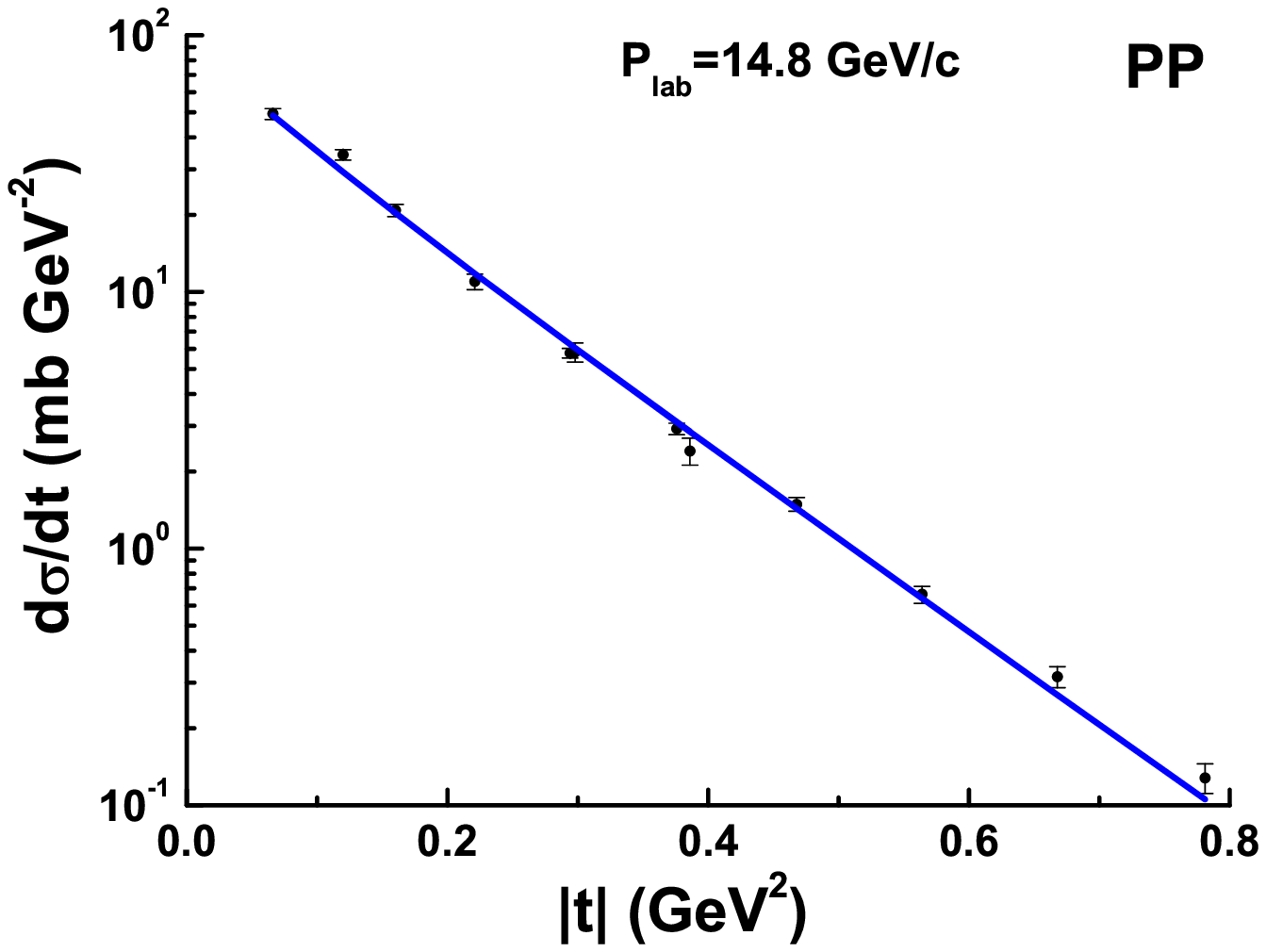}
\begin{minipage}{75mm}
{
\caption{The points are the experimental data by J.V. Allaby et al., Nucl. Phys. {\bf B52} (1973) 316.}
}
\end{minipage}
\hspace{5mm}
\begin{minipage}{75mm}
{
\caption{The points are the experimental data by K.J. Foley et al., Phys. Rev. Lett. {\bf 11} (1963) 425.}
}
\end{minipage}
%-------------------------------------------------------
\end{figure}

\begin{figure}[cbth]
%-------------------------------------------------------
\includegraphics[width=75mm,height=66mm,clip]{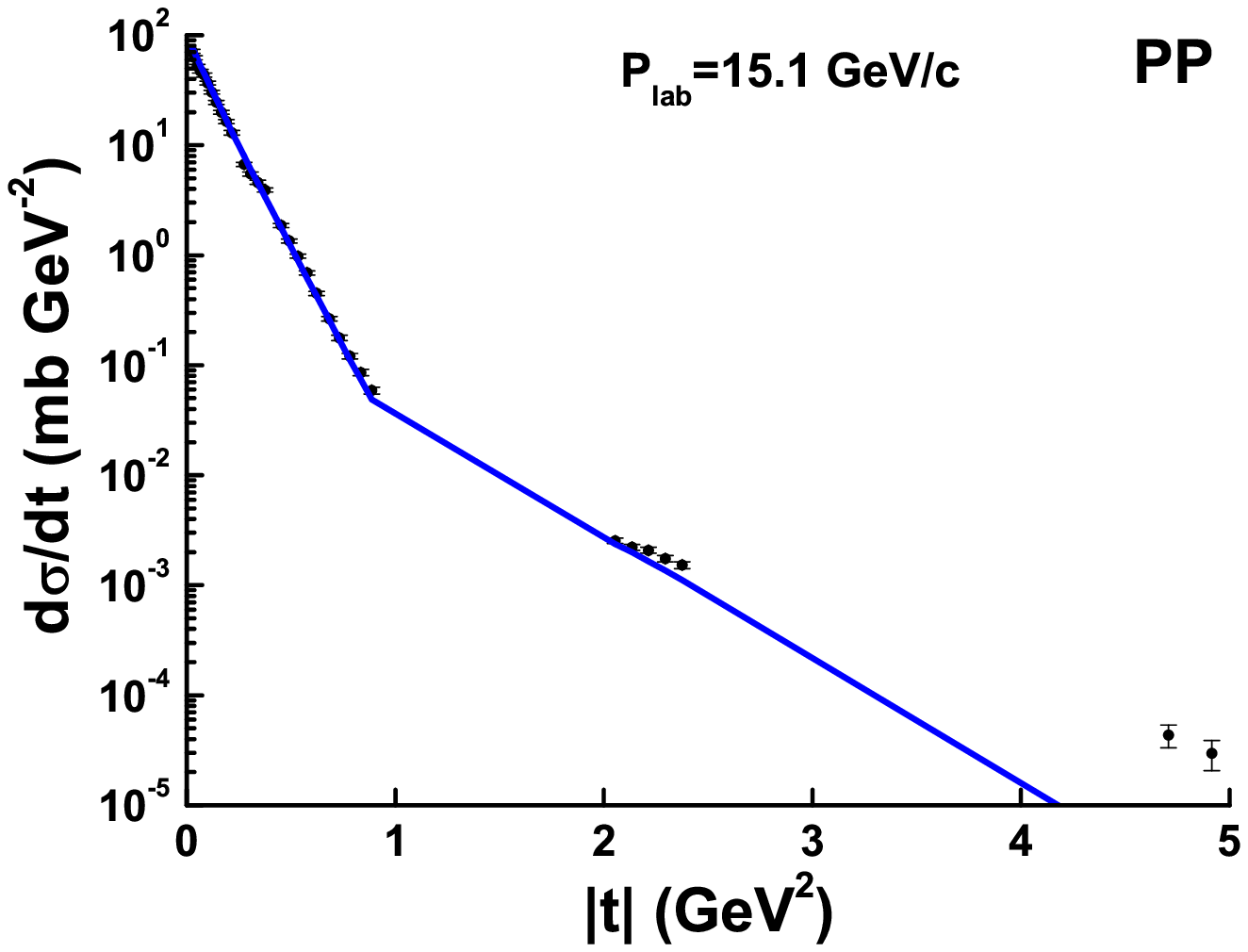}\hspace{5mm}\includegraphics[width=75mm,height=66mm,clip]{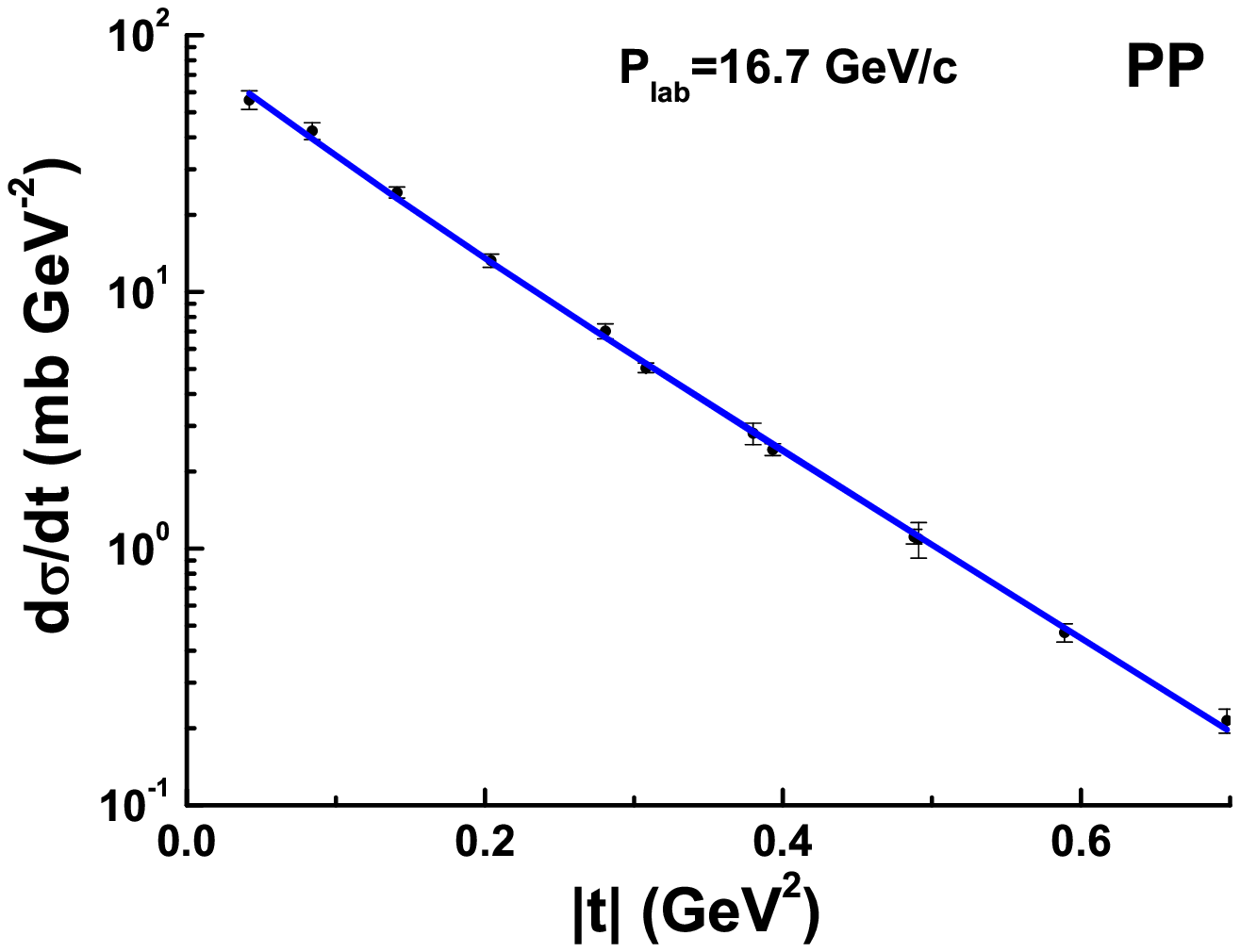}
\begin{minipage}{75mm}
{
\caption{The points are the experimental data by R.M. Edelstein et al., Phys. Rev. {\bf D5} (1972) 1073.}
}
\end{minipage}
\hspace{5mm}
\begin{minipage}{75mm}
{
\caption{The points are the experimental data by .K.J. Foley et al., Phys. Rev. Lett. {\bf 11} (1963) 425}
}
\end{minipage}
%-------------------------------------------------------
\includegraphics[width=75mm,height=66mm,clip]{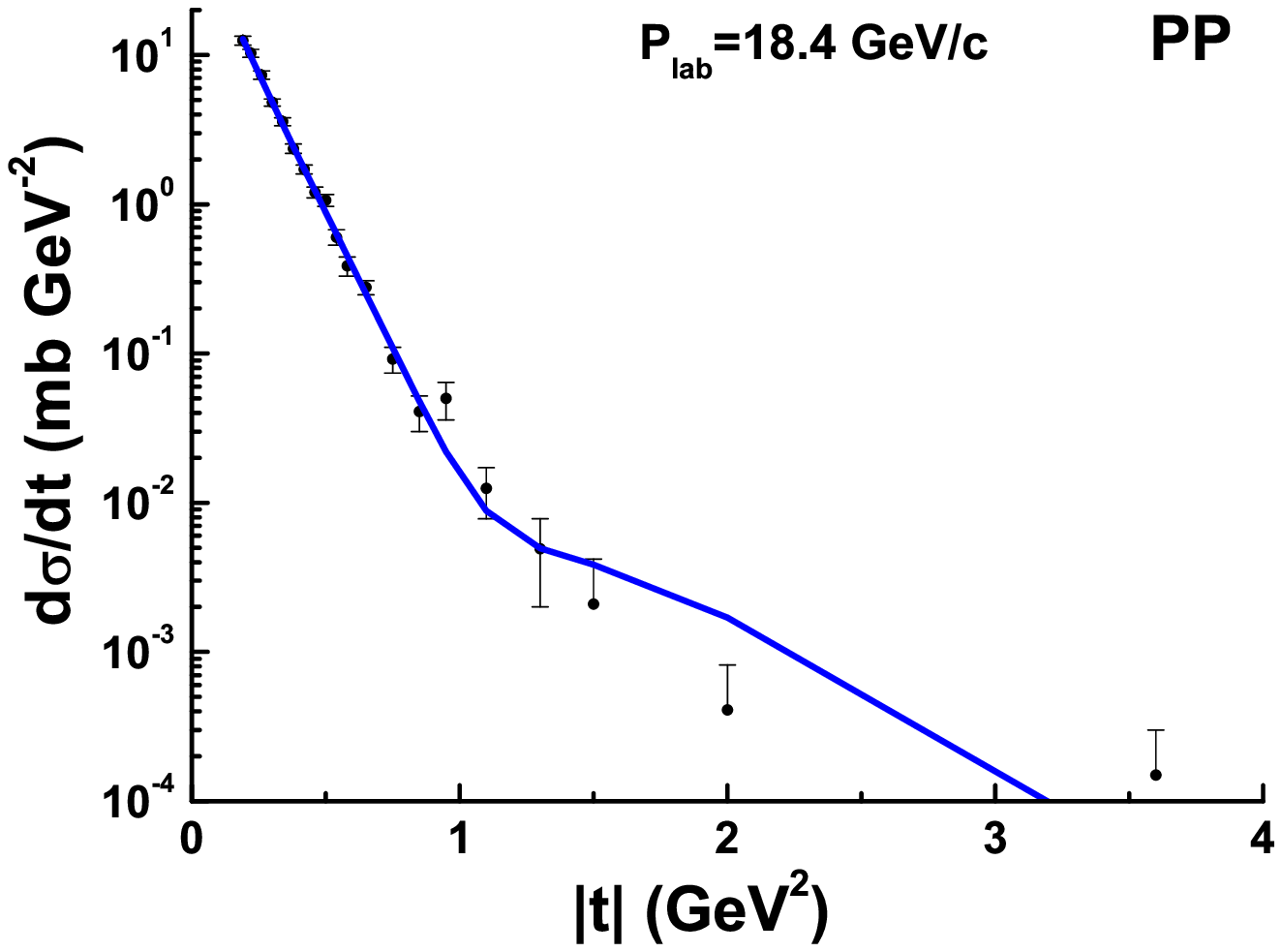}\hspace{5mm}\includegraphics[width=75mm,height=66mm,clip]{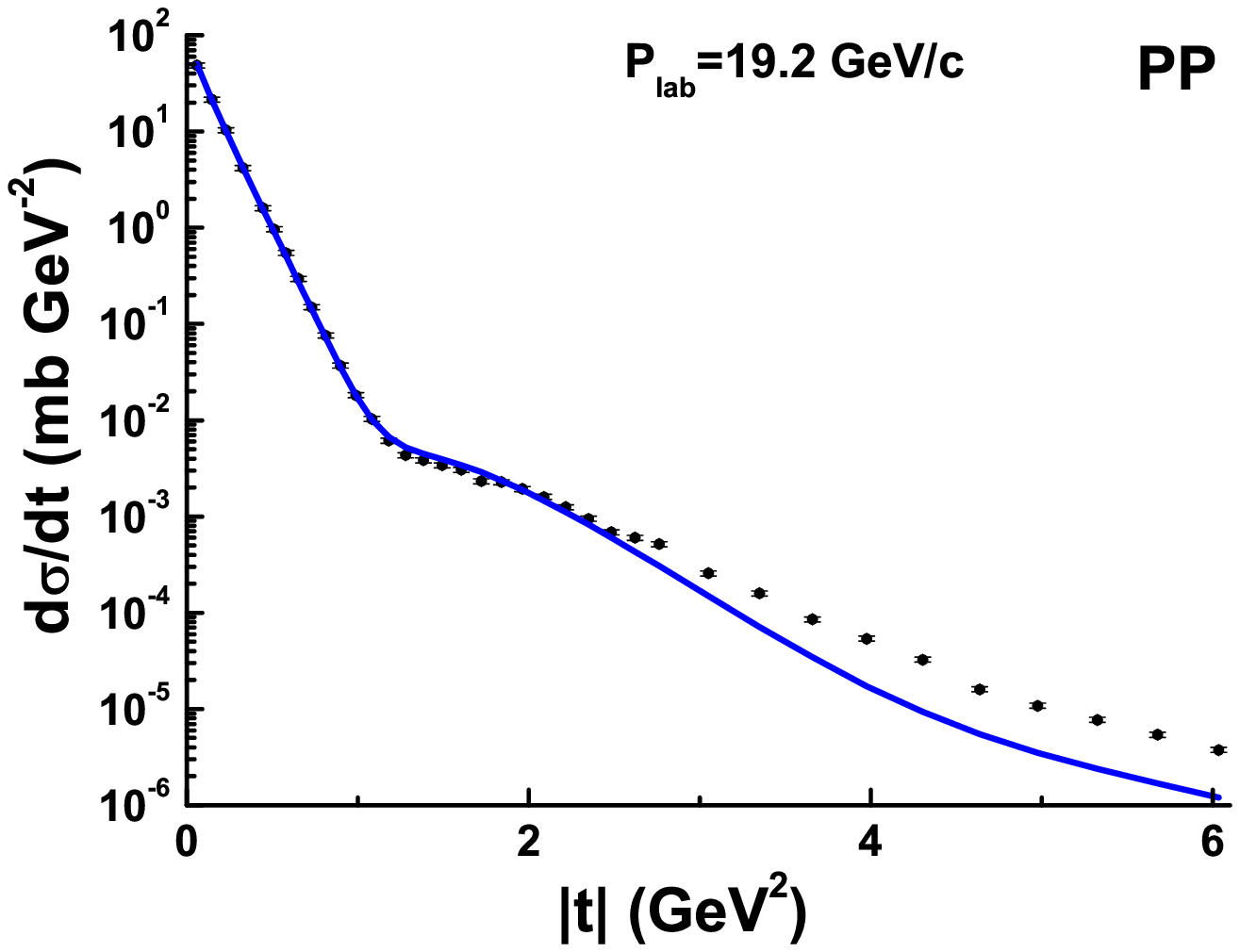}
\begin{minipage}{75mm}
{
\caption{The points are the experimental data by D. Harting, Nuovo Cimento {\bf 38} (1965) 60.}
}
\end{minipage}
\hspace{5mm}
\begin{minipage}{75mm}
{
\caption{The points are the experimental data by J.V. Allaby et al., Phys. Lett. {\bf B28} (1968) 67.}
}
\end{minipage}
%-------------------------------------------------------
\includegraphics[width=75mm,height=66mm,clip]{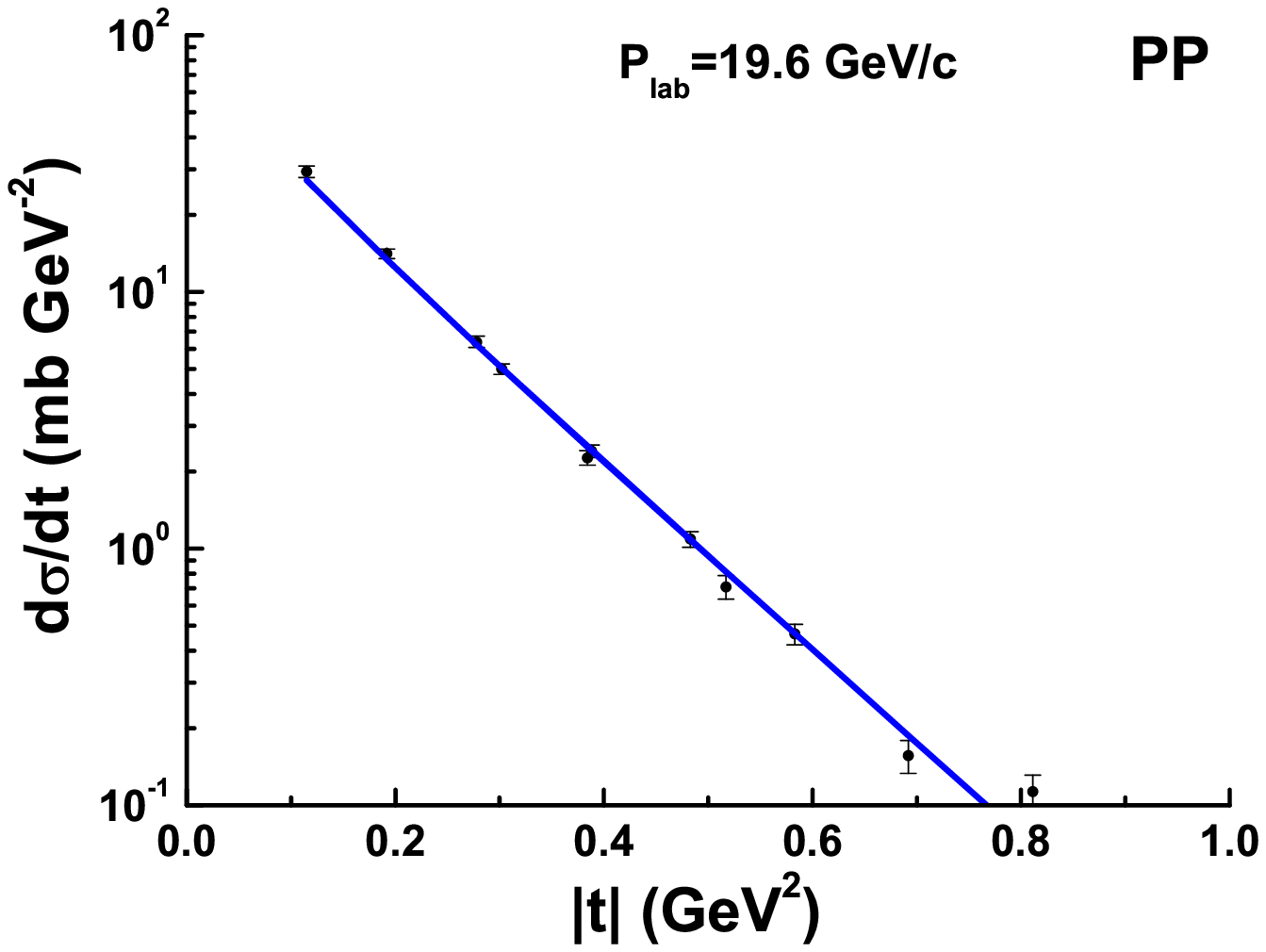}\hspace{5mm}\includegraphics[width=75mm,height=66mm,clip]{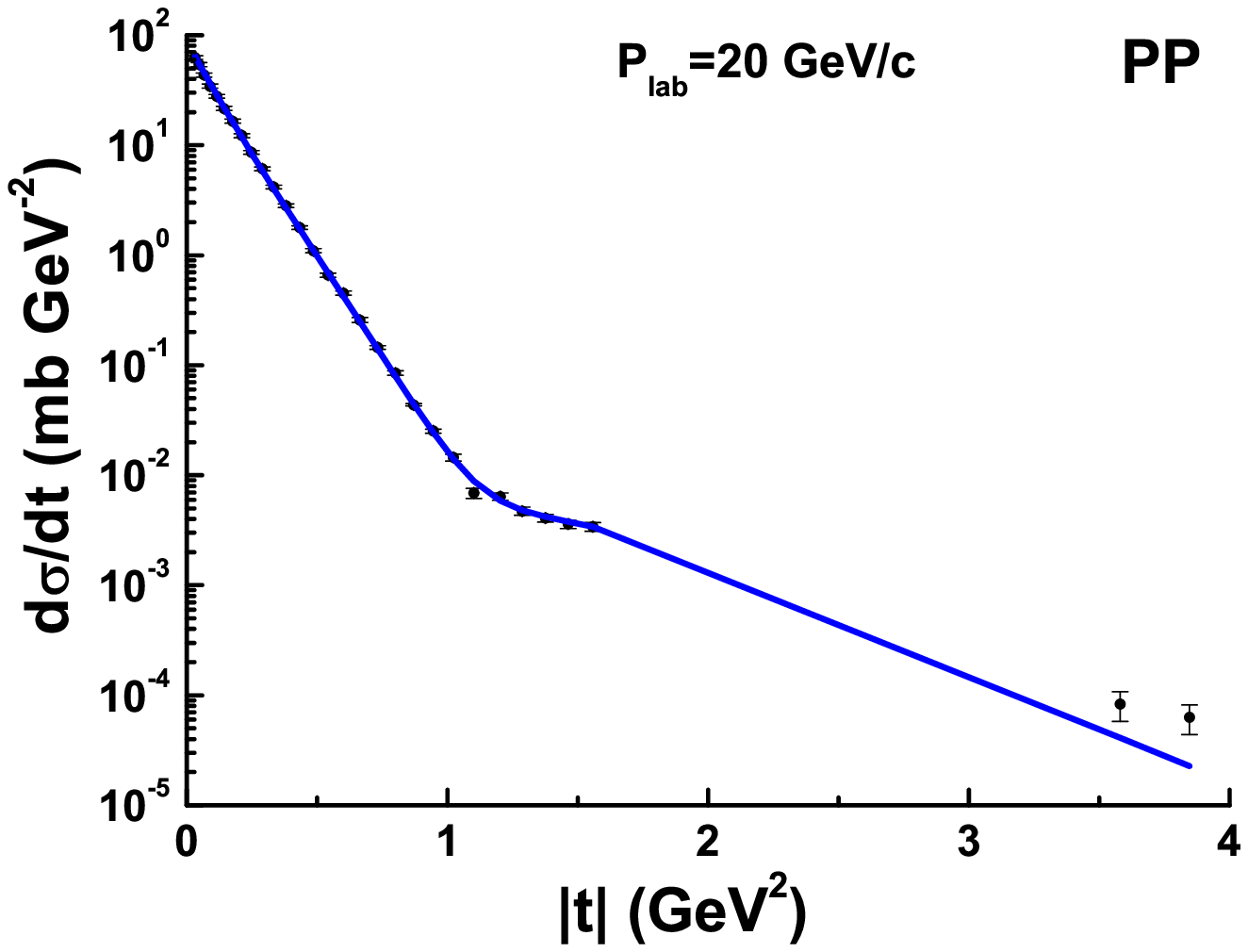}
\begin{minipage}{75mm}
{
\caption{The points are the experimental data by .K.J. Foley et al., Phys. Rev. Lett. {\bf 11} (1963) 425}
}
\end{minipage}
\hspace{5mm}
\begin{minipage}{75mm}
{
\caption{The points are the experimental data by R.M. Edelstein et al., Phys. Rev. {\bf D5} (1972) 1073.}
}
\end{minipage}
\end{figure}

%%%%%%%%%%%%%%%%%%%%%%%%%%%%%%%%%%%
\begin{figure}[cbth]
%-------------------------------------------------------
\includegraphics[width=75mm,height=66mm,clip]{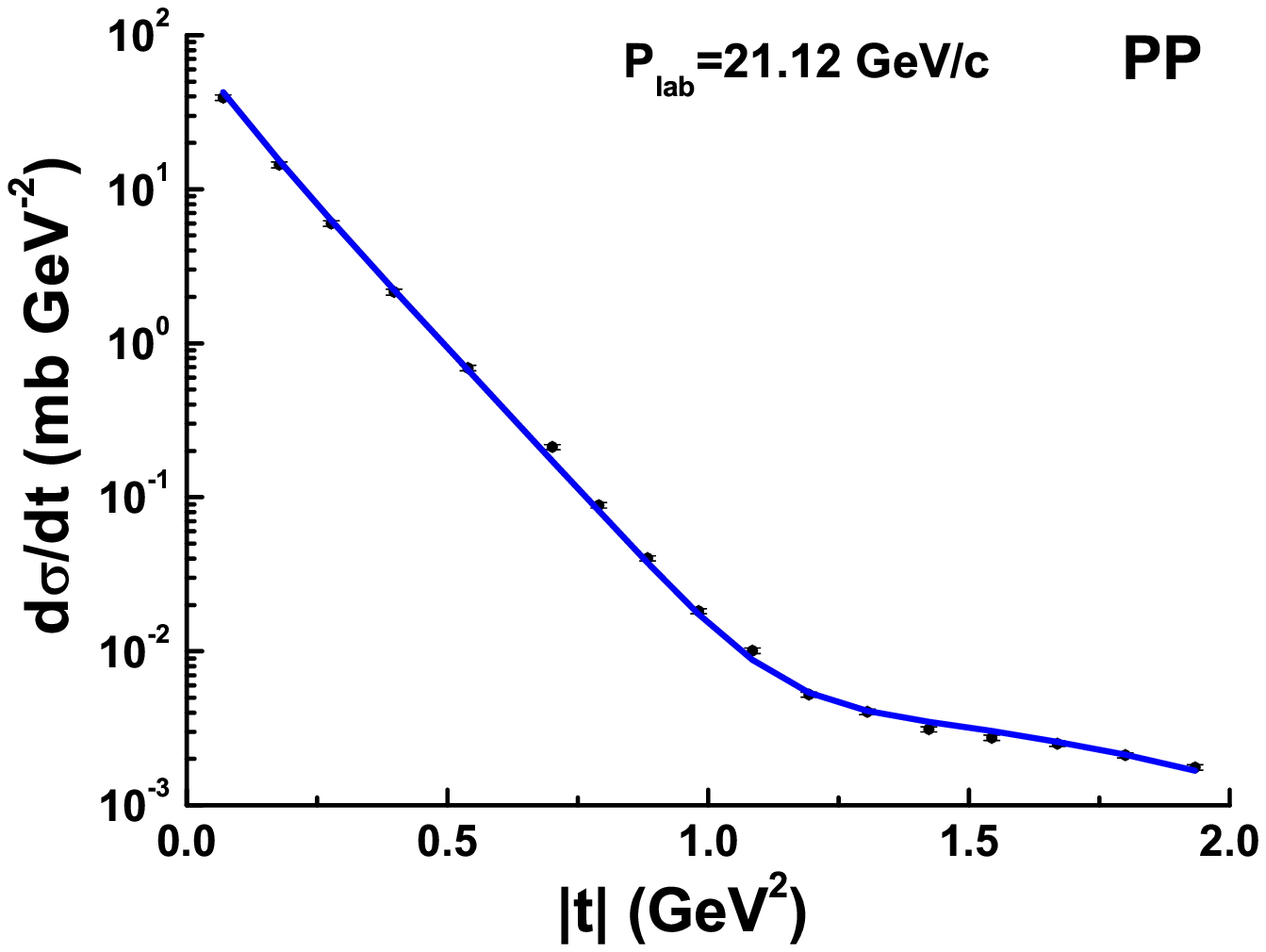}\hspace{5mm}\includegraphics[width=75mm,height=66mm,clip]{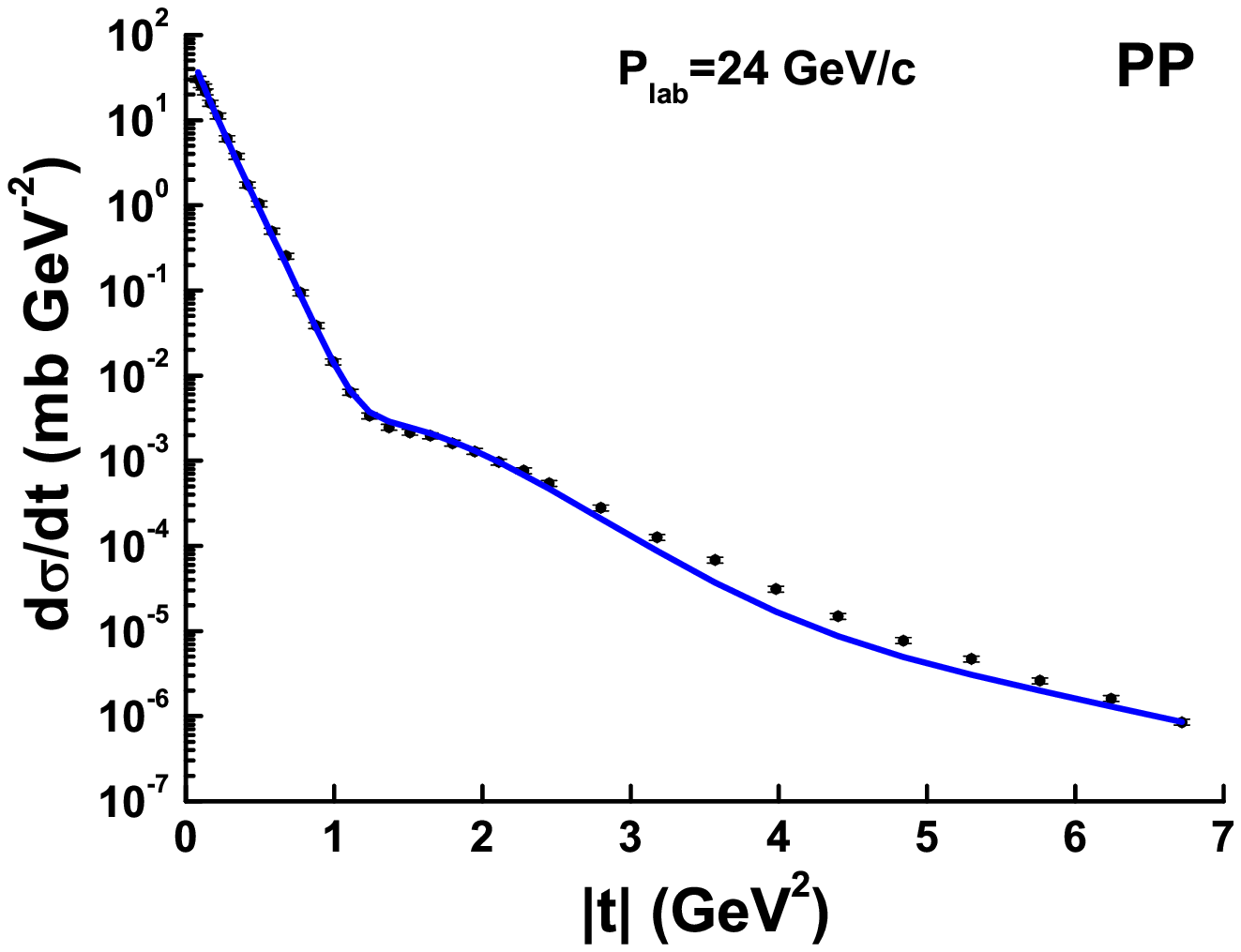}
\begin{minipage}{75mm}
{
\caption{The points are the experimental data by J.V. Allaby et al., Phys. Lett. {\bf B28} (1968) 67.}
}
\end{minipage}
\hspace{5mm}
\begin{minipage}{75mm}
{
\caption{The points are the experimental data by J.V. Allaby et al., Nucl. Phys. {\bf B52} (1973) 316.}
}
\end{minipage}
%-------------------------------------------------------
\includegraphics[width=75mm,height=66mm,clip]{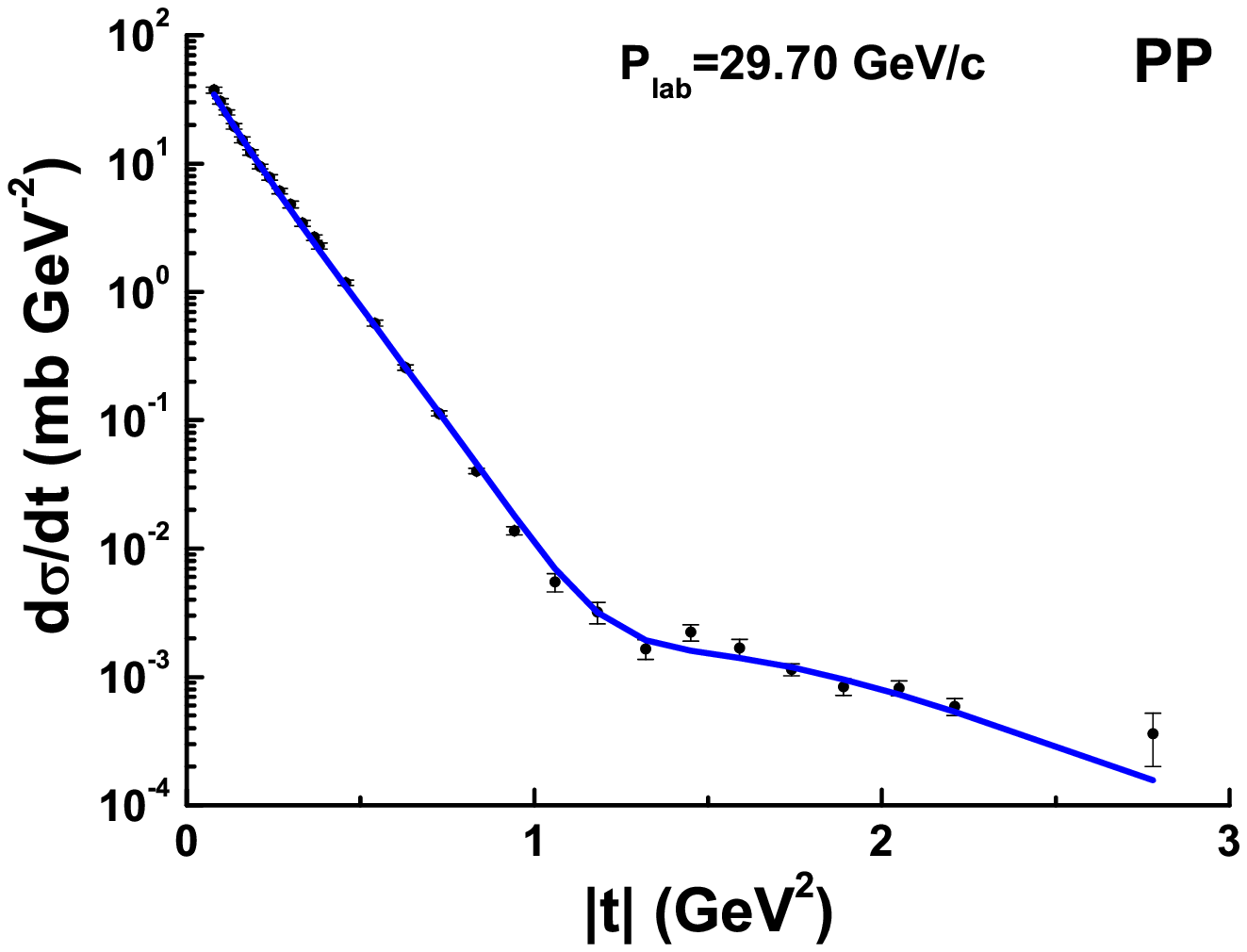}\hspace{5mm}\includegraphics[width=75mm,height=66mm,clip]{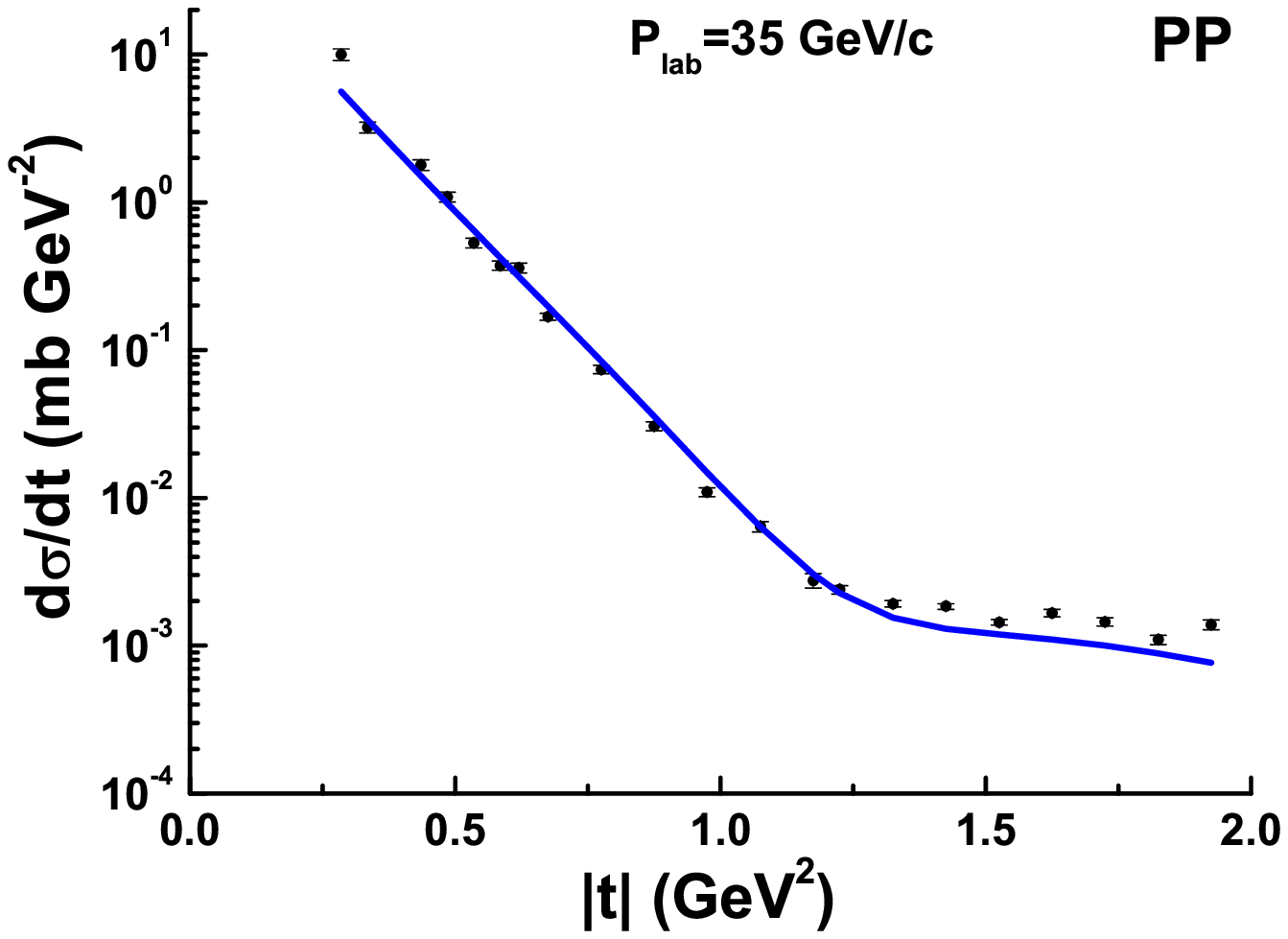}
\begin{minipage}{75mm}
{
\caption{The points are the experimental data by R.M. Edelstein et al., Phys. Rev. {\bf D5} (1972) 1073.}
}
\end{minipage}
\hspace{5mm}
\begin{minipage}{75mm}
{
\caption{The points are the experimental data by R. Rusack et al., Phys. Rev. Lett. {\bf 41} (1978) 1632.}
}
\end{minipage}
%-------------------------------------------------------
%-------------------------------------------------------
\includegraphics[width=75mm,height=66mm,clip]{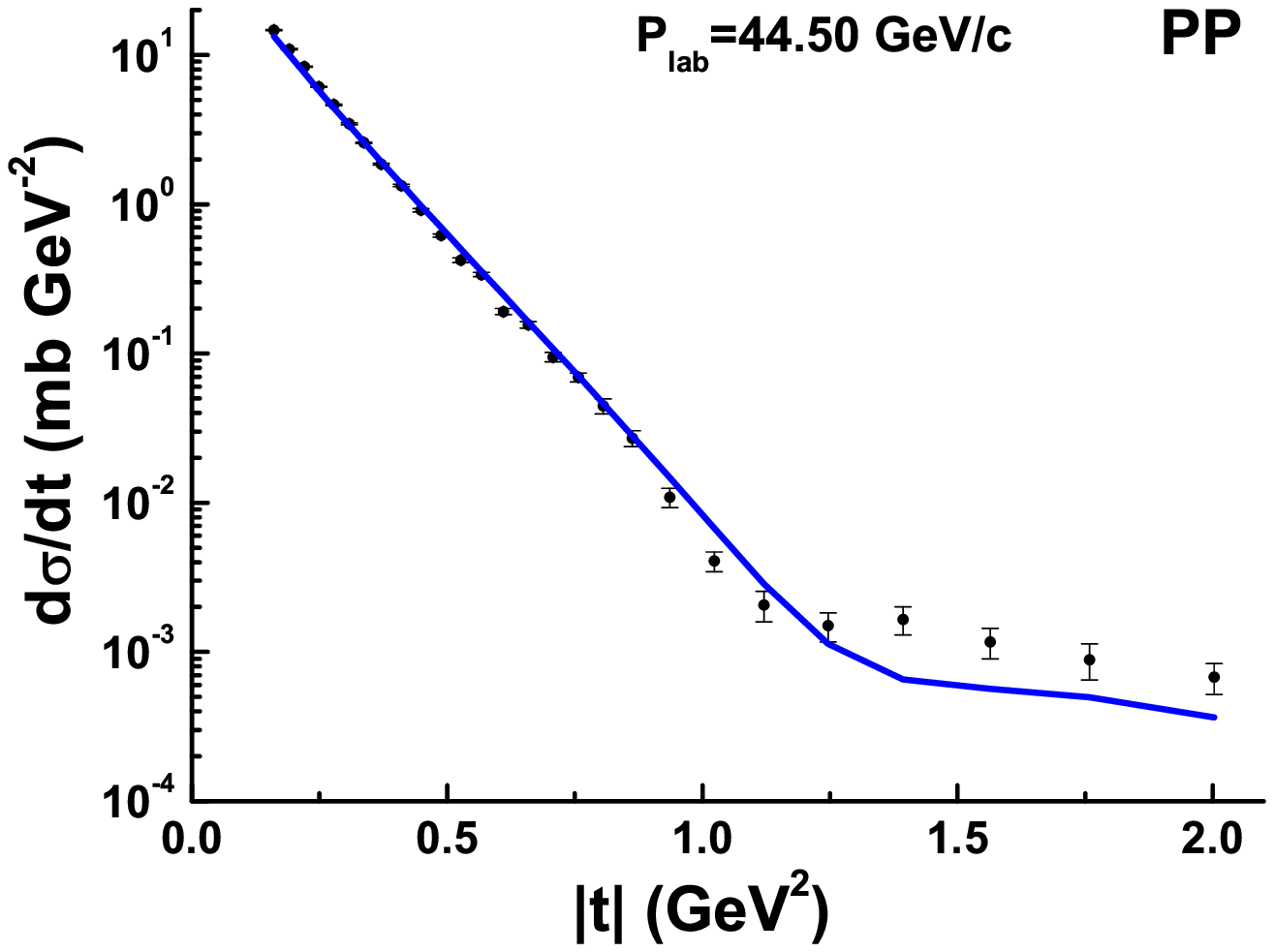}\hspace{5mm}\includegraphics[width=75mm,height=66mm,clip]{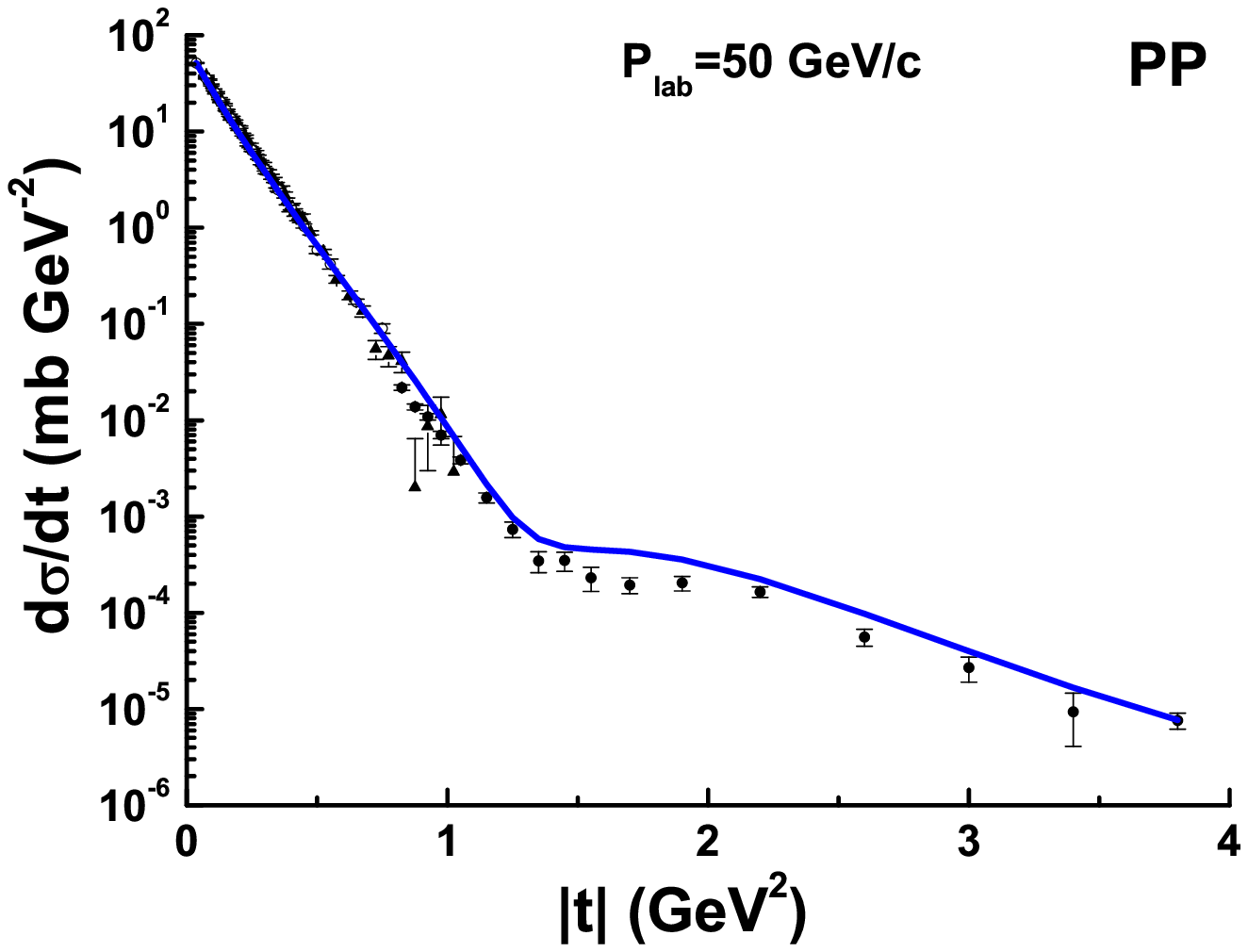}
\begin{minipage}{75mm}
{
\caption{The points are the experimental data by C. Bruneton et al., Nucl. Phys. {\bf B124} (1977) 391.}
}
\end{minipage}
\hspace{5mm}
\begin{minipage}{75mm}
{
\caption{The points are the experimental data by
             Z. Asad et al., Nucl. Phys. {\bf B255} (1984) 273;
             C.W. Akerlof et al., Phys. Rev. {\bf D14} (1976) 2864;
             D.S. Ayres et al., Phys. Rev. {\bf D15} (1977) 3105.}
}
\end{minipage}
\end{figure}

\begin{figure}[cbth]
%-------------------------------------------------------
\includegraphics[width=75mm,height=66mm,clip]{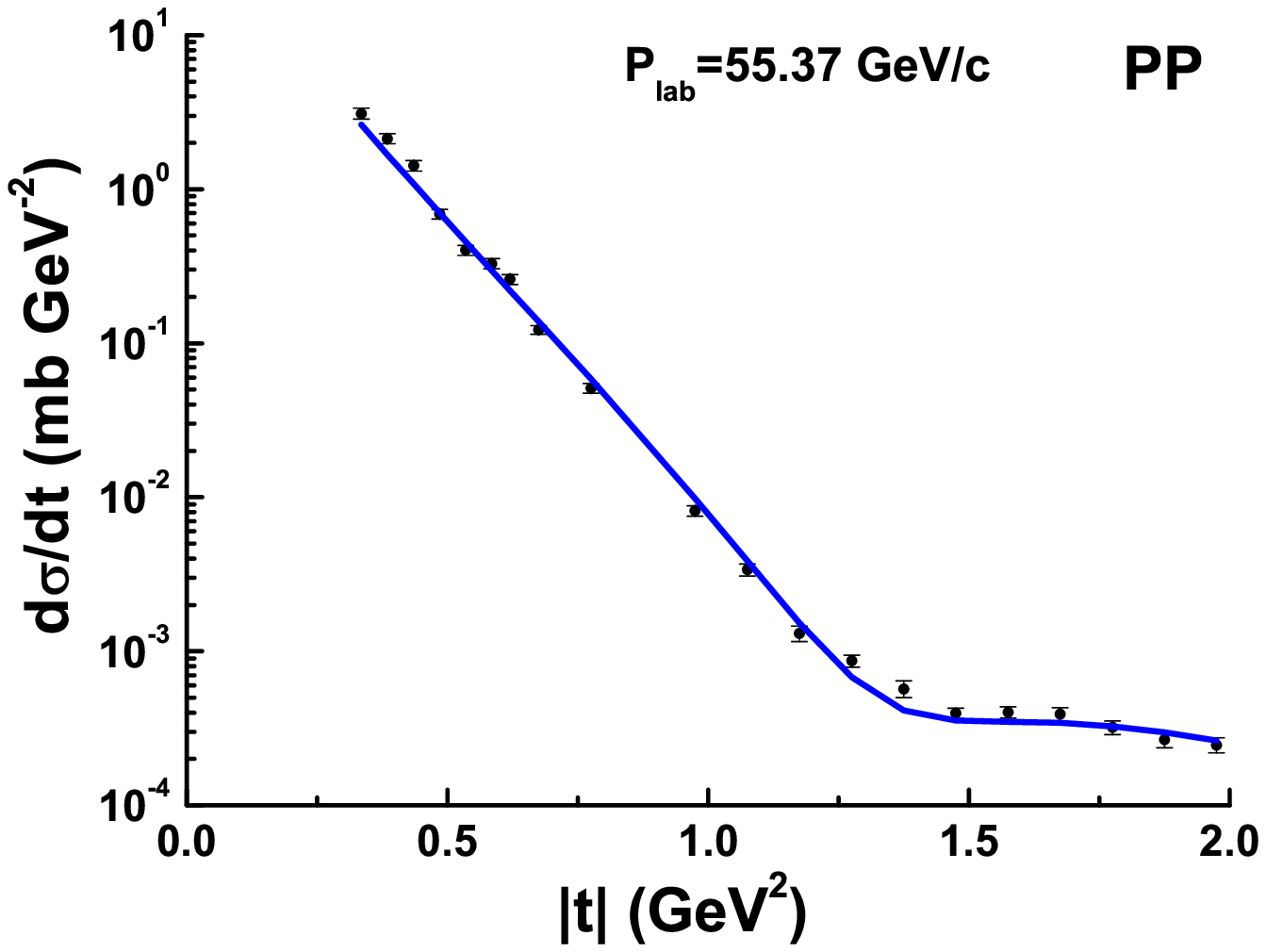}\hspace{5mm}\includegraphics[width=75mm,height=66mm,clip]{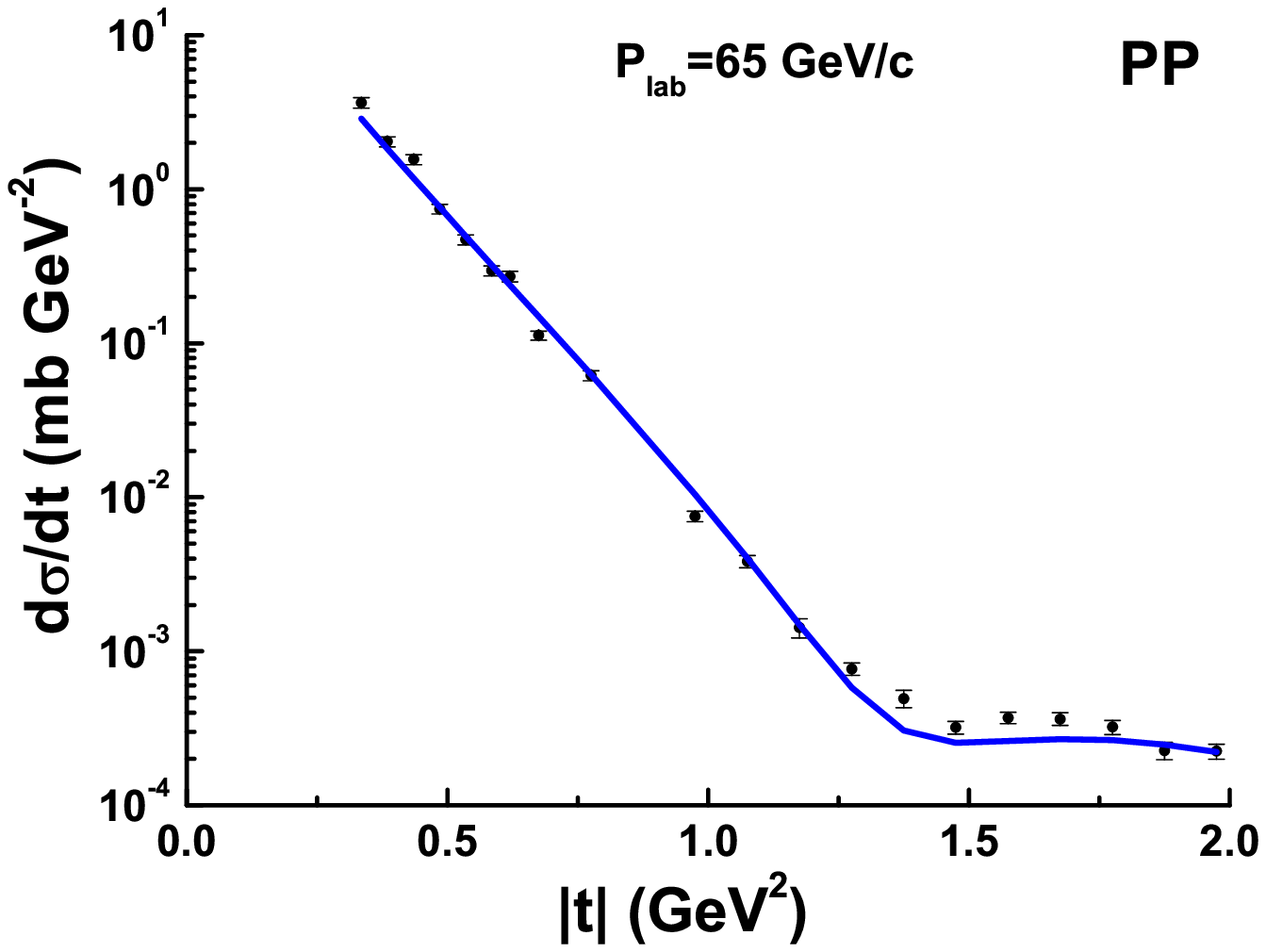}
\begin{minipage}{75mm}
{
\caption{The points are the experimental data by R. Rusack et al., Phys. Rev. Lett. {\bf 41} (1978) 1632.}
}
\end{minipage}
\hspace{5mm}
\begin{minipage}{75mm}
{
\caption{The points are the experimental data by R. Rusack et al., Phys. Rev. Lett. {\bf 41} (1978) 1632.}
}
\end{minipage}

%-------------------------------------------------------
\includegraphics[width=75mm,height=66mm,clip]{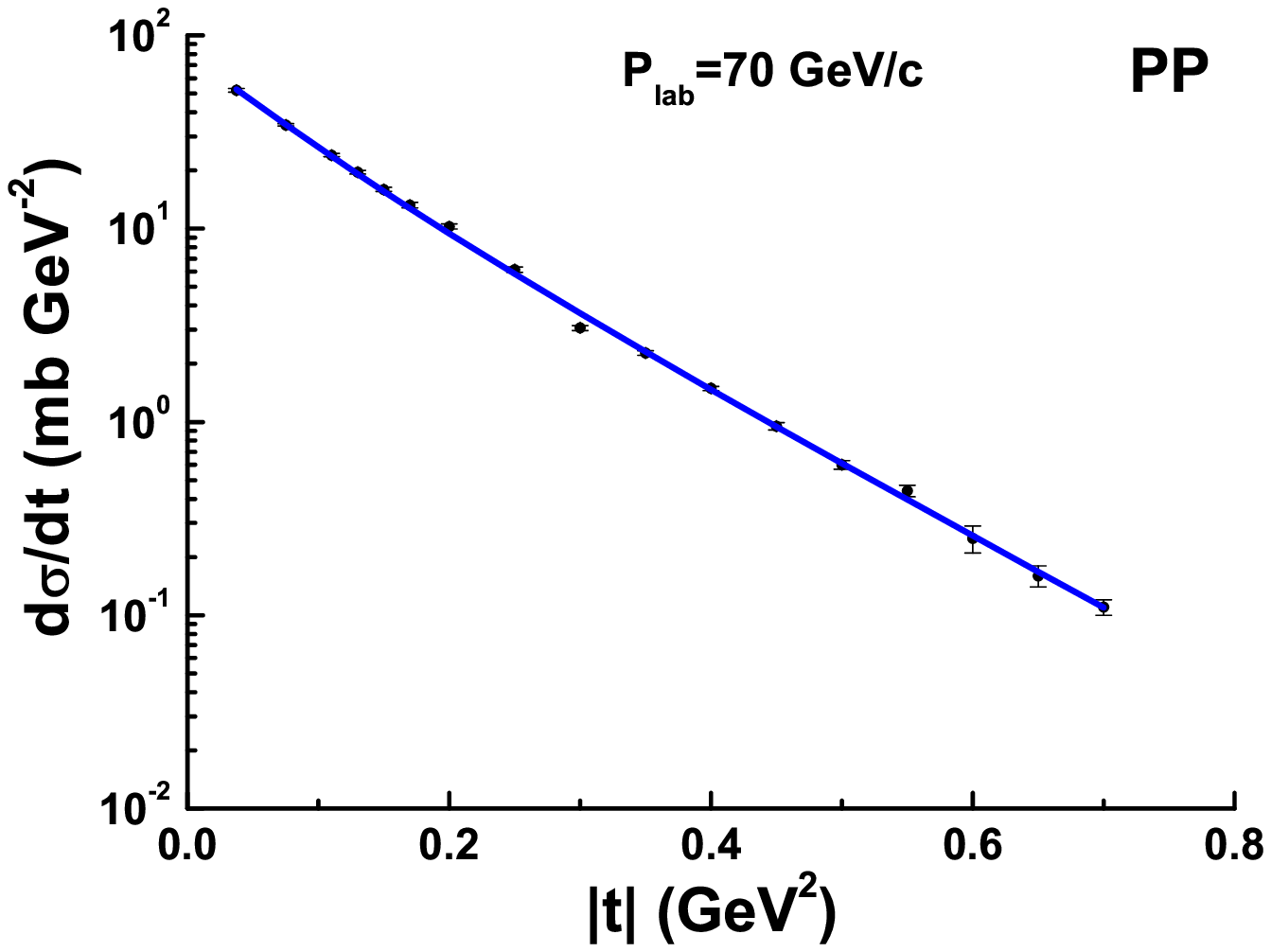}\hspace{5mm}\includegraphics[width=75mm,height=66mm,clip]{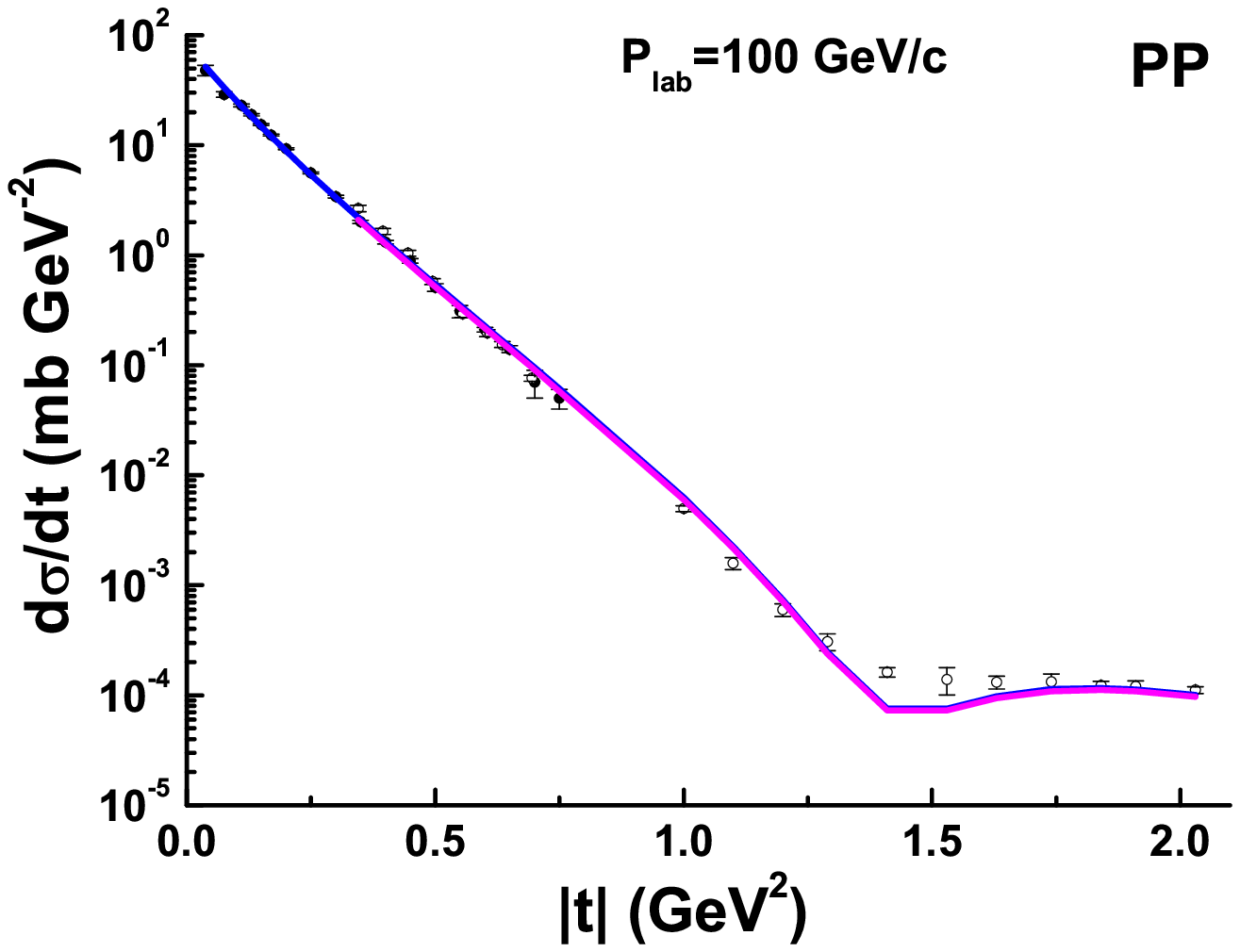}
\begin{minipage}{75mm}
{
\caption{The points are the experimental data by D.S. Ayres et al., Phys. Rev. {\bf D15} (1977) 3105.}
}
\end{minipage}
\hspace{5mm}
\begin{minipage}{75mm}
{
\caption{The points are the experimental data by C.W. Akerlof et al., Phys. Rev. {\bf D14} (1976) 2864;
                                           R. Rubinstein et al., Phys. Rev.{\bf D30} (1984) 1413.}
}
\end{minipage}
%-------------------------------------------------------
\includegraphics[width=75mm,height=66mm,clip]{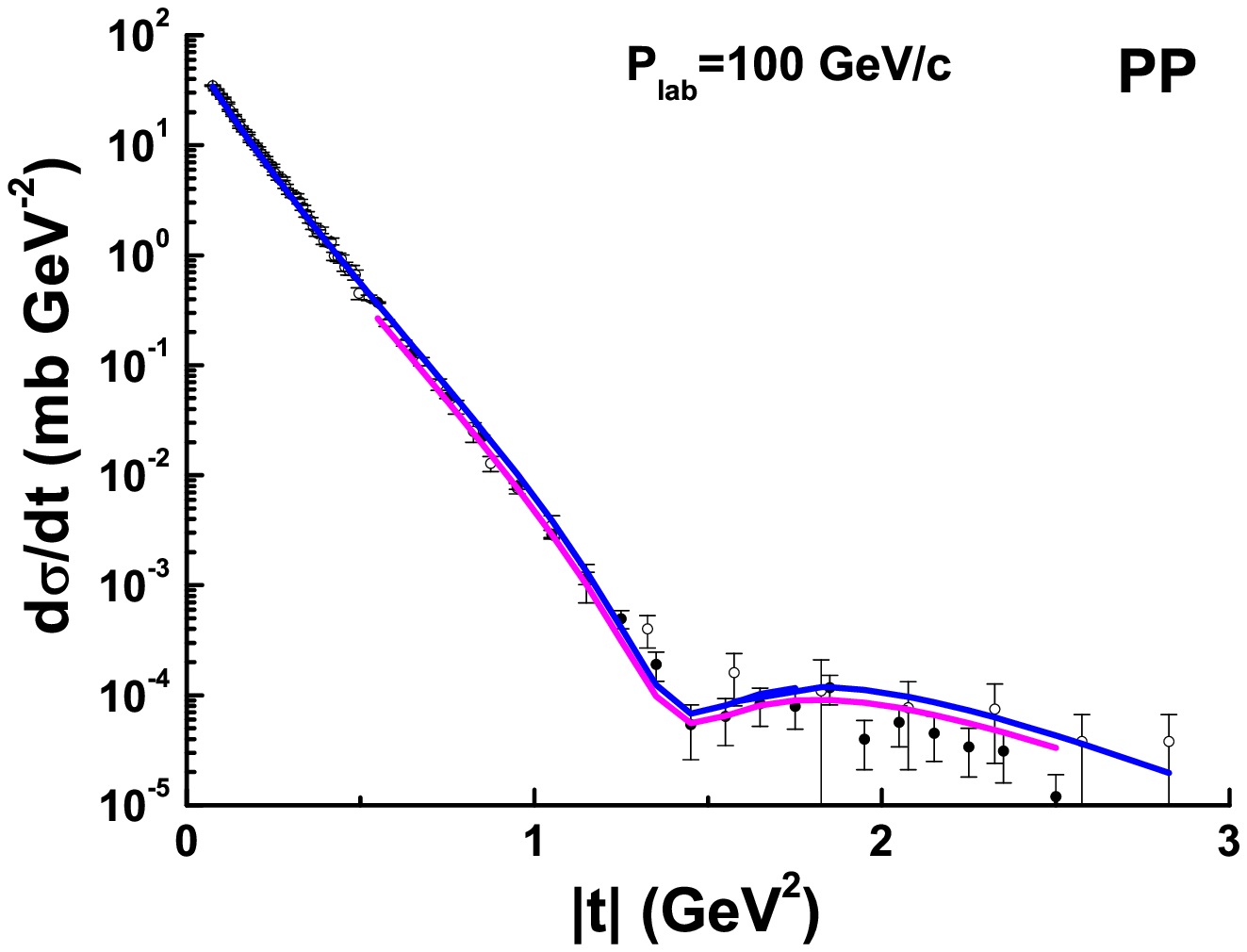}\hspace{5mm}\includegraphics[width=75mm,height=66mm,clip]{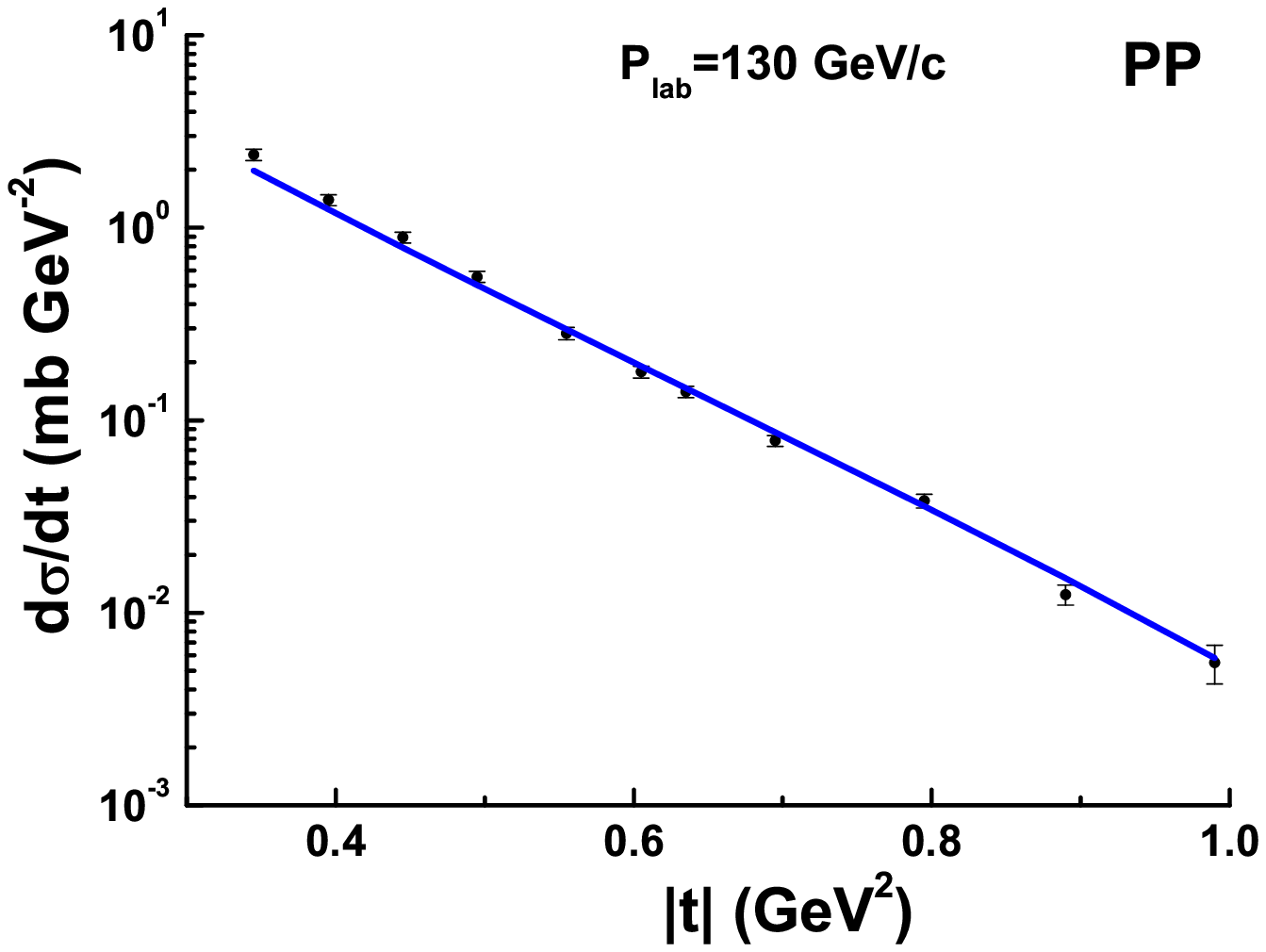}
\begin{minipage}{75mm}
{
\caption{The points are the experimental data by D.S. Ayres et al., Phys. Rev. {\bf D15} (1977) 3105;
R. Rusack et al., Phys. Rev. Lett. {\bf 41} (1978) 1632.}
}
\end{minipage}
\hspace{5mm}
\begin{minipage}{75mm}
{
\caption{The points are the experimental data by R. Rusack et al., Phys. Rev. Lett. {\bf 41} (1978) 1632.}
}
\end{minipage}
\end{figure}

%%%%%%%%%%%%%%%%%%%%%%%%%%%%%%%%%%%
\begin{figure}[cbth]
%-------------------------------------------------------
\includegraphics[width=75mm,height=66mm,clip]{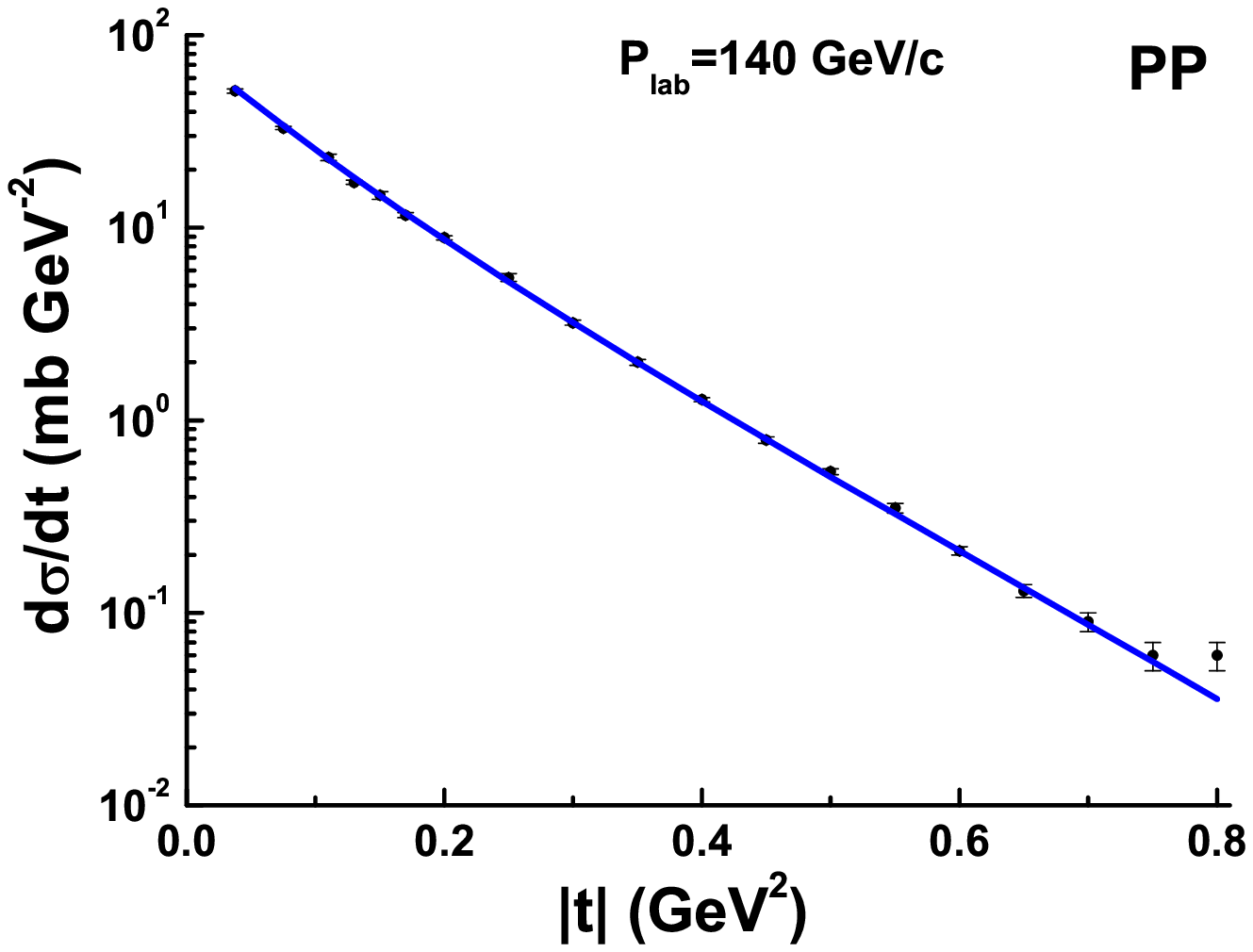}\hspace{5mm}\includegraphics[width=75mm,height=66mm,clip]{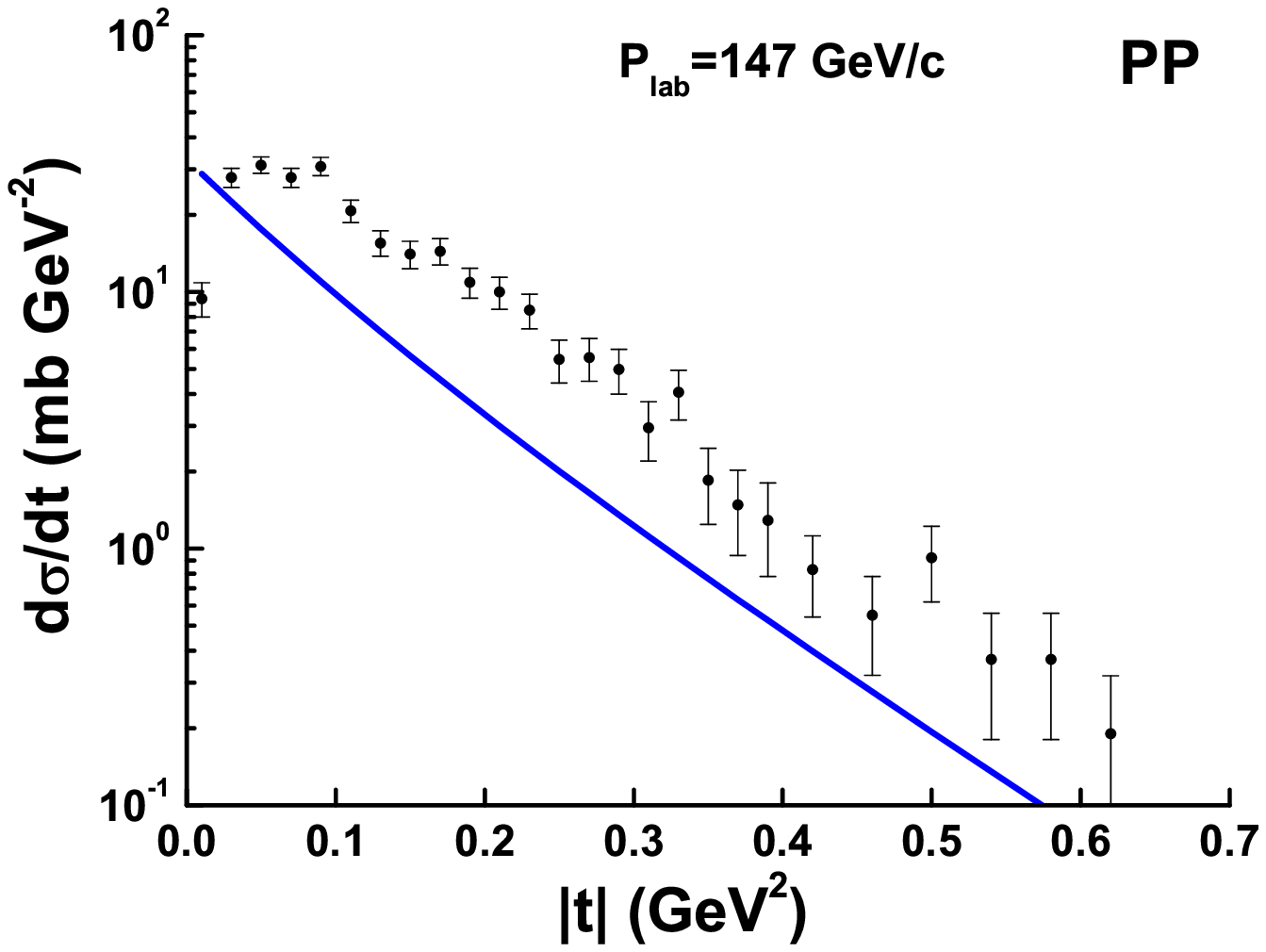}
\begin{minipage}{75mm}
{
\caption{The points are the experimental data by D.S. Ayres et al., Phys. Rev. {\bf D15} (1977) 3105.}
}
\end{minipage}
\hspace{5mm}
\begin{minipage}{75mm}
{
\caption{The points are the experimental data by D. Brick et al., Phys. Rev. {\bf D25} (1982) 2794.}
}
\end{minipage}
%-------------------------------------------------------
\includegraphics[width=75mm,height=66mm,clip]{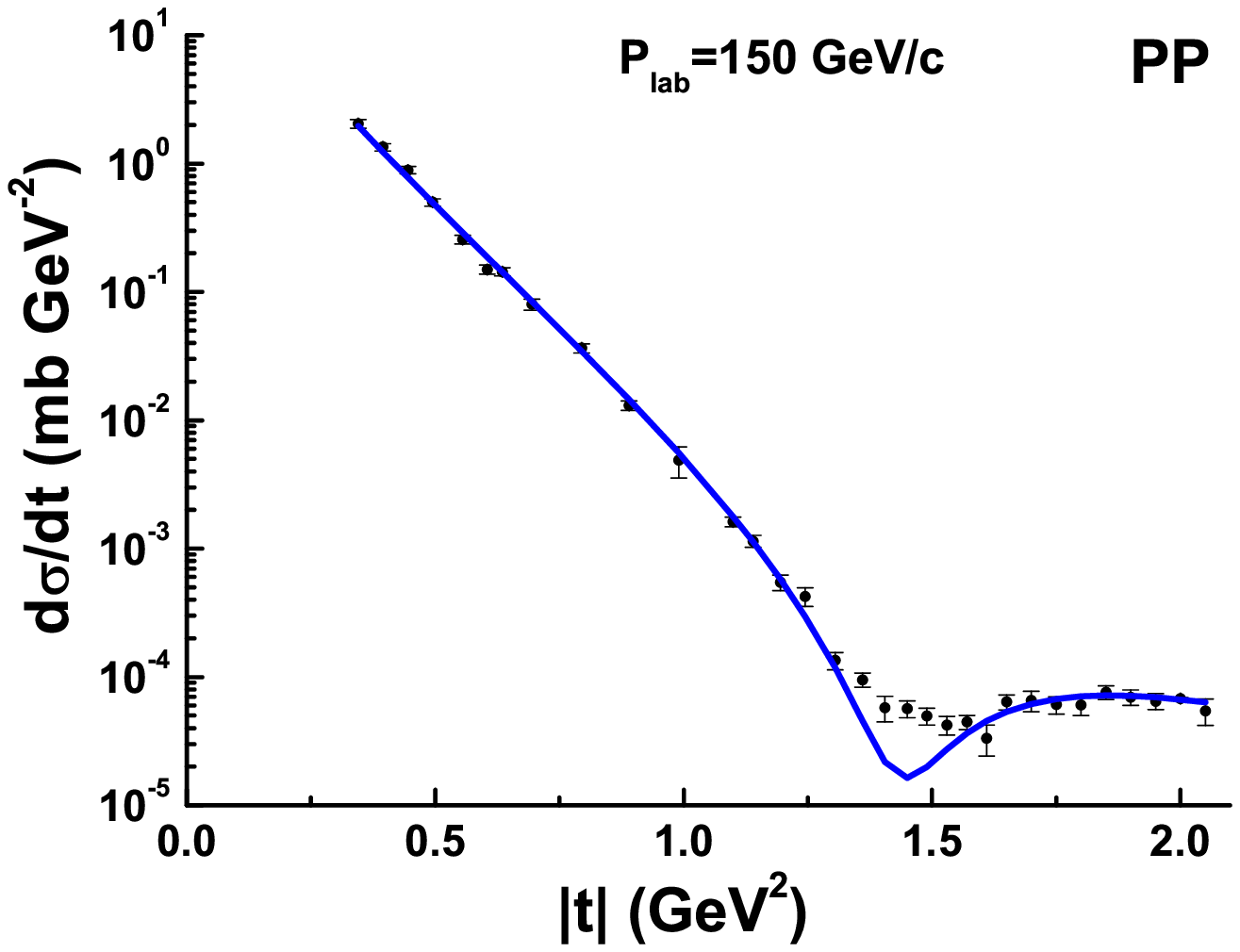}\hspace{5mm}\includegraphics[width=75mm,height=66mm,clip]{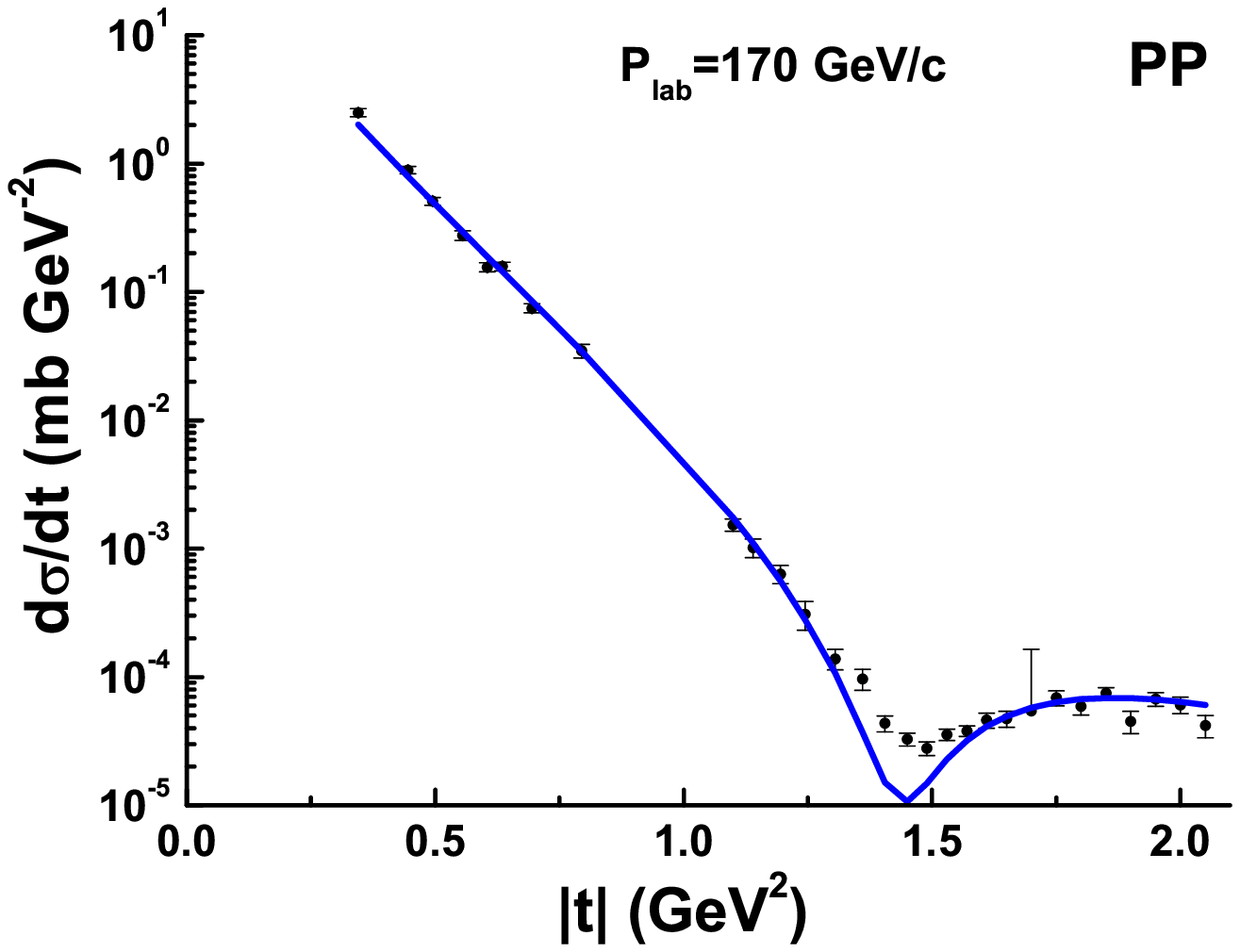}
\begin{minipage}{75mm}
{
\caption{The points are the experimental data by R. Rusack et al., Phys. Rev. Lett. {\bf 41} (1978) 1632.}
}
\end{minipage}
\hspace{5mm}
\begin{minipage}{75mm}
{
\caption{The points are the experimental data by R. Rusack et al., Phys. Rev. Lett. {\bf 41} (1978) 1632.}
}
\end{minipage}
%-------------------------------------------------------
\includegraphics[width=75mm,height=66mm,clip]{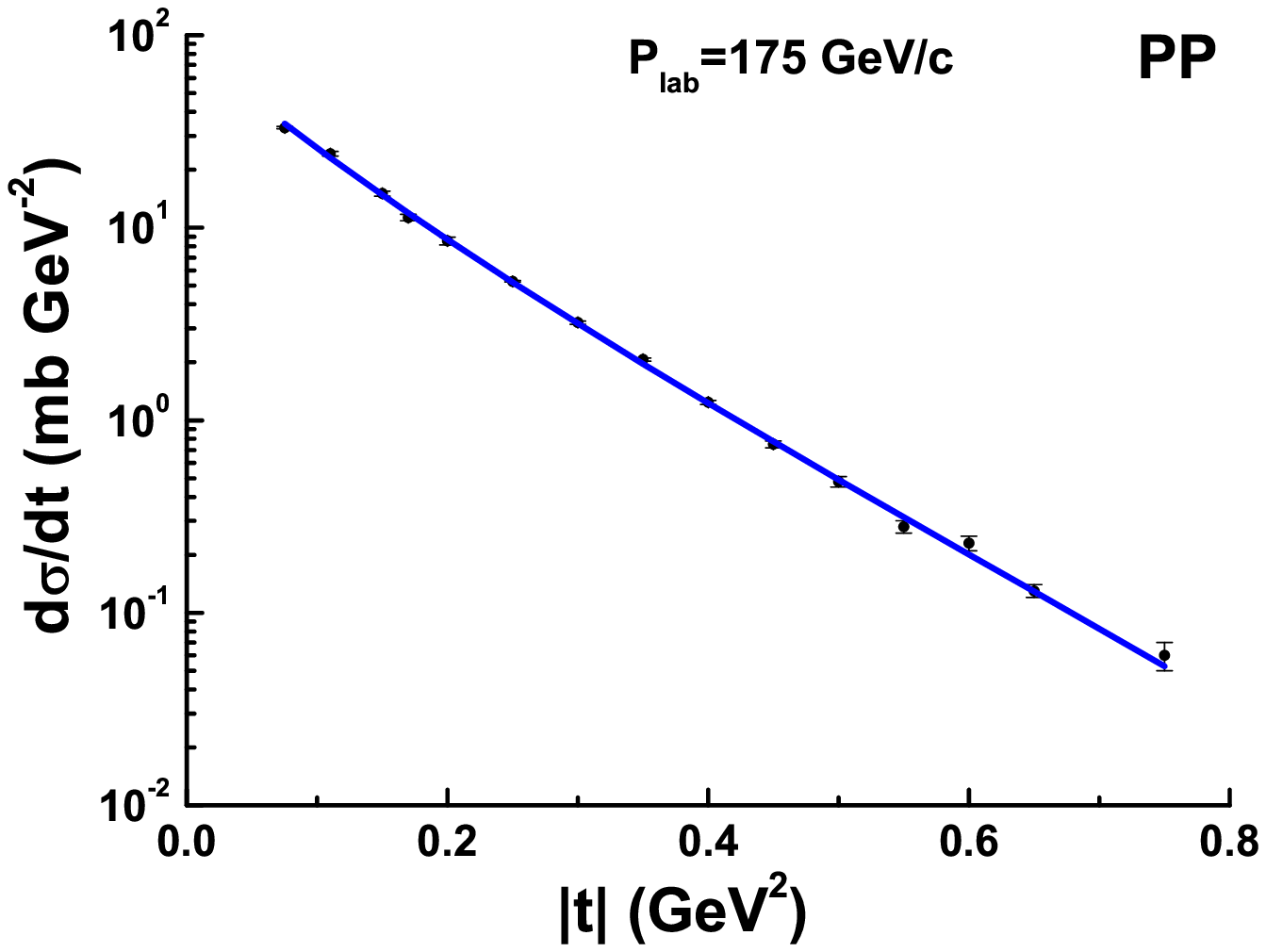}\hspace{5mm}\includegraphics[width=75mm,height=66mm,clip]{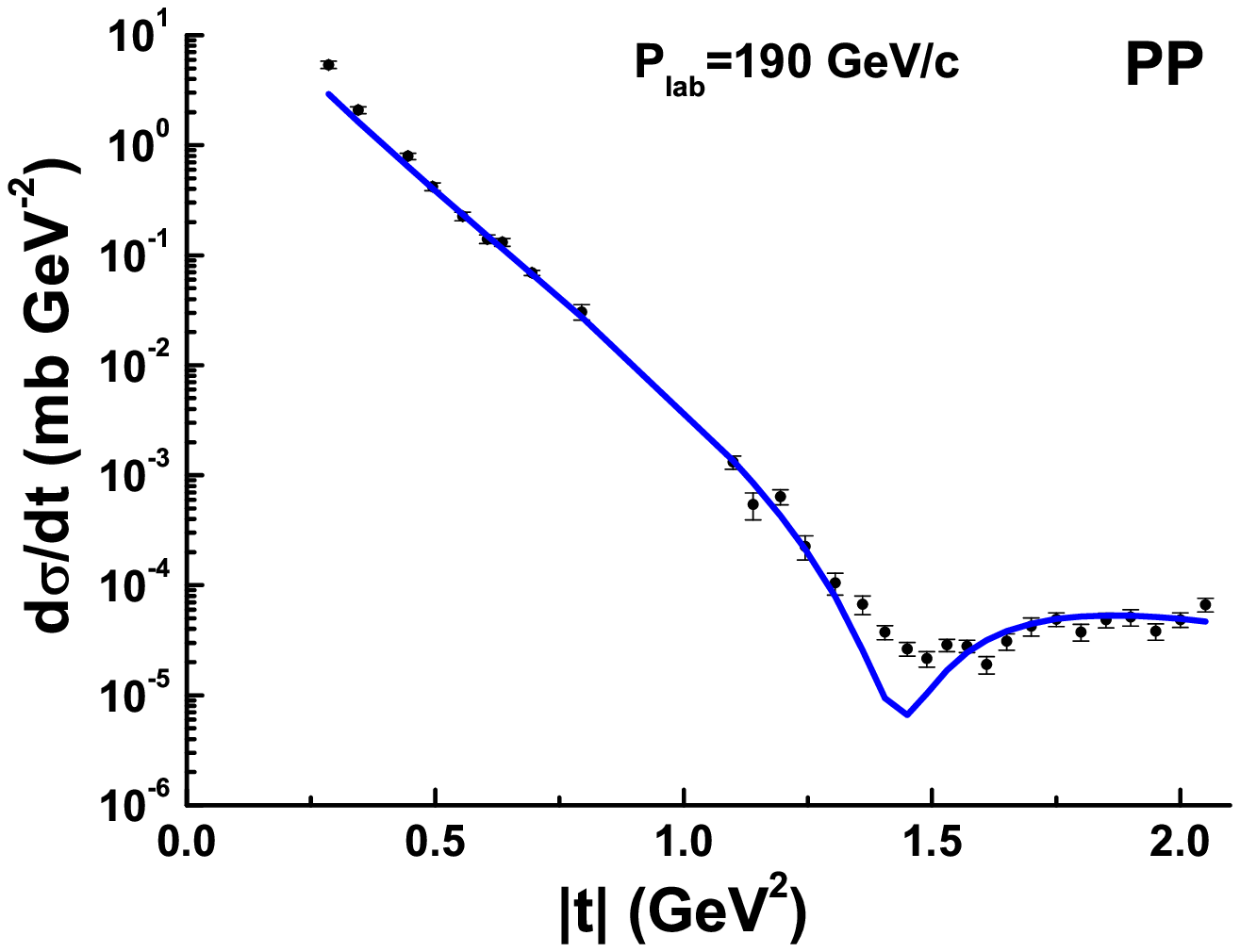}
\begin{minipage}{75mm}
{
\caption{The points are the experimental data by D.S. Ayres et al., Phys. Rev. {\bf D15} (1977) 3105.}
}
\end{minipage}
\hspace{5mm}
\begin{minipage}{75mm}
{
\caption{The points are the experimental data by R. Rusack et al., Phys. Rev. Lett. {\bf 41} (1978) 1632.}
}
\end{minipage}
\end{figure}

%%%%%%%%%%%%%%%%%%%%%%%%%%%%%%%%%%%
\begin{figure}[cbth]

%-------------------------------------------------------
\includegraphics[width=75mm,height=66mm,clip]{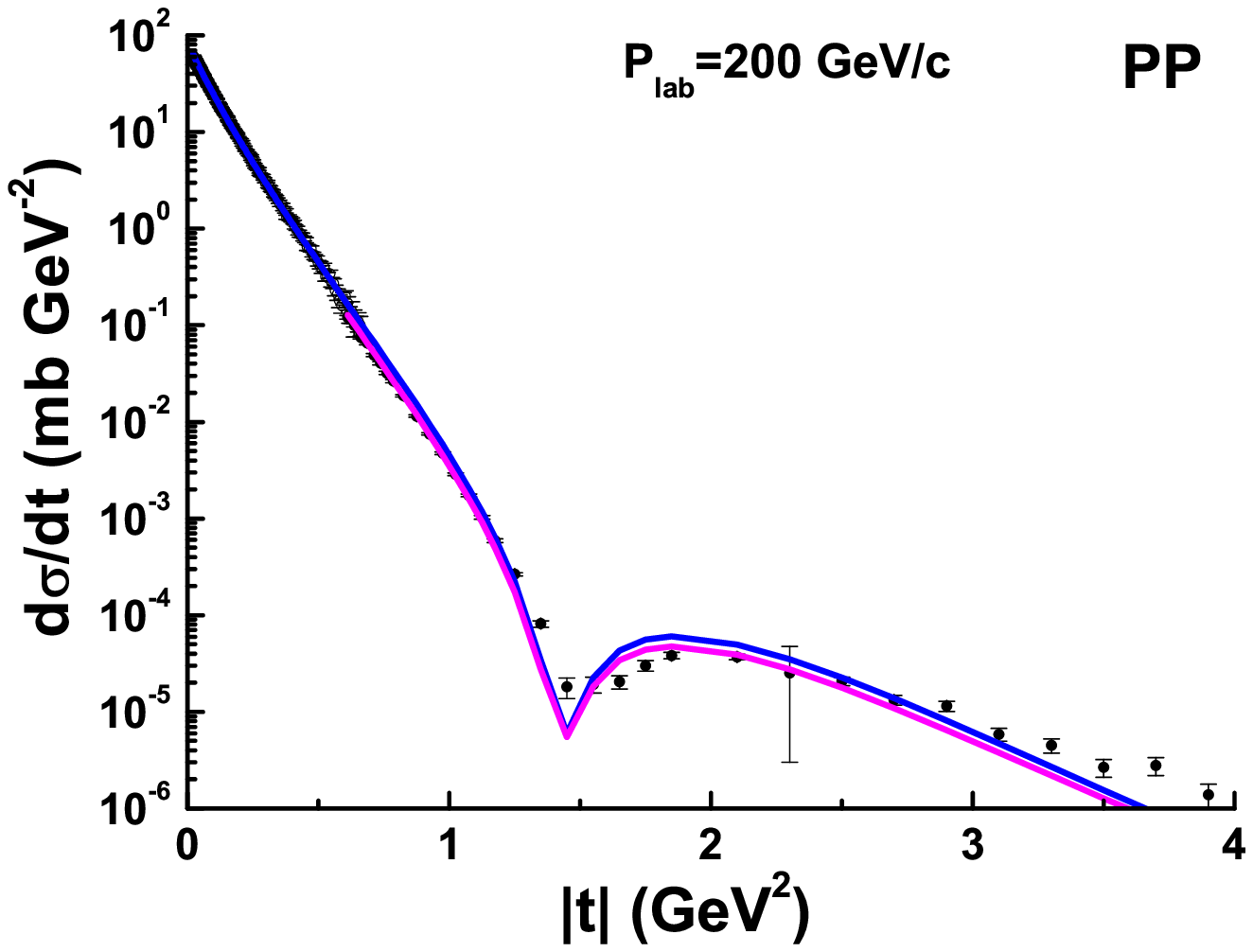}\hspace{5mm}\includegraphics[width=75mm,height=66mm,clip]{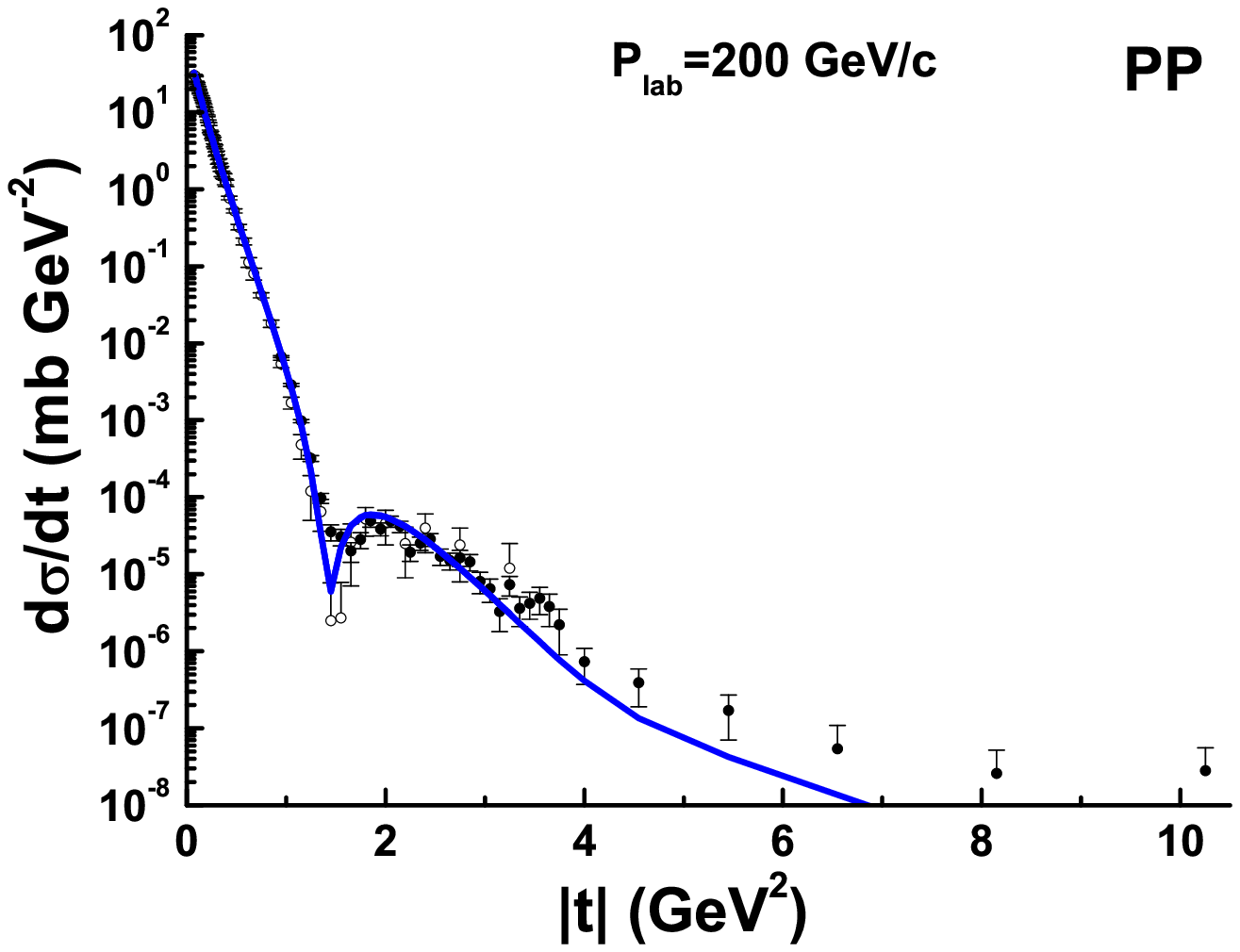}
\begin{minipage}{75mm}
{
\caption{The points are the experimental data by C.W. Akerlof et al., Phys. Rev. {\bf D14} (1976) 2864;
R. Rubinstein et al., Phys. Rev.{\bf D30} (1984) 1413.}
}
\end{minipage}
\hspace{5mm}
\begin{minipage}{75mm}
{
\caption{The points are the experimental data by A. Schiz et al., Phys. Rev. {\bf D24} (1981) 26;
G. Fidecaro et al., Nucl. Phys. {\bf B173} (1980) 513.}
}
\end{minipage}
%-------------------------------------------------------
\includegraphics[width=75mm,height=66mm,clip]{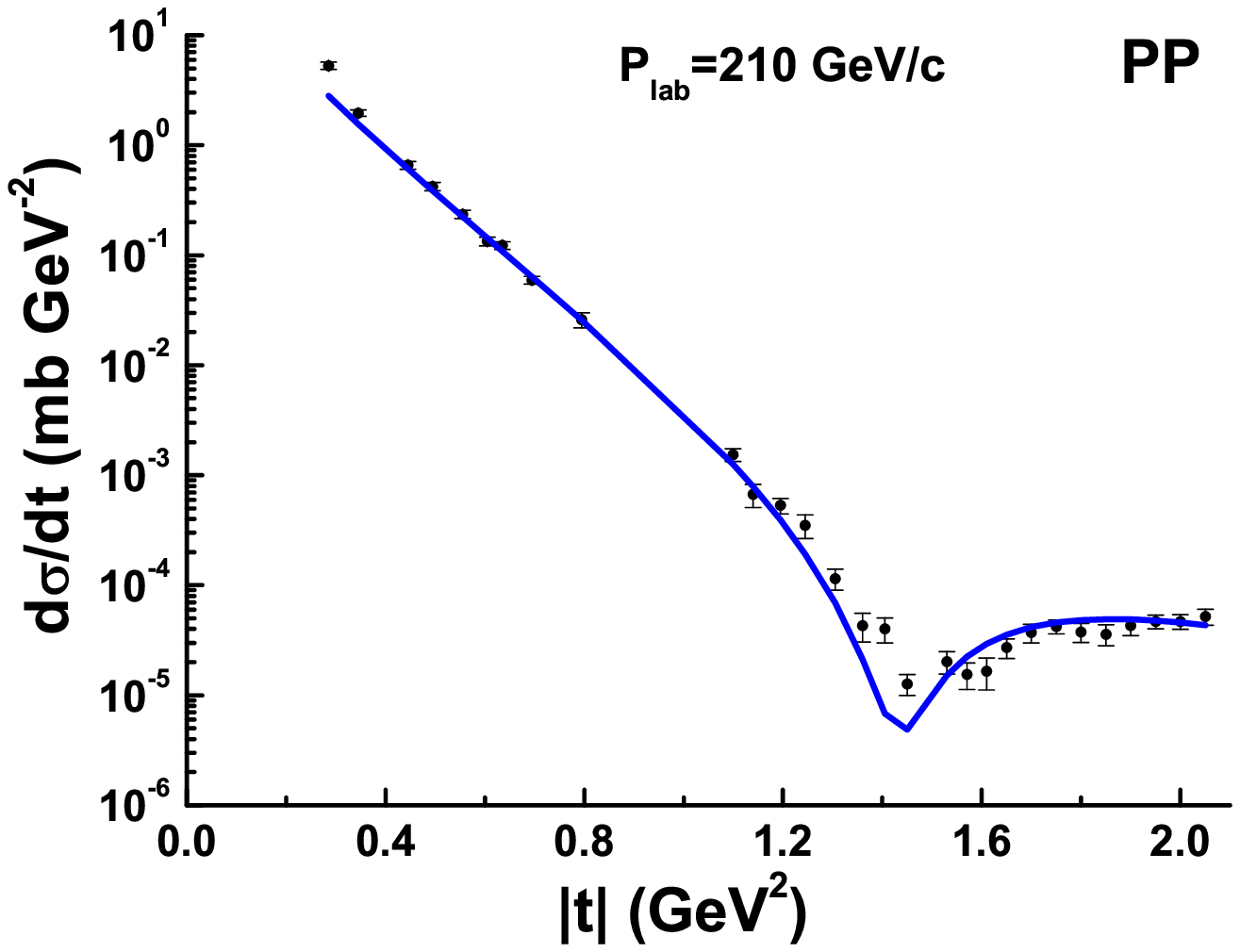}\hspace{5mm}\includegraphics[width=75mm,height=66mm,clip]{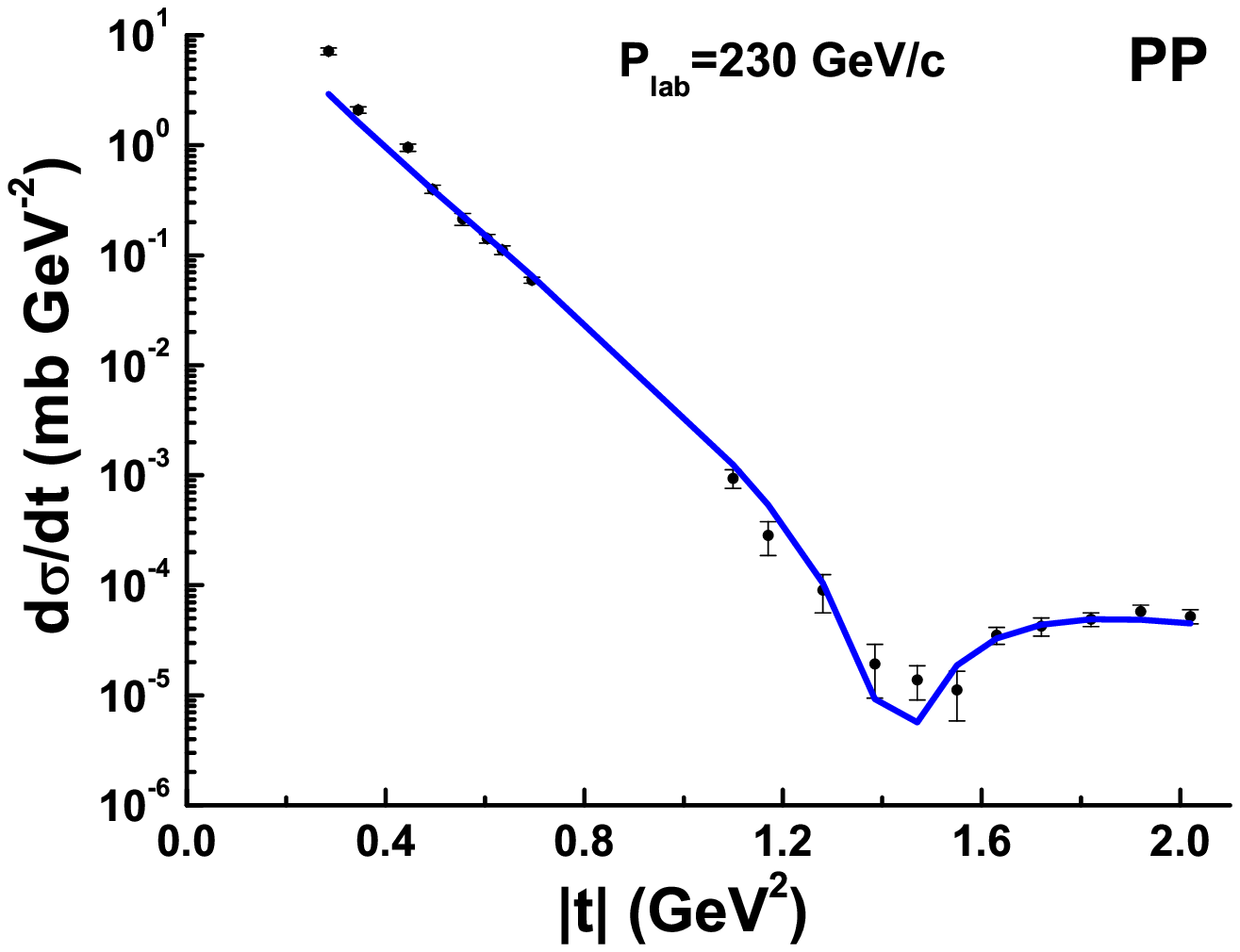}
\begin{minipage}{75mm}
{
\caption{The points are the experimental data by R. Rusack et al., Phys. Rev. Lett. {\bf 41} (1978) 1632.}
}
\end{minipage}
\hspace{5mm}
\begin{minipage}{75mm}
{
\caption{The points are the experimental data by R. Rusack et al., Phys. Rev. Lett. {\bf 41} (1978) 1632.}
}
\end{minipage}
%-------------------------------------------------------
\includegraphics[width=75mm,height=66mm,clip]{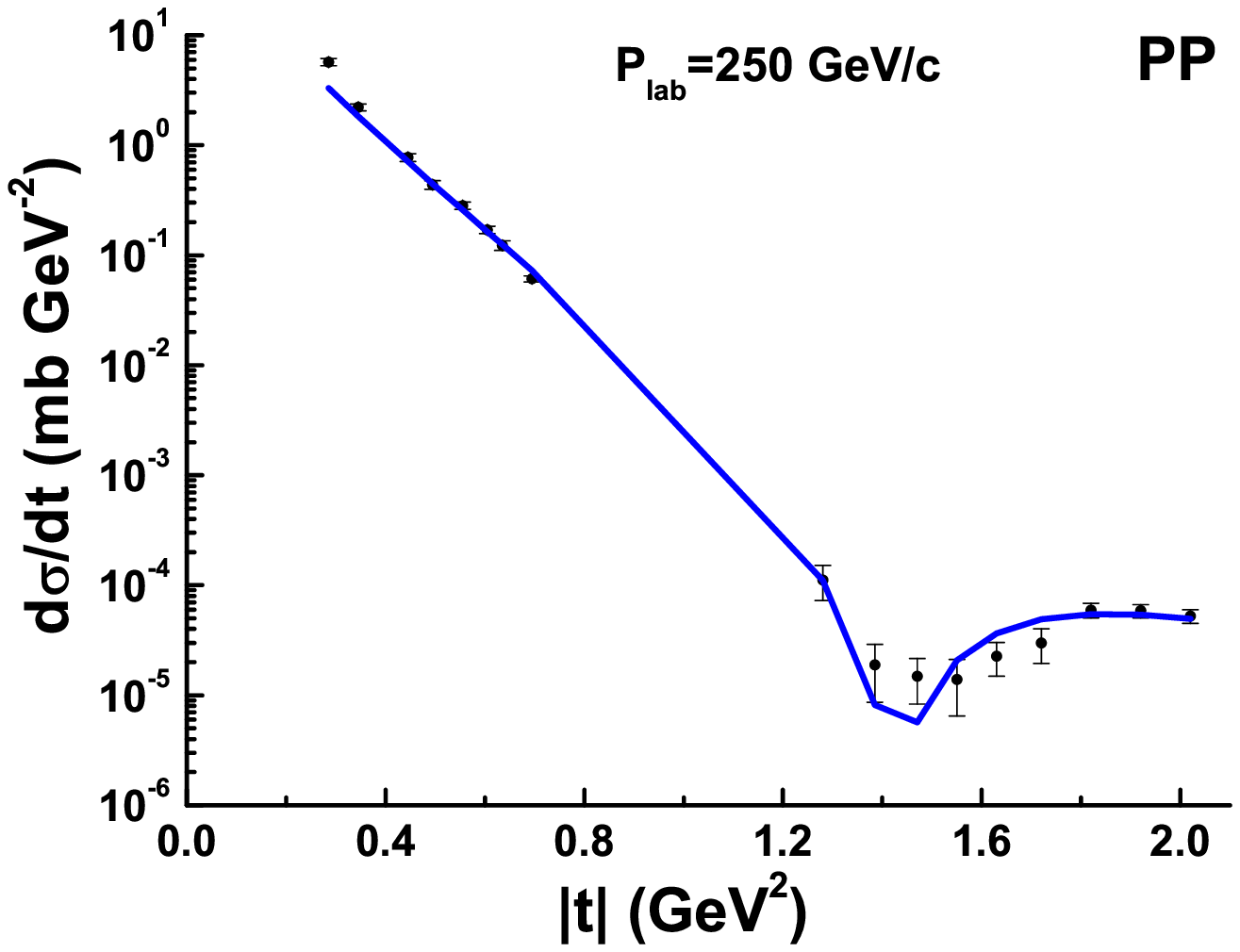}\hspace{5mm}\includegraphics[width=75mm,height=66mm,clip]{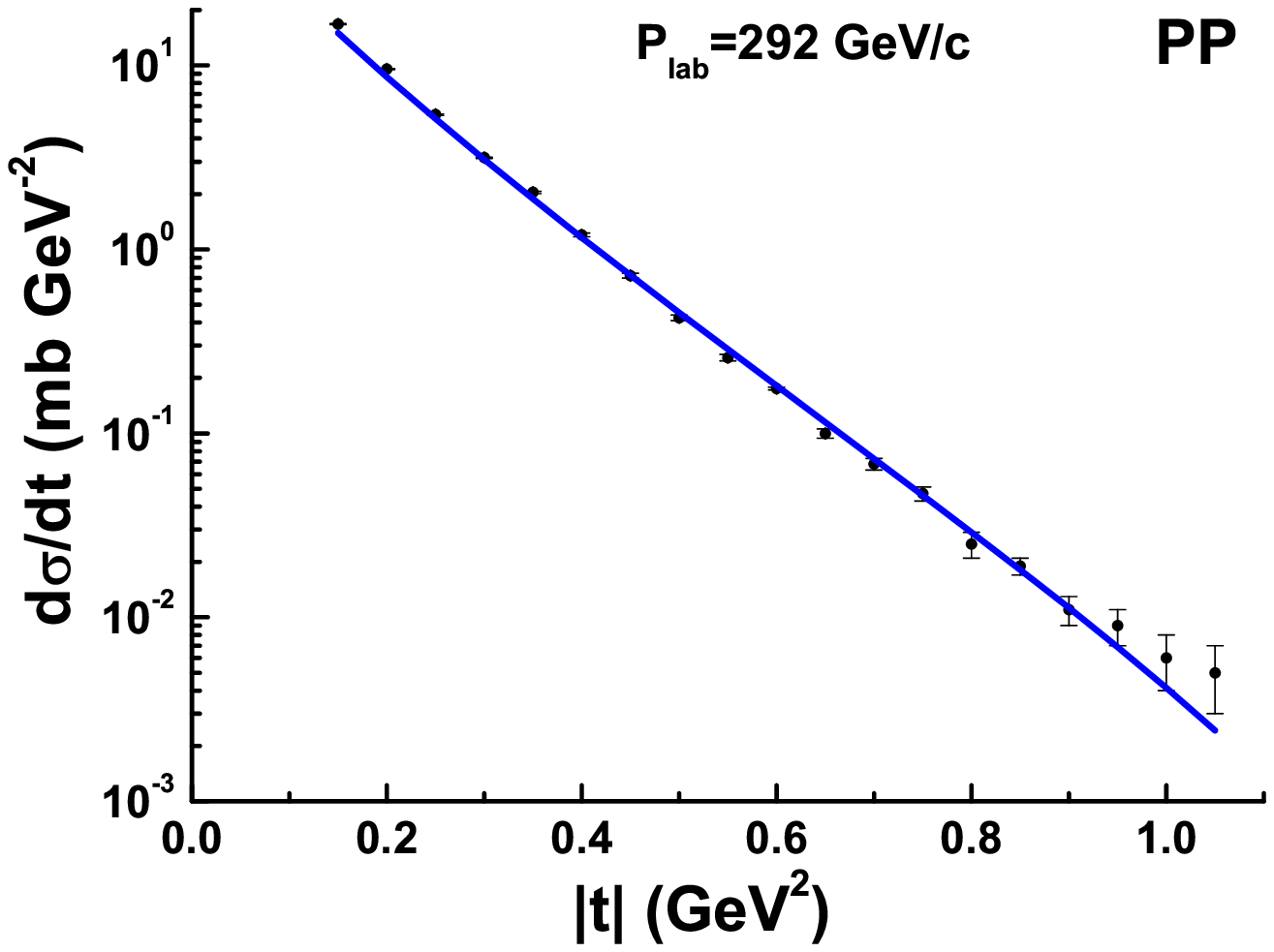}
\begin{minipage}{75mm}
{
\caption{The points are the experimental data by R. Rusack et al., Phys. Rev. Lett. {\bf 41} (1978) 1632.}
}
\end{minipage}
\hspace{5mm}
\begin{minipage}{75mm}
{
\caption{The points are the experimental data by M.G. Albrow et al., Nucl. Phys. {\bf B108} (1976) 1.}
}
\end{minipage}
%-------------------------------------------------------
\end{figure}

\begin{figure}[cbth]
%-------------------------------------------------------
\includegraphics[width=75mm,height=66mm,clip]{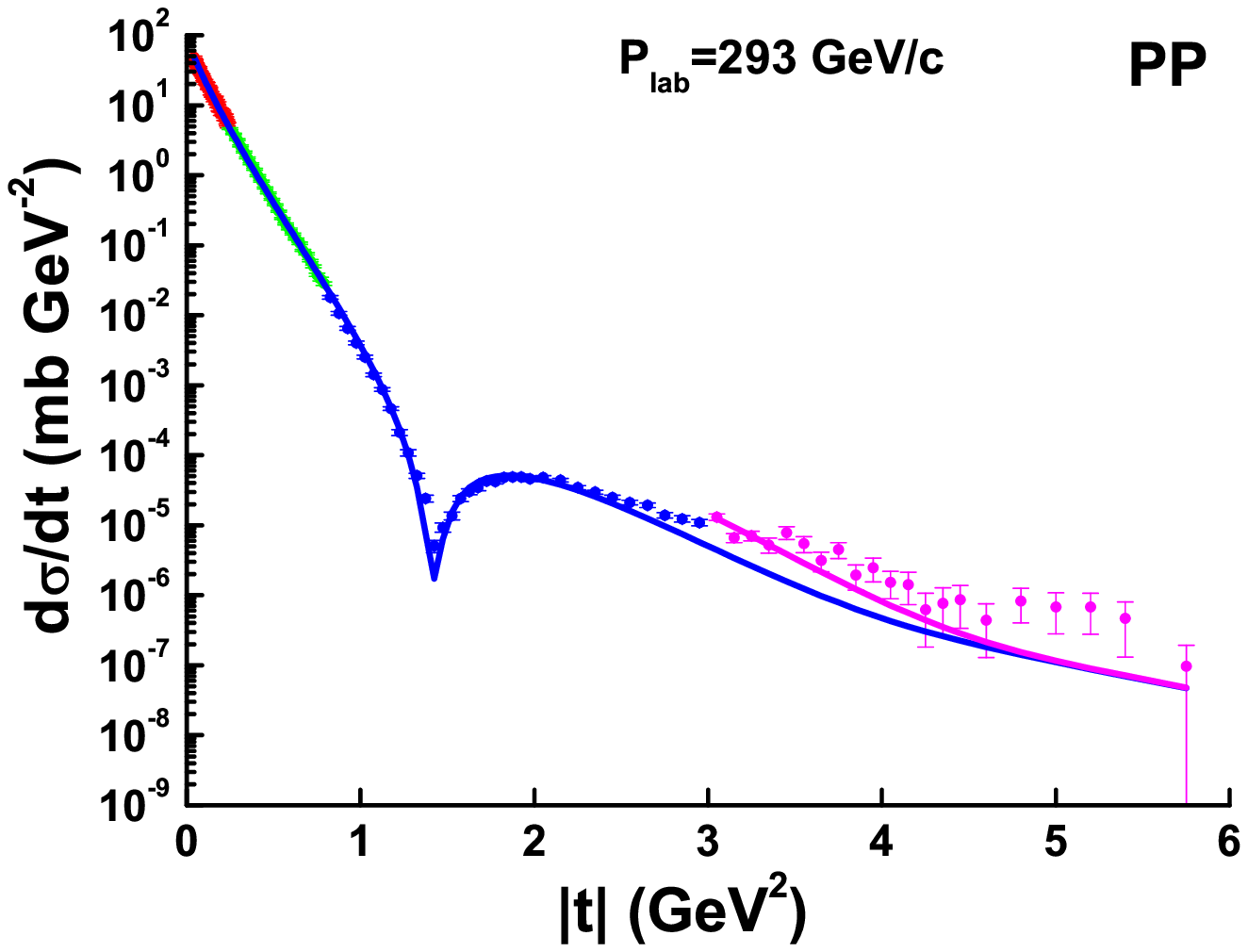}\hspace{5mm}\includegraphics[width=75mm,height=66mm,clip]{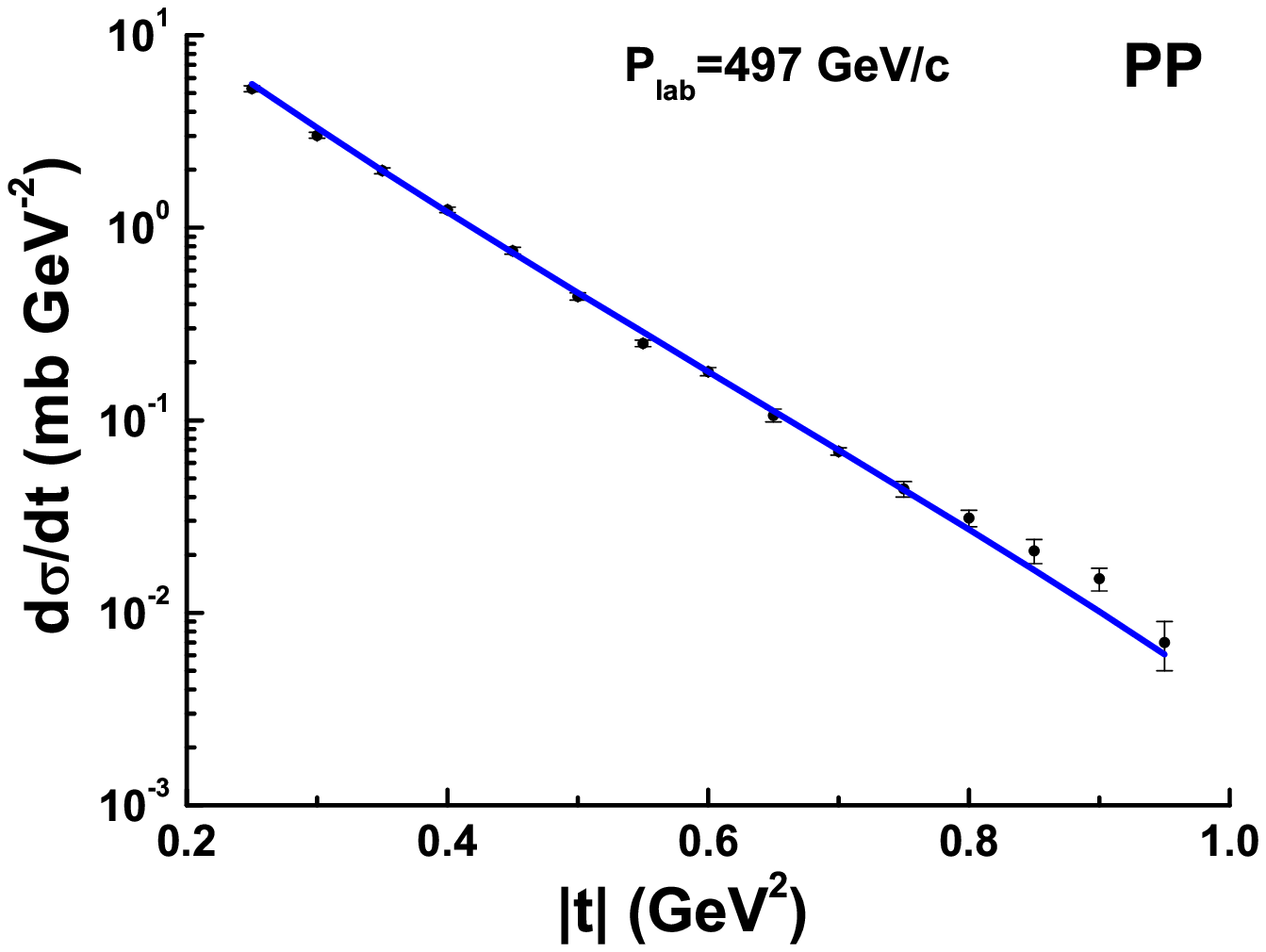}
\begin{minipage}{75mm}
{
\caption{The points are the experimental data by U. Amaldi and K.R. Schubert, Nucl. Phys. {\bf B166} (1980) 301.}
}
\end{minipage}
\hspace{5mm}
\begin{minipage}{75mm}
{
\caption{The points are the experimental data by M.G. Albrow et al., Nucl. Phys. {\bf B108} (1976) 1.}
}
\end{minipage}
%-------------------------------------------------------
\includegraphics[width=75mm,height=66mm,clip]{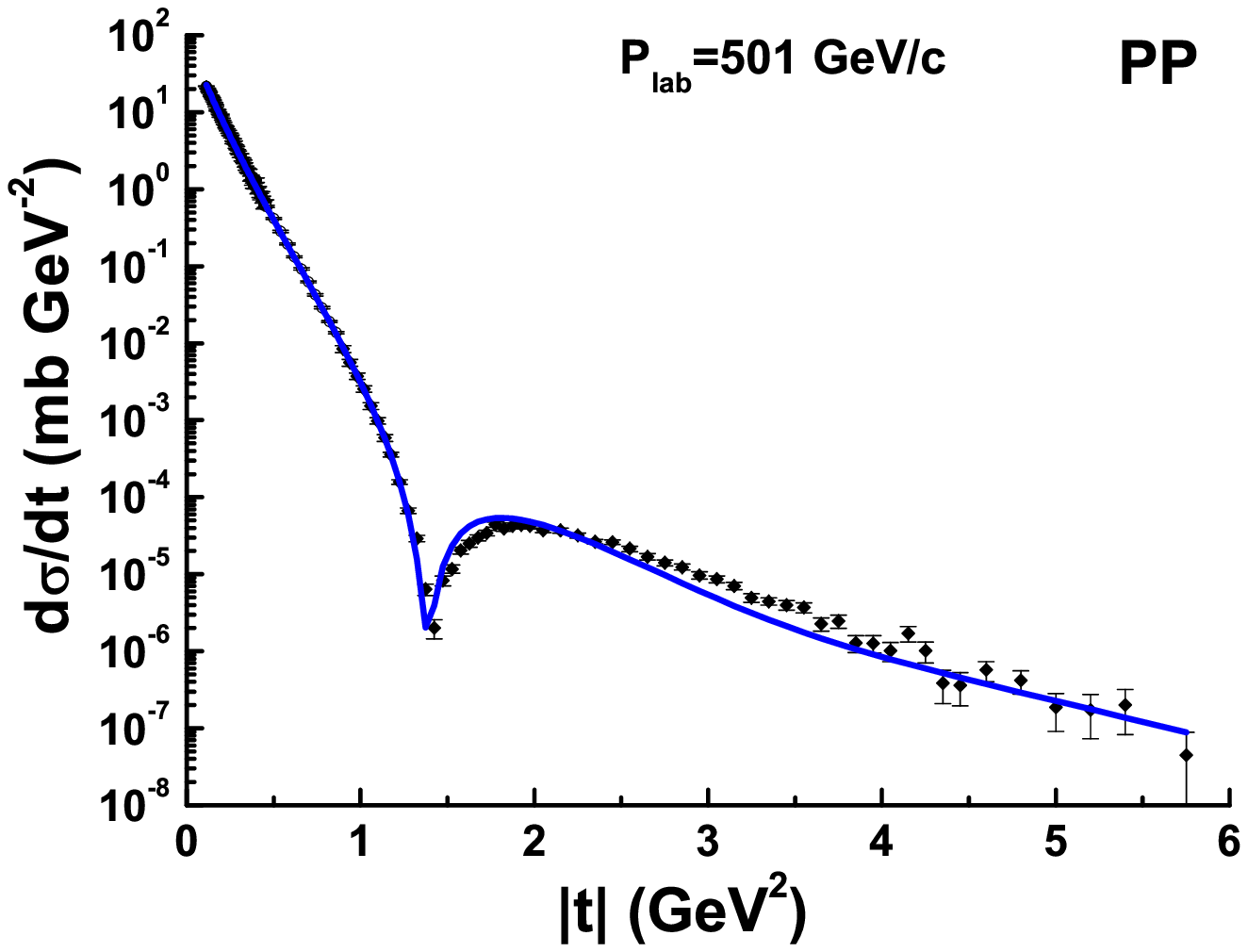}\hspace{5mm}\includegraphics[width=75mm,height=66mm,clip]{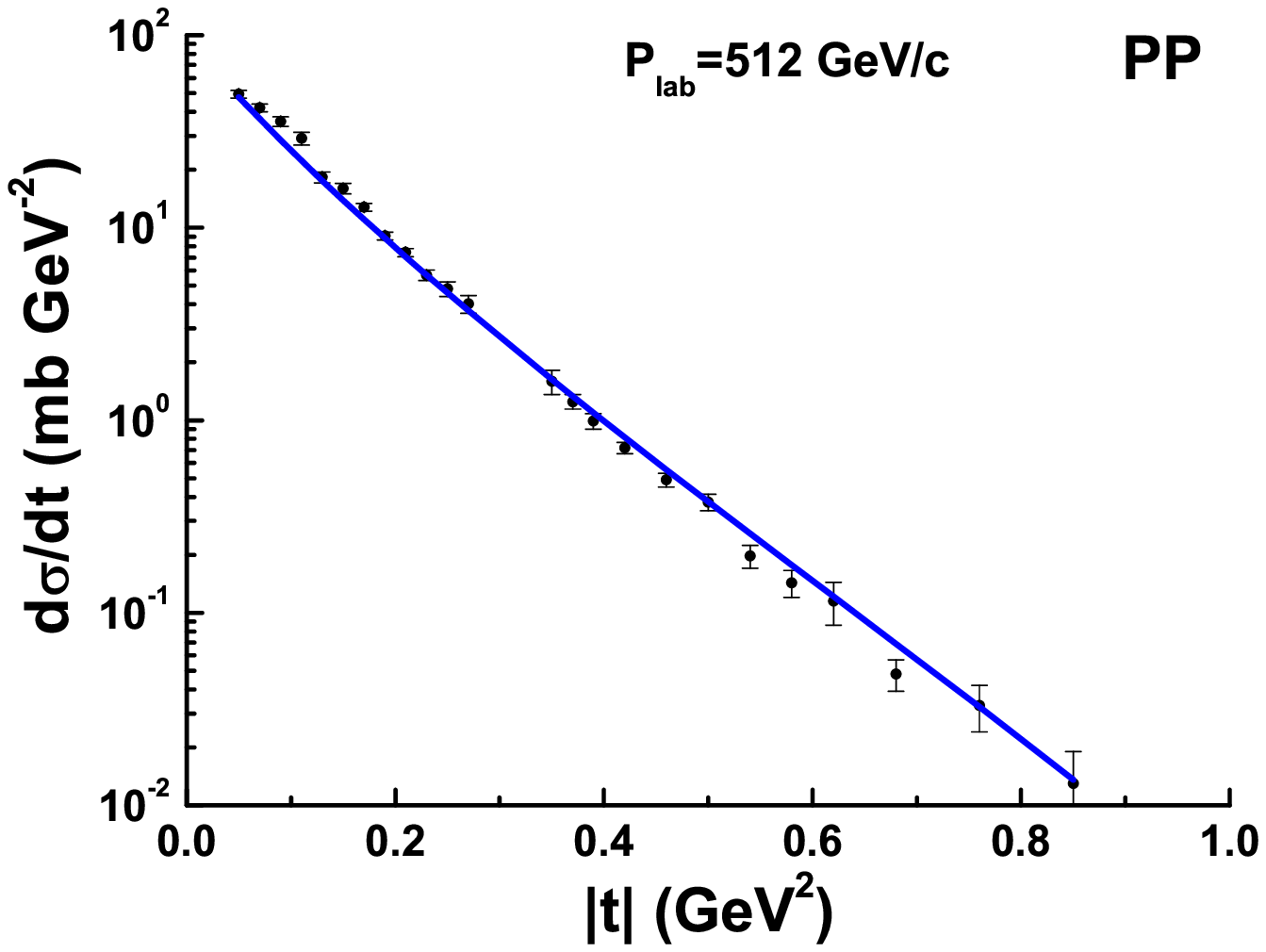}
\begin{minipage}{75mm}
{
\caption{The points are the experimental data by U. Amaldi and K.R. Schubert, Nucl. Phys. {\bf B166} (1980) 301.}
}
\end{minipage}
\hspace{5mm}
\begin{minipage}{75mm}
{
\caption{The points are the experimental data by A. Breakstone et al. Nucl. Phys. {\bf B248} (1984) 253.}
}
\end{minipage}

%-------------------------------------------------------
\includegraphics[width=75mm,height=66mm,clip]{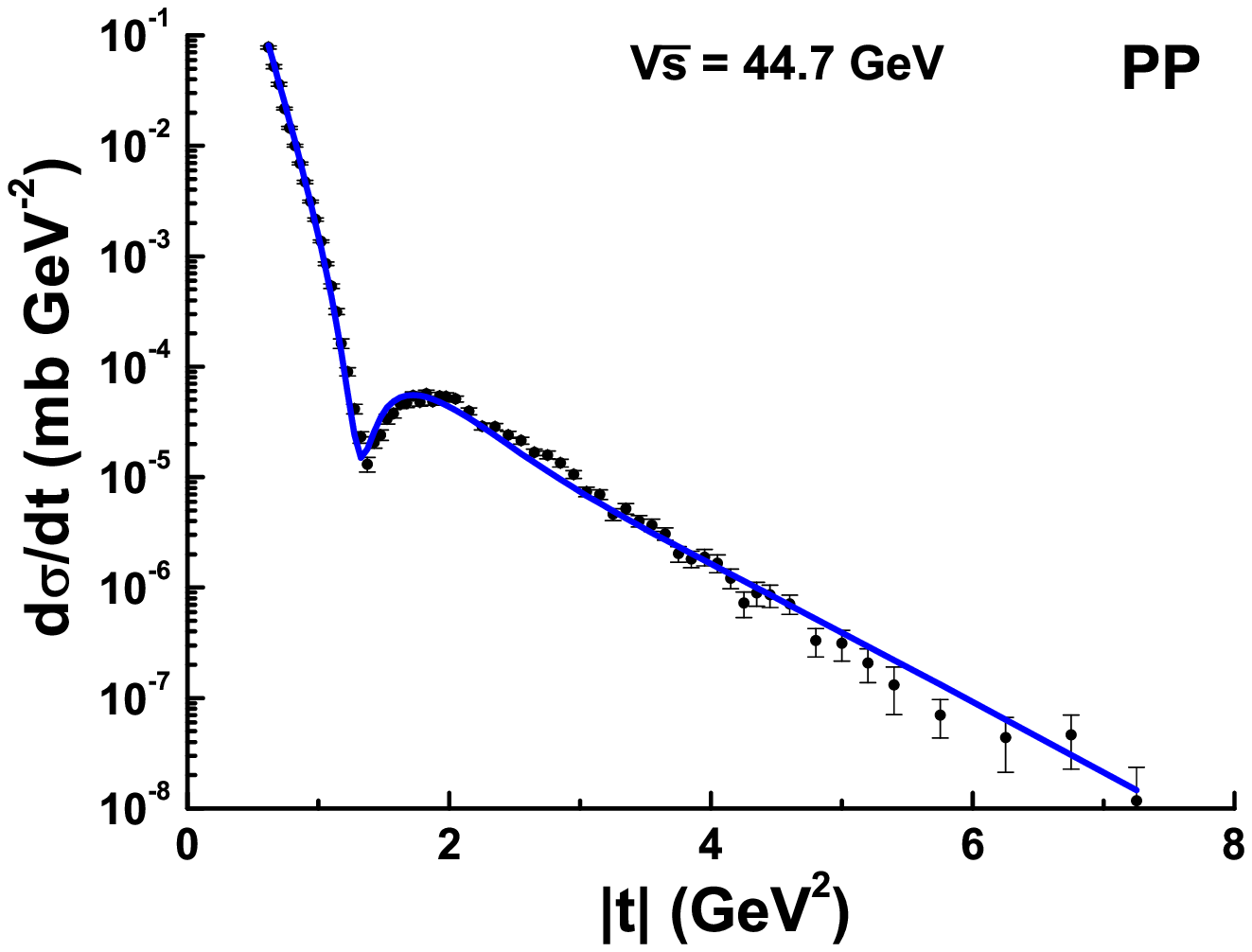}\hspace{5mm}\includegraphics[width=75mm,height=66mm,clip]{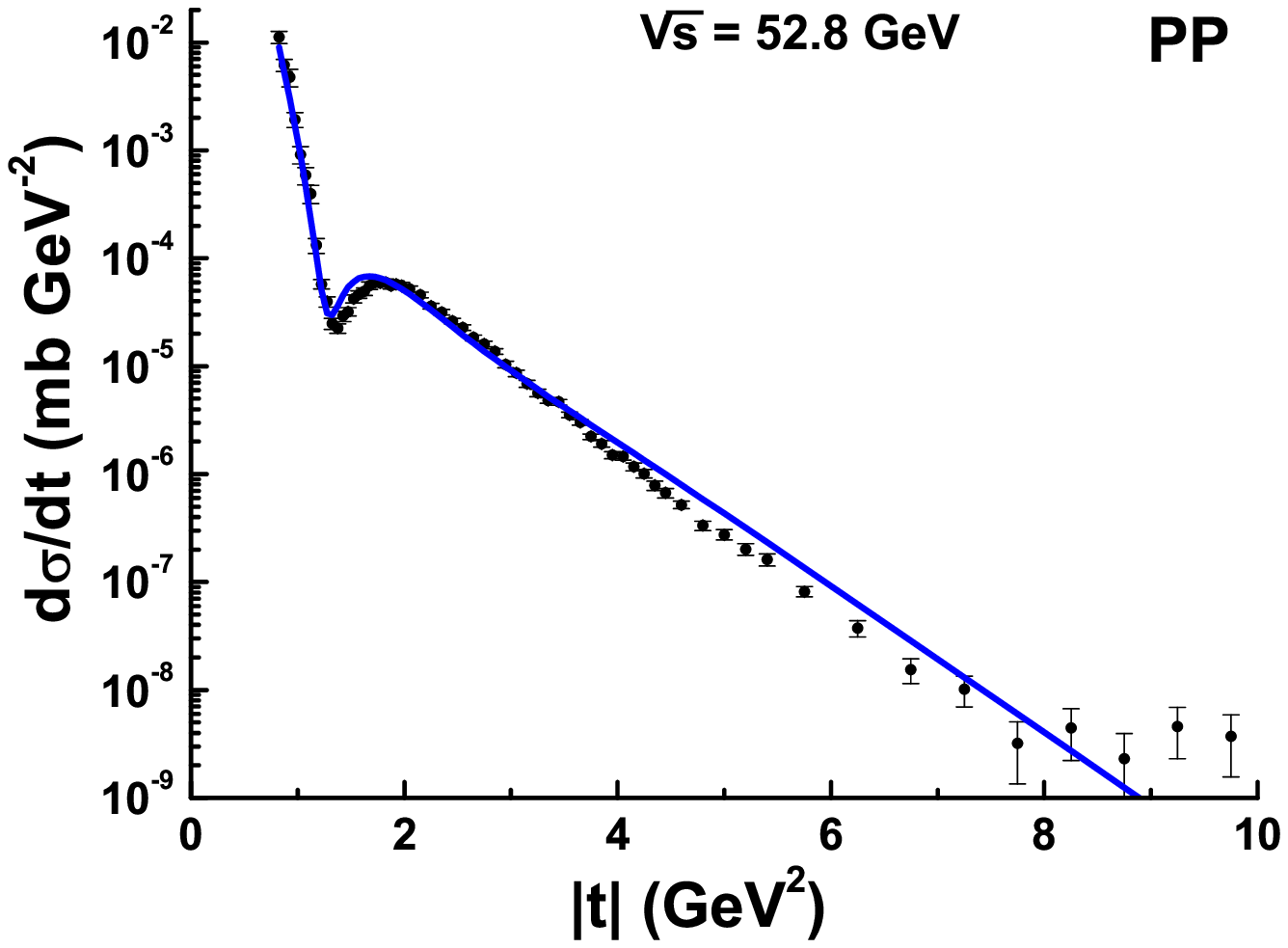}
\begin{minipage}{75mm}
{
\caption{The points are the experimental data by U. Amaldi and K.R. Schubert, Nucl. Phys. {\bf B166} (1980) 301.}
}
\end{minipage}
\hspace{5mm}
\begin{minipage}{75mm}
{
\caption{The points are the experimental data by E. Nagy et al., Nucl. Phys. {\bf B150} (1979) 221.}
}
\end{minipage}
\end{figure}

%%%%%%%%%%%%%%%%%%%%%%%%%%%%%%%%%%%
\begin{figure}[cbth]

%-------------------------------------------------------
\includegraphics[width=75mm,height=66mm,clip]{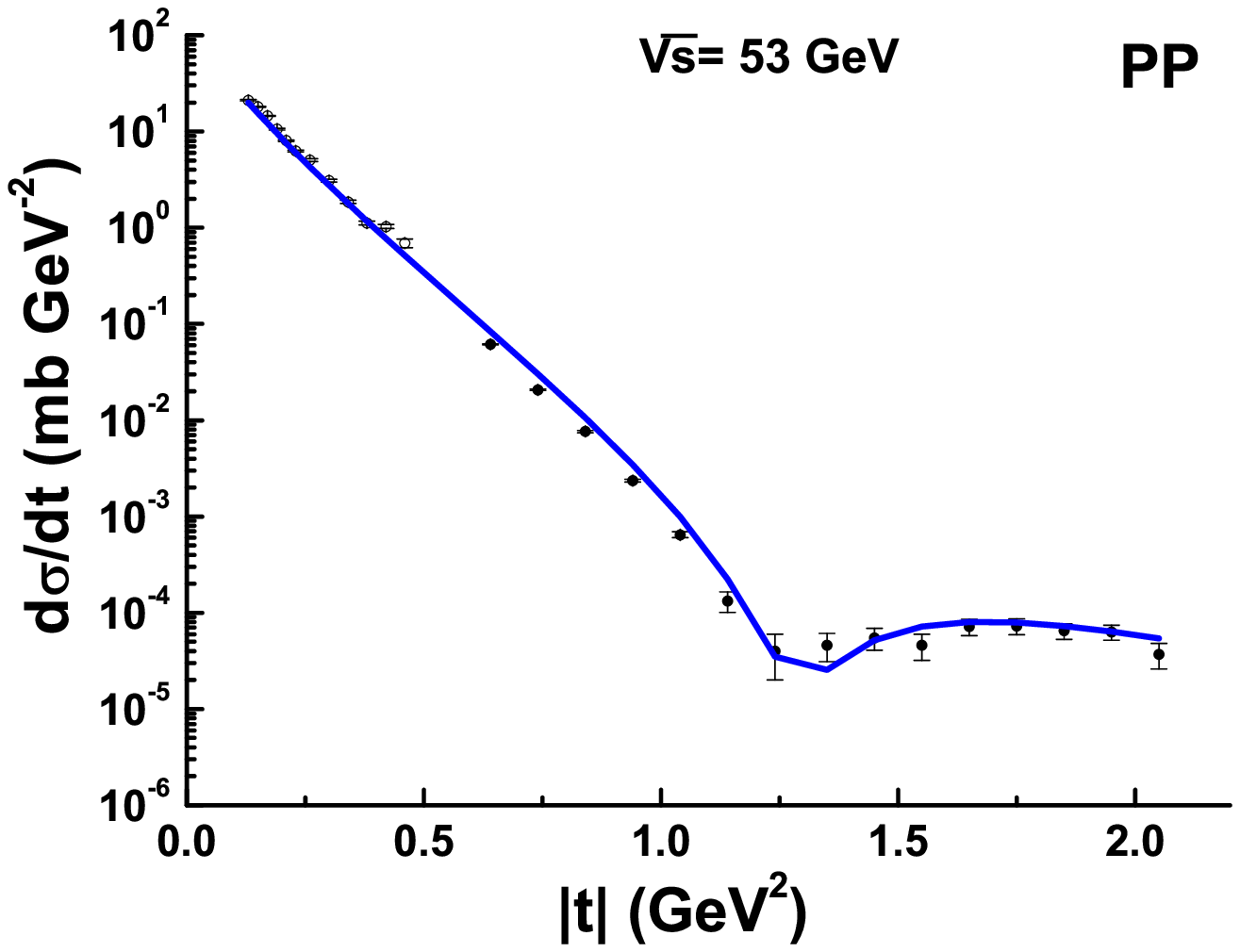}\hspace{5mm}\includegraphics[width=75mm,height=66mm,clip]{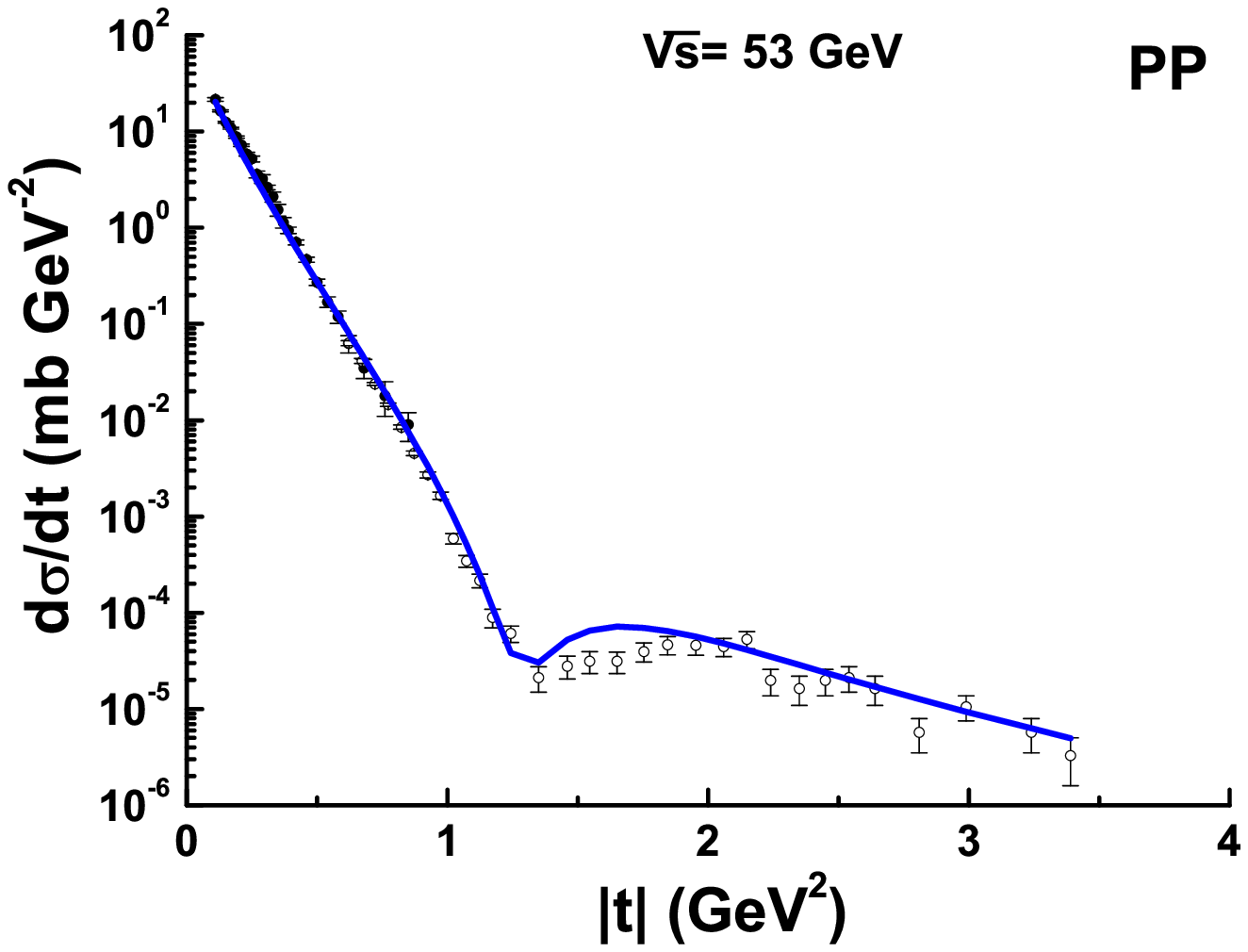}
\begin{minipage}{75mm}
{
\caption{The points are the experimental data by A. Breakstone et al. Nucl. Phys. {\bf B248} (1984) 253;
 Phys. Rev. Lett. {\bf 54} (1985) 2180.}
}
\end{minipage}
\hspace{5mm}
\begin{minipage}{75mm}
{
\caption{The points are the experimental data by S. Erhan et al., Phys. Lett. {\bf B152} (1985) 131;
J.C.M. Armitage et al., Nucl. Phys. {\bf B132} (1978) 365.}
}
\end{minipage}
%-------------------------------------------------------
\includegraphics[width=75mm,height=66mm,clip]{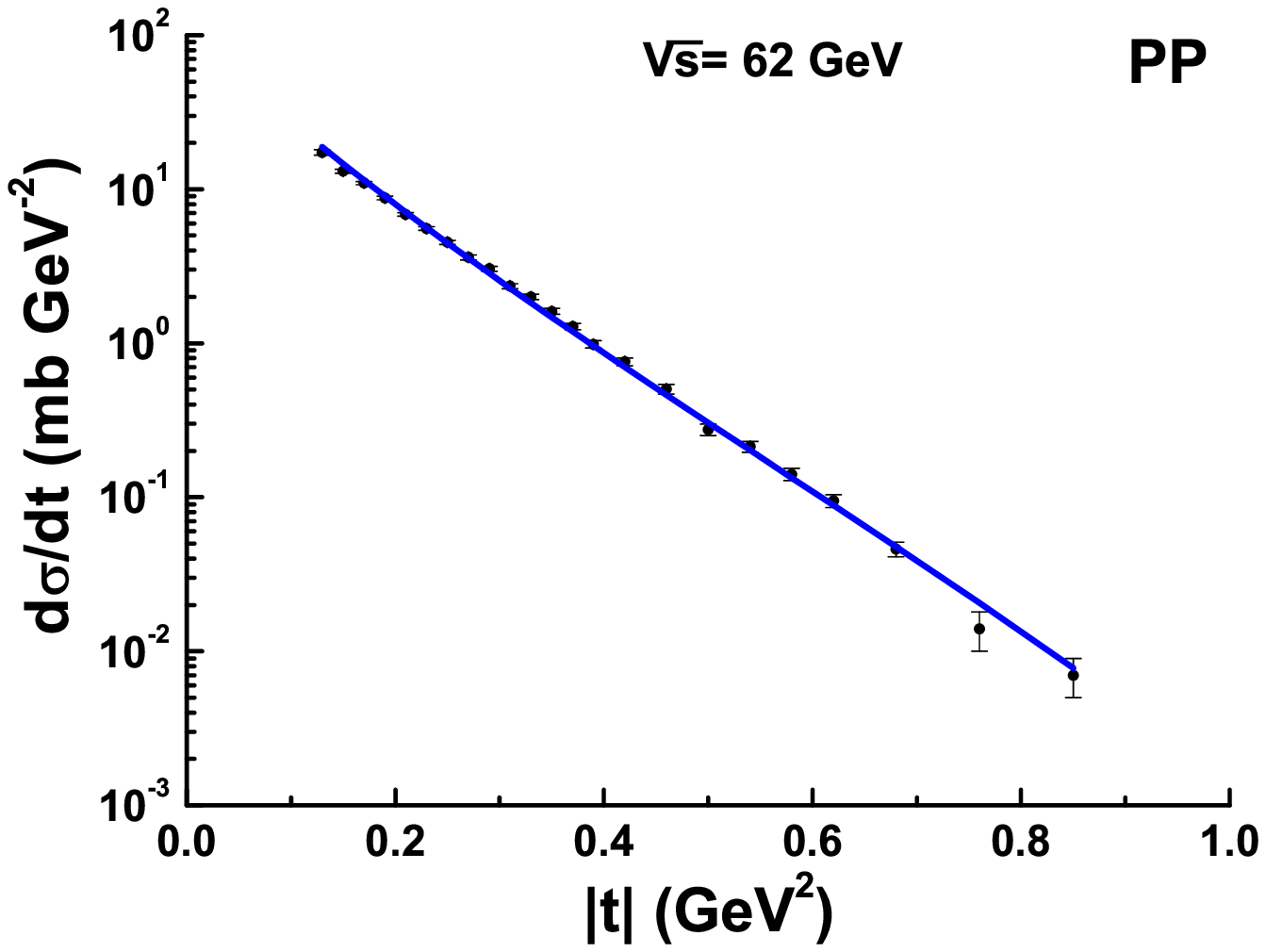}\hspace{5mm}\includegraphics[width=75mm,height=66mm,clip]{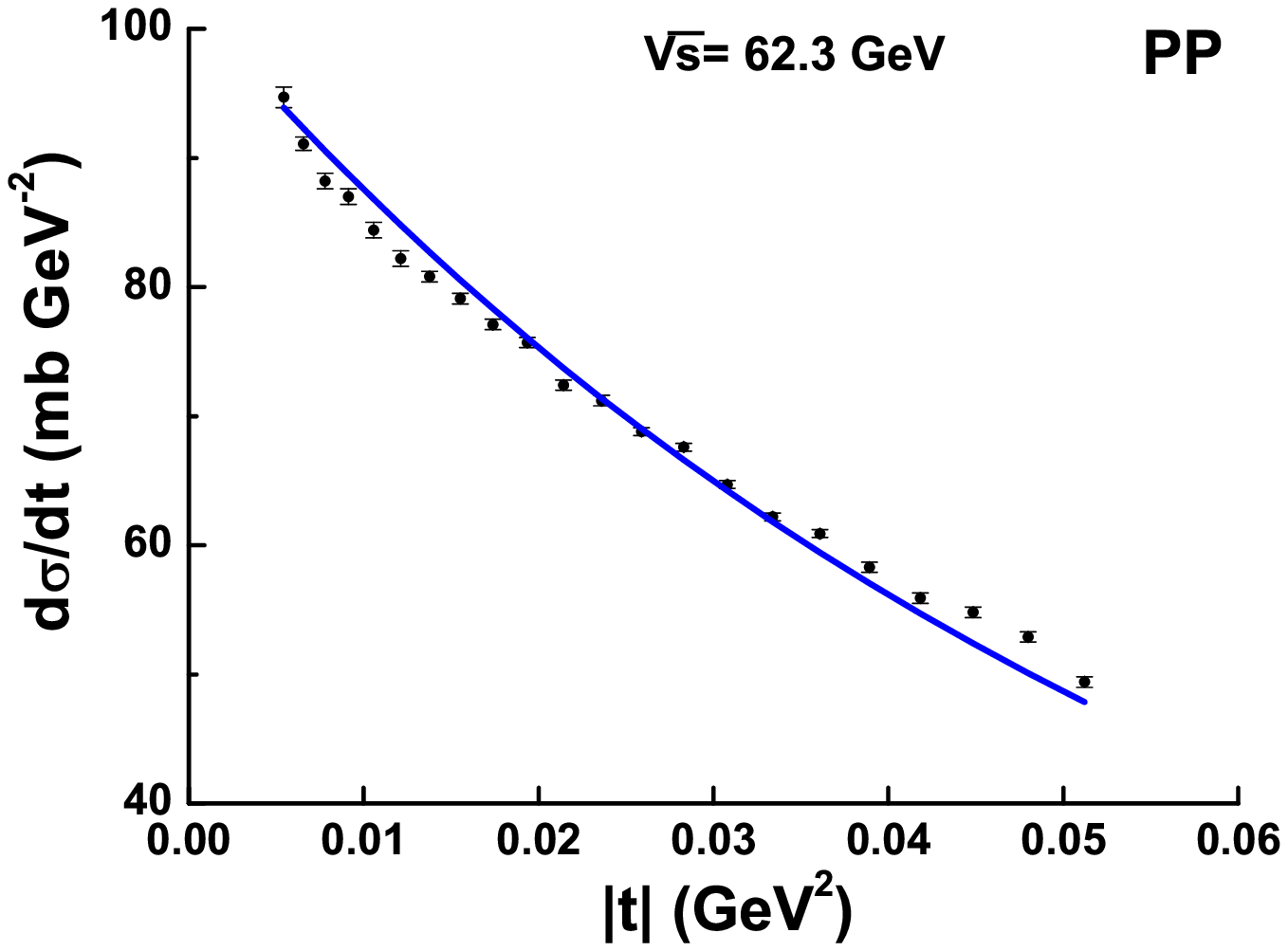}
\begin{minipage}{75mm}
{
\caption{The points are the experimental data by A. Breakstone et al. Nucl. Phys. {\bf B248} (1984) 253.}
}
\end{minipage}
\hspace{5mm}
\begin{minipage}{75mm}
{
\caption{The points are the experimental data by N. Amos et al., Nucl. Phys. {\bf B262} (1985) 689.}
}
\end{minipage}

%-------------------------------------------------------
\includegraphics[width=75mm,height=66mm,clip]{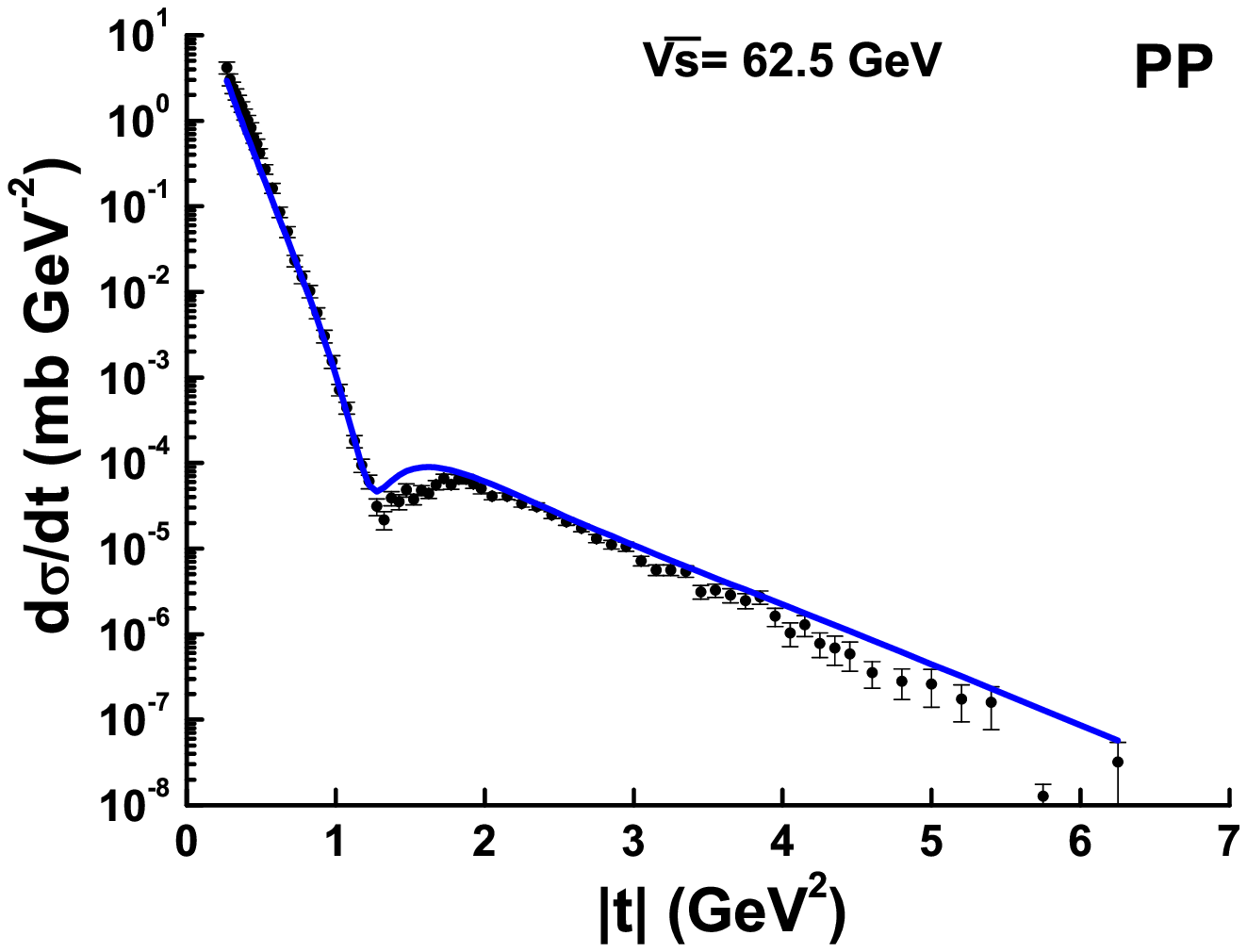}\hspace{5mm}\includegraphics[width=75mm,height=66mm,clip]{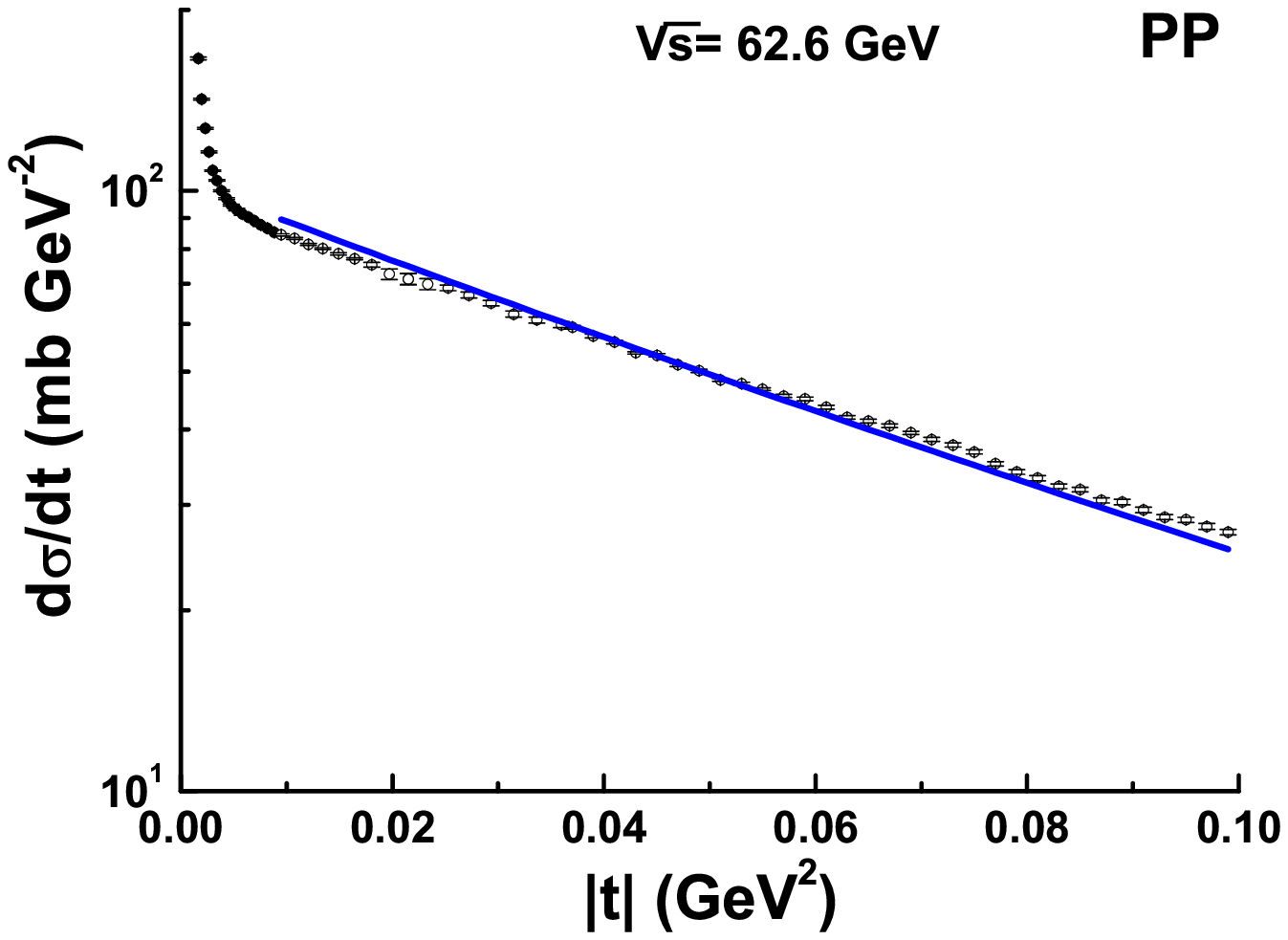}
\begin{minipage}{75mm}
{
\caption{The points are the experimental data by U. Amaldi and K.R. Schubert, Nucl. Phys. {\bf B166} (1980) 301.}
}
\end{minipage}
\hspace{5mm}
\begin{minipage}{75mm}
{
\caption{The points are the experimental data by U. Amaldi and K.R. Schubert, Nucl. Phys. {\bf B166} (1980) 301.}
}
\end{minipage}
%-------------------------------------------------------
\end{figure}

%%%%%%%%%%%%%%%%%%%%%%%%%%%%%%%%%%%
\begin{figure}[cbth]

%-------------------------------------------------------
\includegraphics[width=75mm,height=66mm,clip]{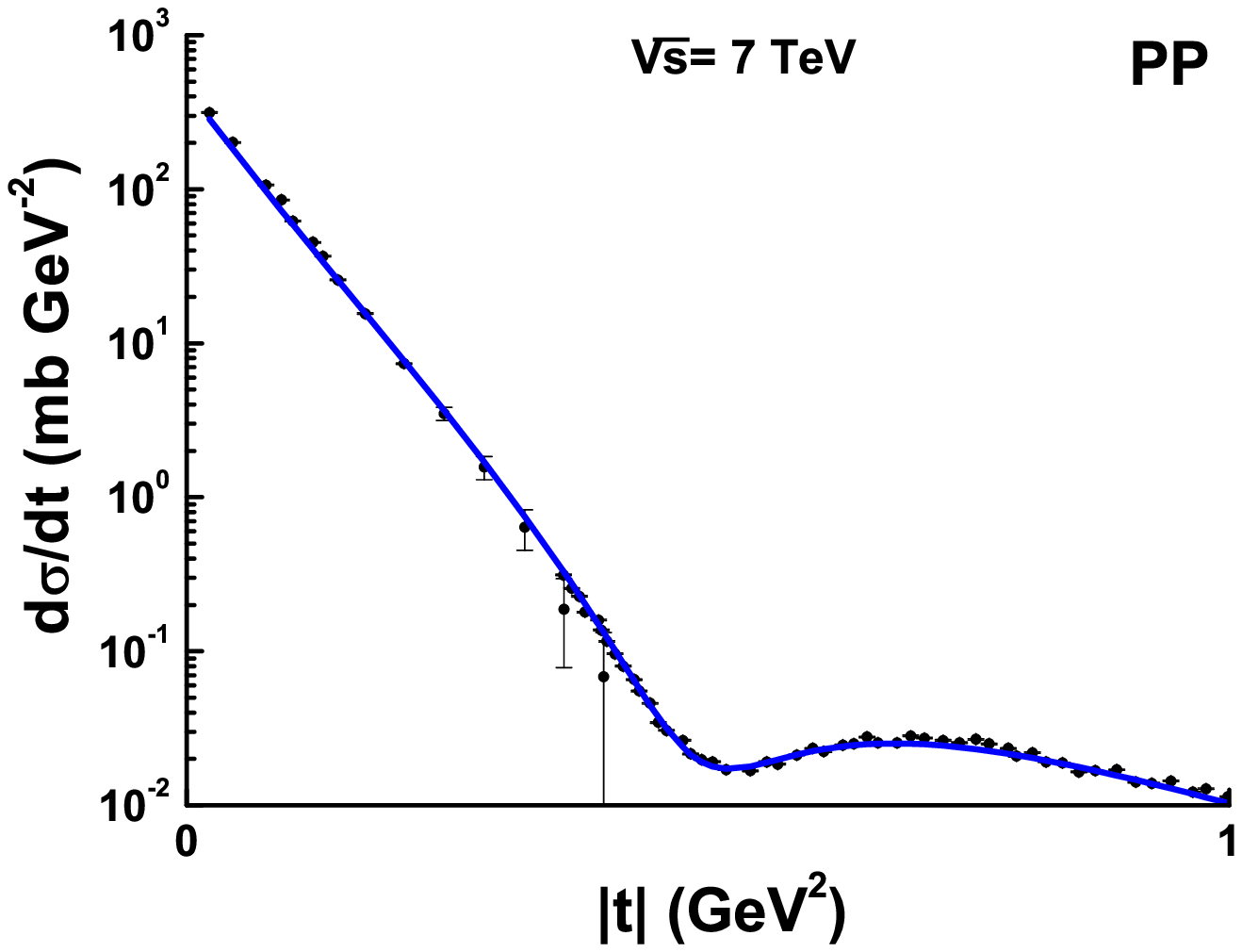}\hspace{5mm}\includegraphics[width=75mm,height=66mm,clip]{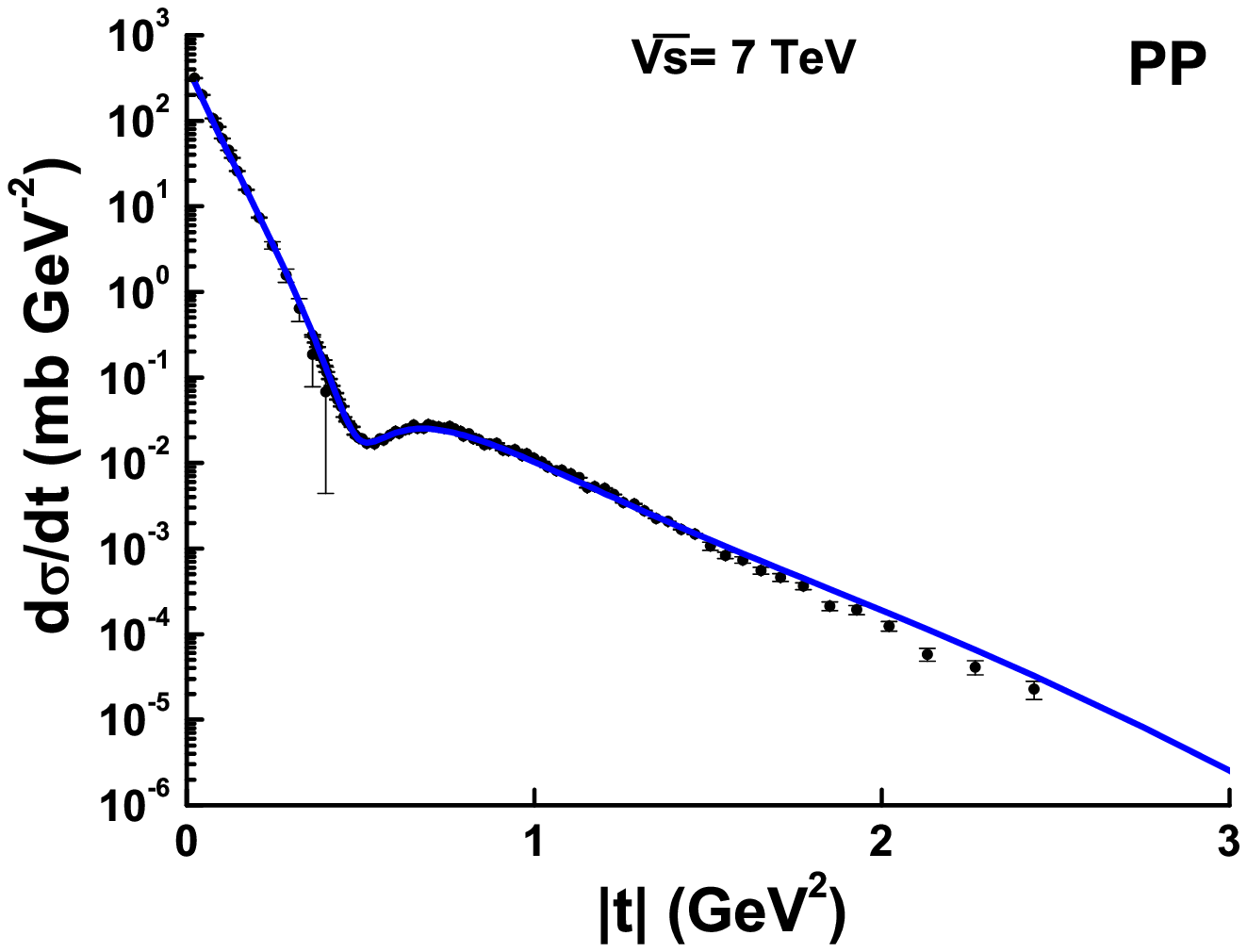}
\begin{minipage}{75mm}
{
\caption{The points are the experimental data by the Totem Collaboration (G. Antchev et al.,
         Europhys. Lett., {\bf 96}, 21002 (2011); Europhys. Lett., {\bf 95}, 41001 (2011)) digitized by us.}
}
\end{minipage}
\hspace{5mm}
\begin{minipage}{75mm}
{
\caption{The points are the experimental data by the Totem Collaboration (G. Antchev et al.,
         Europhys. Lett., {\bf 96}, 21002 (2011); Europhys. Lett., {\bf 95}, 41001 (2011)) digitized by us.}
}
\end{minipage}
%-------------------------------------------------------
\includegraphics[width=150mm,height=120mm,clip]{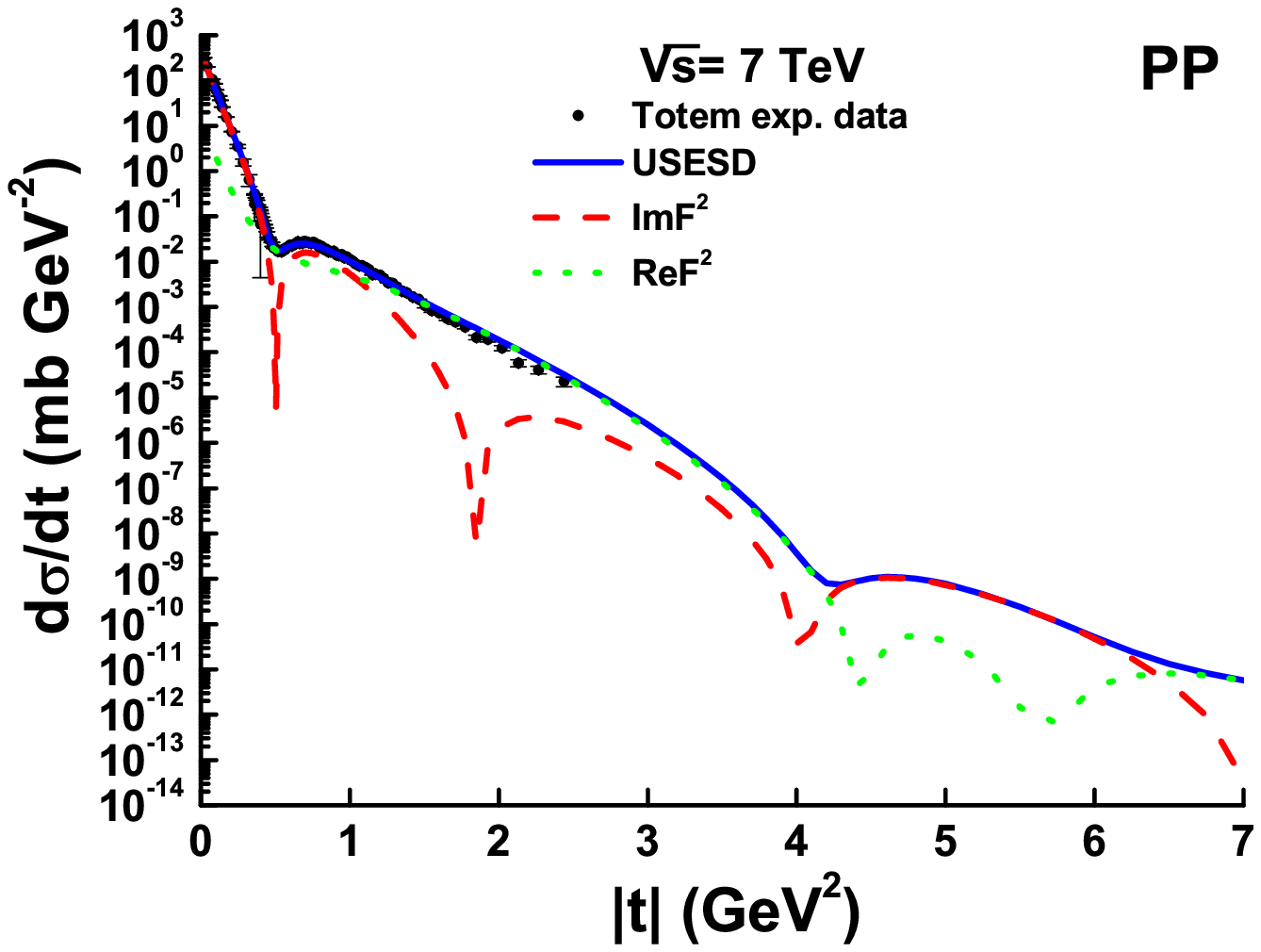}
\caption{The points are the experimental data by the Totem Collaboration (G. Antchev et al.,
         Europhys. Lett., {\bf 96}, 21002 (2011); Europhys. Lett., {\bf 95}, 41001 (2011)) digitized by us.}
%-------------------------------------------------------
\end{figure}


\begin{thebibliography}{1111}
\bibitem{Asymptotica}A. Grau, S. Pacetti, G. Pancheri, and Y.N. Srivastava, Phys. Lett., {\bf B714}, 70 (2012).
\bibitem{TwoExp}R.J.N. Phillips and V.D. Barger, Phys. Lett., {\bf B46}, 412 (1973).

\bibitem{Totem}
The Totem Collaboration (G. Antchev et al.) Europhys. Lett., {\bf 96}, 21002 (2011);\\
The Totem Collaboration (G. Antchev et al.) Europhys. Lett., {\bf 95}, 41001 (2011).

\bibitem{PbarP1}H.B. Crawley, E.S. Hafen and W.J. Kerman, Phys. Rev., {\bf D8}, 2012 (1973).
\bibitem{PbarP2}H.B. Crawley, W.J. Kerman and F. Ogino, Phys. Rev., {\bf D8}, 2781 (1973).
\bibitem{PbarP3}H.B. Crawley, N.W. Dean, E.S. Hafen, W.J. Kerman and F. Ogino, Phys. Rev., {\bf D9}, 189 (1974).
\bibitem{PbarP4}H.B. Crawley, N.W. Dean, and W.J. Kerman, Phys. Rev., {\bf D9}, 3029 (1974).
\bibitem{OurPaperPbarP}A. Galoyan, J. Ritman, A. Sokolov and V. Uzhinsky, arxiv:0809.3804 [hep-ex] (2008).

\bibitem{USESD}V. Uzhinsky and A. Galoyan, arxiv:1111.4984 [hepph] (2011).

\bibitem{SFermi}D.W.L. Sprung and J. Martorell, J. Phys. {\bf A30} (1997) 6525; {\bf A31} (1998) 8973.

\bibitem{DL}A. Donnachie and P.V. Landshoff, Nucl. Phys., {\bf B231}, 189 (1984).
\bibitem{Martynov}E. Martynov, J.R. Cudel and A. Lengyel, arxiv:1201.4458 [hep-ph] (2012).

\bibitem{PDG}Particle Data Group, http://pdg.lbl.gov/2009/hadronic-xsections/hadron.html

\bibitem{Plab} %14.2, 19.2, 24, 200, 293, 501
J.V. Allaby et al., Nucl. Phys. {\bf B52}, 316 (1973).\\
J.V. Allaby et al., Phys. Lett. {\bf B28}, 67 (1968).\\
A. Schiz et al., Phys. Rev. {\bf D24},  26 (1981).\\
G. Fidecaro et al., Nucl. Phys. {\bf B173}, 513 (1980).\\
U. Amaldi and K.R. Schubert, Nucl. Phys. {\bf B166},  301 (1980).

\bibitem{SqrtS}% 44.7, 52.9, 62.5
U. Amaldi and K.R. Schubert, Nucl. Phys. {\bf B166},  301 (1980).\\
E. Nagy et al., Nucl. Phys. {\bf B150},  221 (1979).

\bibitem{PhobosMC}
B. Alver, M. Baker, C. Loizides, and P. Steinberg, arxiv:0805.4411 [nucl-exp] (2005).
\bibitem{Polyaki}
W. Broniowski, M. Rybczynski, and P. Bozek, Comp. Phys. Commun., {\bf 180}, 69 (2009).
\bibitem{GLmc}
M.L. Miller, K. Reygers, S.J. Sanders and P. Steinberg,
                   Ann. Rev. Nucl. Part. Sci., {\bf 57}, 205 (2007).




\end{thebibliography}
\end{document}